\documentclass[11pt]{article}

\usepackage{latexsym}
\usepackage{geometry}
\usepackage{epic}
\usepackage{eepic}
\newcommand{\wrt}{w.r.t.\ }
\newcommand{\If}{\leftarrow}
\newcommand{\eq}{\!=\!}
\newcommand{\deq}{\!\neq\!}
\newcommand{\mgu}{\mathit{mgu}}
\newcommand{\vars}{\mathit{vars}}
\newcommand{\true}{\mathit{true}}

\newtheorem{lemma}{Lemma}
\newtheorem{theorem}{Theorem}
\newenvironment{proof}{\begin{trivlist}
                       \item[]{\em Proof\/}:}{\hfill $\Box$
                       \end{trivlist}
                      }
\newtheorem{definition}{Definition}

\newtheorem{example}{Example}
\newtheorem{proposition}{Proposition}
\newtheorem{trule}{Rule}

\newcommand{\feq}{$\begin{array}{rlll}}
\newcommand{\nex}{\end{array}$\par $\begin{array}{rlll}}
\newcommand{\eeq}{\end{array}$}
\newcommand{\num}[1]{\hspace*{-.5cm}\makebox[.8cm][r]{#1}&}
\newcommand{\Feq}[1]{\feq\num{#1}}
\newcommand{\Nex}[1]{\nex\num{#1}}
\newcommand{\nc}{&\hspace*{-.2cm}} %advance one column
\newcommand{\bdy}{\\ \nc\nc\nc}
\newcommand{\bpr}{\par\smallskip}
\newcommand{\epr}{\par\smallskip}
\newcommand{\fpr}[2]
     {\par\smallskip\fbox{
     \begin{minipage}[t]{15.5cm}
     {\bf Program} #1 \par\smallskip
     #2 \end{minipage}}\par\smallskip}
\newcommand{\oeq}[2]{\par\smallskip
$\begin{array}{rl}\hspace*{-.5cm}
\makebox[.8cm][r]{#1}&{#2}\end{array}$
\par\smallskip}
\begin{document}

\title{Derivation of Efficient Logic Programs by Specialization and
Reduction of  Nondeterminism\footnote{A preliminary version of this paper
appears as: Reducing Nondeterminism while Specializing Logic
Programs. {\em Proceedings of the 24th Annual ACM Symposium on
Principles of Programming Languages, Paris, France, January 15--17, 1997},
ACM Press, 1997, pp. 414--427.}}

\author{{\bf \large Alberto Pettorossi}\\
{\normalsize DISP, University of Roma Tor Vergata, Roma, Italy}\\
\texttt{adp@iasi.rm.cnr.it}\\
\\
{\bf \large Maurizio Proietti}\\
{\normalsize IASI-CNR, Roma, Italy}\\
\texttt{proietti@iasi.rm.cnr.it}\\
\\
{\bf \large Sophie Renault}\\
{\normalsize European Patent Office,
Rijswijk,
The Netherlands}\\
\texttt{srenault@epo.org}
}

\maketitle

\begin{abstract}
\noindent
Program specialization is a program transformation methodology which
improves program efficiency by exploiting the information
about the input data which are available at compile time.
We show that current techniques for program
specialization based on partial evaluation do not perform well on
nondeterministic logic programs.
We then consider a set of transformation rules
which extend the ones used for partial evaluation, and
we  propose a strategy for guiding the application of these extended
rules so  to derive very efficient specialized programs. The
efficiency improvements which sometimes are exponential,
are due to the reduction of nondeterminism and to the fact that
the computations which are performed
by the initial programs in different branches of the
computation trees, are performed by the specialized programs
within single branches. In order to reduce nondeterminism
we also make use of mode information for guiding the unfolding
process.
To exemplify our technique, we show that we can automatically derive
very efficient matching programs and parsers for regular languages.  The
derivations we have performed could not have been done by previously
known partial evaluation techniques.
\end{abstract}

\section{Introduction}
\label{intro}

\normalsize

The goal of {\it program specialization}~\cite{Jo&93} is the
 adaptation of a generic program to a specific context of use.
{\it Partial evaluation}~\cite{Da&96,Jo&93}
is a well established technique for program
specialization which from a program and its {\em static input} (that is,
the portion of the input which is known at compile time), allows us to
derive a new, more efficient program in which the portion of the
output which depends on the static input, has already been computed.
Partial evaluation has been applied
in several areas of computer science, and it has been applied  also
to logic programs~\cite{Gall93,Le&98b,LlS91}, where
it is also called {\em partial deduction}. In this paper we follow a
{\it rule-based} approach to the
specialization of logic programs~\cite{Bo&90,Pre93b,PrP93b,Sah93}.  In
particular, we consider definite logic programs~\cite{Llo87} and we propose
new program specialization techniques based on \mbox{unfold/}\mbox{fold}
transformation rules~\cite{BuD77,TaS84}. In our approach, the process  of
program specialization can be viewed as the construction of a sequence, say
$P_0, \ldots, P_n$, of programs, where $P_0$ is the program to be
specialized, $P_n$ is the derived, specialized program,  and every program 
of the sequence is obtained from the previous one by applying a
transformation rule.

As shown in~\cite{Pre93b,Sah93}, partial deduction can be viewed as a 
particular rule-based program transformation technique using the
definition, unfolding, and folding rules~\cite{TaS84} with the following
two restrictions:  (i) each new predicate introduced by the definition rule
is defined by precisely one non-recursive clause whose body consists of
precisely one atom (in this sense, according to the terminology
of~\cite{GlS96}, partial deduction is said to be {\em
monogenetic}), and (ii) the folding rule uses only clauses
introduced by the definition rule. In what follows
the definition and folding rules which comply with restrictions
(i) and (ii), are called {\em atomic definition}
and {\em atomic folding}, respectively.

In Section~\ref{rules_stra_pe_sect} we will see that the use of these
restricted transformation rules makes it easier to automate the partial
deduction process, but it may limit the program improvements
which can be achieved during  program specialization.
In particular, when we perform partial deduction of nondeterministic
programs using  atomic definition, unfolding, and atomic folding,
it is impossible to combine information present in  different
branches of the computation trees, and as a consequence, it is  often
the case that we cannot reduce the nondeterminism of the programs.

This weakness of partial deduction is demonstrated in
Section~\ref{limitations_pd}  where we revisit the familiar problem of
looking for occurrences of a pattern in a string. It has been
shown in~\cite{Fuj87,Gall93,GlK93} that by partial deduction
of a string matching program, we may
derive a deterministic finite automaton (DFA, for short), similarly to what 
is done by the Knuth-Morris-Pratt algorithm~\cite{Kn&77}.  However,
in~\cite{Fuj87,Gall93,GlK93} the string matching program to which
partial deduction is applied, is {\em deterministic}. We show that by
applying partial deduction to a {\em nondeterministic}
version of the matching program, one cannot derive a specialized
program which is deterministic, and thus, one cannot get a program
which corresponds to a DFA.

{\it Conjunctive partial deduction}~\cite{De&99} is a program
specialization technique which extends partial deduction by
allowing the specialization of logic programs w.r.t.~conjunctions
of atoms, instead of a single atom. Conjunctive partial
deduction can be realized by the definition, unfolding, and folding
rules where  each new
predicate introduced by the definition rule is defined by precisely
one non-recursive clause whose body is a conjunction of atoms
(in this sense conjunctive partial deduction is said to be
{\em polygenetic}).

Conjunctive partial deduction may sometimes reduce nondeterminism.
In particular, it may transform generate-and-test programs
into programs where the generation phase and the test phase are
interleaved. However, as shown in Section~\ref{limitations_pd},
conjunctive partial deduction is not capable to derive from the
nondeterministic version of the matching program a new
program which corresponds to a DFA.

In our paper, we propose a
specialization technique which enhances both partial deduction and
conjunctive partial deduction by making use of more powerful transformation
rules. In particular, in Section~\ref{rules_sect} we consider a version of
the definition introduction rule so that a new predicate may be introduced
by means of several non-recursive clauses whose bodies consist of
conjunctions of atoms, and we allow folding steps which use these predicate
definitions consisting of several clauses. We also consider the following
extra rules: {\em head generalization}, {\em case split},
{\em equation elimination}, and {\em disequation replacement}.
These rules may introduce, replace, and eliminate equations and negated
equations between terms.

Similarly to~\cite{GeK94,TaS84,Ro&99a}, our extended set of
program transformation rules preserves the least Herbrand model semantics.
For the logic language with equations and
negated equations considered in this paper, we adopt the usual
Prolog operational semantics with the left-to-right selection rule,
where equations are evaluated by using unification.
Unfortunately, the unrestricted use of the extended set of
transformation rules may not preserve the Prolog
operational semantics. To overcome this problem,
we consider:
(i)~the class of {\em safe programs} and
(ii)~suitably restricted transformation rules, called {\em safe
transformation rules}.
Through some examples we show that the class of
safe programs and the safe transformation rules are general enough to
allow significant program specializations.

Our notions of safe programs and transformation rules, and also
the notion of determinism are based on the {\em modes}
which are associated with predicate calls~\cite{Mel87,War77}.
We describe these notions in Section~\ref{modes},
where we also prove that the application of safe transformation rules
preserve the operational semantics of safe programs.

Then, in Section~\ref{strategy}, we introduce a strategy, called {\em
Determinization}, for applying our safe transformation rules in an
automatic way, so to specialize programs and  reduce their nondeterminism.
The new features of our strategy w.r.t.~other specialization
techniques are: (i)~the use of mode information for unfolding
and producing deterministic programs, (ii)~the use of
the case split rule for deriving {\em mutually exclusive} clauses (e.g.
from the clause $H \leftarrow {\it Body}$ we may derive the
two clauses: $(H \leftarrow {\it Body})\{X/t\}$ and
$H \leftarrow X\deq t, {\it Body}$), and
(iii)~the use of the enhanced
definition and folding rules for replacing many
clauses by one clause only, thereby reducing nondeterminism.

{F}inally, in Section~\ref{examples}, we show by means of some examples which
refer to parsing and matching problems, that our strategy is more powerful
than both partial deduction and conjunctive partial deduction. In
particular, given a nondeterministic version of the matching program, one
can derive by using our strategy a specialized program which corresponds to
a DFA.

\section{Logic Programs with Equations and Disequations between Terms}

In this section we introduce an extension of definite logic
programs with equations and negated equations between terms. Negated
equations will also be called {\em disequations}. The introduction of
equations and disequations during program specialization allows us to
derive mutually exclusive clauses.
The declarative semantics we consider, is a straightforward
extension of the usual least Herbrand model of definite logic programs.
The operational semantics essentially is SLD-resolution as implemented by
most Prolog systems: atoms are selected from left to right, and equations
are evaluated by using unification. This operational semantics is sound
w.r.t.~the declarative semantics (see Theorem~\ref{soundness} below).
However, since non-ground disequations can be selected, a goal evaluated
according to our operational semantics can fail, even if it is true
according to the declarative semantics. In this sense, the operational
semantics is not complete w.r.t.~the declarative semantics.

{F}or the notions of
{\em substitution},
{\em composition} of substitutions,
{\em identity} substitution,
{\em domain} of a substitution,
{\em restriction} of a substitution,
{\em instance},
{\em most general unifier} (abbreviated as {\it mgu}),
{\em ground expression},
{\em ground substitution},
{\em renaming substitution},
{\em variant}, and
for other notions not defined here, we refer to~\cite{Llo87}.

\subsection{Syntax} \label{syntax}

The syntax of our language is defined starting from
the following infinite and pairwise disjoint sets:
\\
(i) \emph{variables\/}:
\( X,Y,Z,X_{1},X_{2},\ldots,  \)
\\
(ii) \emph{function symbols} (with arity): \( f,f_{1},f_{2},\ldots,  \) and
\\
(iii) \emph{predicate symbols} (with arity):
{\it true}, =, $\neq$, \( p,p_{1},p_{2},\ldots  \) The
predicate symbols {\it true}, =, and $\neq$ are said to be {\em
basic}, and the other predicate symbols are said to be {\em non-basic}.
Predicate symbols will also be called {\em predicates}, for short.

\medskip
\noindent Now we introduce the following sets:
(iv)~\textit{Terms}: $t,t_{1},t_{2},\ldots,$ \ \
(v)~\textit {Basic atoms}: $B, B_1, B_2, \ldots,$  \ \
(vi)~\textit{Non-basic atoms}: $A,A_{1},A_{2},\ldots,$ and
(vii)~\textit{Goals}: $G,G_{1},G_{2},\ldots$
Their syntax is as follows:

\smallskip

$\begin{array}{lcl}
\textit{Terms}: & $\hspace{0.2cm}$ &  t::=X\ |\
f(t_{1},\ldots ,t_{n})  \\
\textit{Basic Atoms}: & &   B::= \mathit{true}\ |\
	t_{1}\! =\! t_{2}\ |\
	t_{1}\! \neq\! t_{2} \\
\textit{Non-basic Atoms}: & & A::= p(t_{1},\ldots
,t_{m}) \\
\textit{Goals}: & &  G::= B\ |\
	A\ |\
        G_{1},G_{2}
\end{array}
$

\smallskip

\noindent Basic and non-basic atoms are collectively called
{\em atoms}. Goals made out of basic atoms only are said
to be {\em basic goals}.  Goals with at least one  non-basic
atom are said to be   {\em non-basic goals}.
The binary operator `,' denotes
\emph{conjunction} and it is assumed to be associative with neutral
element \emph{true}. Thus, a goal \( G \) is
the same as goal \( (\mathit{true},G) \), and it is also the same as
goal \( (G,\mathit{true}) \).

\smallskip

\noindent \emph{Clauses}: \( C,C_{1},C_{2},\ldots  \)
have the following syntax:

\smallskip
\( C::=A\leftarrow G \)
\smallskip

\noindent
Given a clause \( C \) of the form: $A\leftarrow G$, the non-basic atom
$A$ is called the \emph{head} of $C$ and it is
denoted by $hd(C)$, and the goal \( G \) is called the \emph{body} of
\( C \) and it is denoted by $bd(C)$. A clause
\( A \leftarrow G \) where $G$ is a basic goal, is called a
{\em unit clause}.
We write a unit clause of the form:
\( A \leftarrow \textit {true} \) also as: \( A\leftarrow  \).

\noindent We say that $C$ {\em is a clause for\/} a predicate
$p$ iff $C$ is a clause of the form
$p(\ldots)\If G$.

\smallskip

\noindent \textit{Programs}: \( P,P_{1},P_{2},\ldots  \)  are {\em
sets} of clauses.

\smallskip \noindent
In what follows we will feel free to use different meta-variables to denote
our syntactic expressions, and
in particular, we will also denote non-basic atoms by $H,H_1,\ldots$,
and goals by $K,K_1,\mathit{Body},\mathit{Body}_1,\ldots$

Given a program $P$, we consider the relation ${\it \delta_P}$
over pairs of predicates such that
${\it \delta_P}(p,q)$ holds iff there exists in $P$ a clause for $p$ whose
body contains an occurrence of $q$. Let ${\it \delta^+_P}$ be the
transitive closure of ${\it \delta_P}$. We say that
$p$ {\it depends on $q$ in\/} $P$ iff ${\it \delta^+_P}(p,q)$ holds.
We say that a predicate $p$ depends on a clause $C$ in a program $P$
iff {\em either} $C$ is a clause for $p$ {\em or}
$C$ is a clause for a predicate
$q$ and $p$ depends on $q$ in $P$.

Terms, atoms, goals, clauses, and programs
are collectively called {\em expressions},
ranged over by $e, e_1, e_2, \ldots$
By $\mathit{vars}(e)$  we denote the set of variables occurring in an
expression $e$.
We say that $X$ is a {\em local}
variable of a goal $G$ in a clause $C:$ $H\leftarrow G_1, G, G_2$
iff
$X\in {\it vars}(G)\!-\! {\it vars}(H,G_1,G_2)$.

The application of a renaming substitution to an expression is
also called a {\em renaming of variables}.
A renaming of variables can be applied to a clause
whenever needed, because it preserves
the least Herbrand model semantics which we define below.
Given a clause $C$, a {\em renamed apart} clause
$C^\prime$ is any clause  obtained from $C$ by a
renaming of variables, so that each variable of $C^\prime$ is a
fresh new variable. (For a formal definition of this concept, see the
definition of {\em standardized apart} clause in~\cite{Apt90,Llo87})

{F}or any two unifiable terms $t_1$ and $t_2$, there exists
at least one mgu $\vartheta$ which is
{\em relevant} (that is, each variable
occurring in $\vartheta$ also occurs in
${\it vars}(t_1)\cup {\it vars}(t_2)$) and
{\em idempotent} (that is, $\vartheta
\vartheta = \vartheta$)~\cite{Apt90}. Without loss of
generality, we assume that all mgu's considered in this paper
are relevant and idempotent.

\subsection{Declarative Semantics}\label{decl_sem}

In this section we extend the definition of least 
Herbrand model of definite logic programs~\cite{Llo87} to 
logic programs 
with equations and disequations between terms. We follow the approach
usually taken when defining the least $\mathcal{D}$-model of 
CLP programs (see, for instance, \cite{Ja&98}). 
According to this approach, we consider a class of Herbrand models, 
called $\mathcal{H}$-{\em models}, where the
predicates {\em true}, $=$, and $\neq$ have a 
fixed interpretation.
In particular, the predicate $=$ is interpreted as the
identity relation over the Herbrand universe and the predicate~$\neq$
is interpreted as the complement of the identity relation.
Then we define the least Herbrand model of a logic program
with equations and disequations between terms as the least 
$\mathcal{H}$-model of the program.

The {\em Herbrand base} $\mathcal{HB}$
is the set  of all ground {\it non-basic} atoms.
An  $\mathcal{H}$-interpretation is a subset of $\mathcal{HB}$.
Given an $\mathcal{H}$-interpretation
$I$ and a ground goal, or ground clause, or program $\varphi$, the relation
$I \models \varphi$, read as $\varphi$ {\em is true in} $I$, is inductively
defined as follows (as usual, by $I\not\models \varphi$ we indicate
that $I\models \varphi$ does not hold):

\smallskip
%\begin{itemize}
%\item[(i)]
(i) $I \models \true$

%\item[(ii)]
(ii) for every ground term $t$, $I \models t\!=\!t$

%\item[(iii)]
(iii) for every pair of distinct ground
terms $t_1$ and $t_2$, $I \models t_1 \deq t_2$

%\item[(iv)]
(iv) for every non-basic ground atom $A$, $I\models A$ iff $A \in
I$

%\item[(v)]
(v) for every pair of ground goals $G_1$ and $G_2$, $I\models G_1,G_2$ iff
$I\models G_1$ and $I\models G_2$

%\item[(vi)]
(vi) for every ground clause $C$, $I\models C$ iff
either $I\models hd(C)$ or $I\not\models bd(C)$

%\item[(vii)]
(vii) for every program $P$, $I\models P$ iff for every ground
instance $C$ of a clause in $P$, $I\models C$.
%\end{itemize}

\smallskip
\noindent
As a consequence of the above definition, a ground basic goal is
true in an $\mathcal{H}$-interpretation iff it is true in all 
$\mathcal{H}$-interpretations. We say that a ground basic goal 
{\em holds} iff it is
true in all $\mathcal{H}$-interpretations.

An $\mathcal{H}$-interpretation $I$ is said to be an $\mathcal{H}$-{\em model}
of a program $P$ iff $I\models P$.
Since the model
intersection property holds for $\mathcal{H}$-models, 
similarly to~\cite{Ja&98,Llo87}, we can prove the
following important result.

\begin{theorem}
{\rm
{F}or any program $P$ there exists an $\mathcal{H}$-model of $P$ which is
the least (\wrt set inclusion) $\mathcal{H}$-model.
} %end rm
\end{theorem}

\noindent The {\em least Herbrand model} of a program $P$ is defined
as the least $\mathcal{H}$-model of $P$ and is denoted by
$M(P)$.

\subsection{Operational Semantics}\label{op_sem}

We define the  operational semantics of our programs
by introducing, for each program $P$, a
relation $G_1 \stackrel\vartheta\longmapsto_P G_2$, where $G_1$ and $G_2$
are goals and $\vartheta$ is a substitution, defined as follows:

\smallskip
$\begin{array}{llcl}
(1) & (t_1\!=\!t_2,\ G)\ \stackrel\vartheta\longmapsto_P \ G\vartheta  & {\rm
iff} & t_1 {\rm \ and\ } t_2 {\rm \ are\ unifiable\ via\ an\ mgu\ }
\vartheta\\
(2) & (t_1\neq t_2,\ G)\ \stackrel\varepsilon\longmapsto_P \ G  &
{\rm iff} & t_1 {\rm \ and\ } t_2 {\rm \ are\ not\ unifiable\ and\
\varepsilon\ is\ the\ identity\ substitution}\\
(3) & (A,G)\ \stackrel\vartheta\longmapsto_P \
(bd(C),G)\vartheta & {\rm iff} & {\rm (i)\ } A {\rm \ is\ a\  non}$-${\rm
basic\ atom,}\\ & & & {\rm (ii)\  } C {\rm \ is\ a\ renamed\
apart\ clause\ in\ } P, {\rm \ and} \\
& & & {\rm (iii)\  } A {\rm \ and\ } hd(C)  {\rm \ are\ unifiable\ via\
an\ mgu\ } \vartheta.
\end{array}
$
\smallskip

\noindent

A sequence $G_0\stackrel{\vartheta_1}\longmapsto_P \dots
\stackrel{\vartheta_n}\longmapsto_P G_n$, with $n\!\geq\!0$, is called a
{\em derivation} using $P$. If $G_n$ is {\it true} then the derivation is
said to be {\em successful}. If there exists a successful derivation
$G_0\stackrel{\vartheta_1}\longmapsto_P \dots
\stackrel{\vartheta_n}\longmapsto_P {\it true}$
and $\vartheta$ is the substitution obtained by
restricting the composition $\vartheta_1 \dots \vartheta_n$ to the
variables of $G_0$, then we
say that the goal $G_0$ {\em succeeds} in $P$ with {\em  answer
substitution} $\vartheta$.

When denoting derivations, we will feel free to omit their associated
substitutions.
In particular, given two goals
$G_1$ and $G_2$, we write $G_1 \longmapsto_P G_2$ iff there exists a
substitution $\vartheta$ such that $G_1 \stackrel\vartheta\longmapsto_P G_2$.
We say that
$G_2$ is {\em derived in one step} from $G_1$ ({\em using} $P$) iff
$G_1 \longmapsto_P G_2$ holds.
In particular, if $G_2$ is derived in one step from $G_1$ according to
Point (3) of the operational semantics by using a clause $C$,
then we say that $G_2$ is derived in one step from $G_1$ {\em using} $C$.
The relation $\longmapsto^*_P$ is the
reflexive and transitive closure of $\longmapsto_P$. Given two goals
$G_1$ and $G_2$ such that $G_1 \longmapsto^*_P G_2$ holds, we say that
$G_2$ is {\em derived} from $G_1$ (using $P$).
We will feel free to omit the reference to program $P$
when it is understood from the context.

The operational semantics presented above
can be viewed as an
abstraction of the usual Prolog semantics, because: (i)~given a goal $G_1$,
in order to derive a goal $G_2$ such that $G_1 \longmapsto_P G_2$, we consider
the leftmost atom in $G_1$, (ii)~the predicate $=$  is interpreted as
unifiability of terms, and (iii)~the predicate $\neq$ is
interpreted as non-unifiability of terms. Similarly
to~\cite{Llo87}, we have the following relationship between the
declarative and the operational  semantics.

\begin{theorem} \label{soundness}
{\rm
{F}or any program $P$ and ground goal $G$,
if $G$ succeeds in $P$ then $M(P)\models G$.
} % end rm
\end{theorem}

\noindent The converse of Theorem~\ref{soundness} does not hold. Indeed,
consider the program $P$ consisting of the clause
$p(1) \leftarrow X\!\neq \!0$\ \  only. We have that $M(P)\models
p(1)$ because there exists a value for $X$, namely 1, which is
syntactically different from 0. However, $p(1)$ does not succeed in
$P$, because $X$ and $0$ are unifiable terms.

% \smallskip
%
% \noindent
% {\bf Remark} \ Let us observe that given a program $P$, a function
% symbol $f$, and a ground term $r$, (i) in the declarative semantics
% $r\!\neq\!f(X)$ holds, that is, $M(P) \models r\!\neq\! f(X)$, iff {\em there
% exists\/} a ground term $s$ such that $r$ is different from
% $f(s)$, and (ii)~in the operational semantics,  $r\!\neq\!f(X)$ succeeds in
% $P$, that is, $r\!\neq\! f(X) \longmapsto_P {\mathit true}$, iff {\em for
% all\/} ground terms $s$ we have that $r$ is different from
% $f(s)$.

\subsection{Deterministic Programs}\label{determ_sec}

Various notions of determinism have been proposed
for logic programs in the literature (see, for instance,
\cite{Dev90,Fe&96,Mel85,SaTa85}). They capture various properties such as:
``the program succeeds at most once", or
``the program succeeds exactly once", or
``the program will never backtrack to find alternative solutions".

Let us now present the definition of
deterministic program used in this paper. This definition is based
on the operational semantics described in Section~\ref{op_sem}.

We first need the following notation.
Given a program $P$, a clause $C\in P$, and two goals $(A_0,G_0)$ and
$(A_n,G_n)$, where $A_0$ is a non-basic atom,
we write $(A_0,G_0)\Rightarrow_C (A_n,G_n)$  iff
there exists a derivation
$(A_0,G_0)\longmapsto_P \dots \longmapsto_P (A_n,G_n)$, such that:
(i)~$n\!>\!0$,
(ii)~$(A_1,G_1)$ is derived in one step from $(A_0,G_0)$ using $C$,
(iii)~for $i=1,\ldots,n-1$, $A_i$ is a basic atom, and
(iv)~either $A_n$ is a non-basic atom or $(A_n,G_n)$ is the basic atom
{\it true}.
We write $G_0\Rightarrow^*_P G_n$ iff there exist clauses
$C_1,\ldots,C_n$ in $P$
such that $G_0\Rightarrow_{C_1} \ldots \Rightarrow_{C_n} G_n$.

\begin{definition}  [Determinism] \label{determ_def}
{\rm
A program $P$ is {\em deterministic} for a non-basic atom  $A$
iff for each goal $G$ such that $A \Rightarrow^*_P G$, there exists
{\em at most one} clause $C$ such
that $G\Rightarrow_C G'$ for some goal $G'$.
} %end rm
\end{definition}

We say that a program $P$ is {\em nondeterministic} for a non-basic atom
$A$ iff it is not the case that $P$ is deterministic for $A$, that is,
there exists a goal $G$ derivable from $A$, and there exist at least two
goals $G_1$ and $G_2$, and two distinct clauses $C_1$ and $C_2$ in $P$,
such that $G\Rightarrow_{C_1} G_1$ and $G\Rightarrow_{C_2} G_2$.

\sloppy

According to
Definition~\ref{determ_def}, the following program is deterministic for
any atom of the form ${\it non\_zero}({\it Xs},{\it Ys})$ where {\it Xs}
is a ground list.

\fussy

\bpr
\Feq{1.} {\it non\_zero}([~],[~]) \If
\Nex{2.} {\it non\_zero}([0|{\it Xs}],{\it Ys}) \If
                {\it non\_zero}({\it Xs},{\it Ys})
\Nex{3.} {\it non\_zero}([X|{\it Xs}],[X|{\it Ys}]) \If X\deq 0,
                {\it non\_zero}({\it Xs},{\it Ys})
\eeq
\epr

\noindent
Notice that the above definition of a deterministic program for a non-basic
atom $A$ allows some search  during the
construction of a derivation starting from $A$. Indeed, there may be a goal
$G$ derived from $A$ such that from $G$ we can derive in one step two or
more new goals using distinct clauses. However, if the program is
deterministic for $A$, after evaluating the basic atoms occurring at
leftmost positions in these new goals, at most one derivation can be
continued and at most one successful derivation can be constructed. For
instance, from the goal ${\it non\_zero}([0,0,1],{\it Ys})$ we can derive
in one step two distinct goals: (i)~${\it non\_zero}([0,1],{\it Ys})$
(using clause~2), and (ii)~$0\deq 0, {\it non\_zero}([0,1],{\it Ys}')$
(using clause~3). However, there exists only one clause $C$ (that is,
clause~2) such that
${\it non\_zero}([0,0,1],{\it Ys})\Rightarrow_C G'$ for some goal
$G'$ (that is, ${\it non\_zero}([0,1],{\it Ys}')$).

\section{Partial Deduction via Unfold/Fold Transformations}
\label{rules_stra_pe_sect}

In this section we recall the rule-based approach to partial
deduction. We also point out some limitations of  partial
deduction~\cite{Pre93b,Sah93} and conjunctive partial
deduction~\cite{De&99}. These limitations motivate the introduction of the
new, enhanced rules and strategies for program specialization presented in
Sections~\ref{rules_sect},~\ref{modes}, and~\ref{strategy}.

\subsection{Transformation Rules and Strategies for
 Partial Deduction}
\label{rules_stra_pe}

In the rule-based approach, partial deduction
can be viewed as the construction of a sequence $P_0, \ldots, P_n$
of programs, called a {\em transformation sequence}, where
$P_0$ is the initial program to be specialized, $P_n$ is the
final, specialized program, and for $k=0,\ldots,n-1$,
program $P_{k+1}$ is {\em derived} from program $P_k$ by
by applying one of the following {\em transformation rules} PD1--PD4.

\medskip
\noindent
{\bf Rule PD1 (Atomic Definition Introduction)}
We introduce
a clause $D$, called {\it atomic definition
clause}, of the form

\smallskip
${\it newp}(X_1,\dots,X_h) \leftarrow A$

\smallskip
\noindent
where
(i) {\it newp\/} is a non-basic predicate symbol not occurring in
$P_0, \ldots, P_k$,
(ii) $A$ is a non-basic atom whose predicate occurs in program $P_0$, and
(iii) $\{X_1,\dots,X_h\} = \mathit{vars}(A)$.

\noindent
Program $P_{k+1}$ is the program $P_k \cup
\{D\}$.

We denote by $\mathit{Defs_k}$ the set of atomic definition
clauses which have been introduced by
the definition introduction rule during the construction of the
transformation sequence $P_0, \ldots, P_k$. Thus, in particular,
we have that $\mathit{Defs}_0 = \emptyset$.

\medskip
\noindent
{\bf Rule PD2 (Definition Elimination)}.
Let $p$ be a predicate symbol. By {\em definition elimination
\wrt}$\!p$ we derive the program $P_{k+1}= \{C \in P_k ~|~ p$ depends on
$C\}$.

\medskip
\noindent
{\bf Rule PD3 (Unfolding)}.
Let $C$ be a renamed apart clause of $P_k$ of the
form: $H \leftarrow G_1, A, G_2$,
where~$A$ is a non-basic atom.  Let
$C_1,\dots, C_m$, with $m\geq 0$, be the clauses of
$P_k$ such that, for $i = 1,\dots, m$, $A$ is unifiable with the head of
$C_i$ via  the mgu\ $\vartheta_i$.
By {\em unfolding $C$ \wrt $A$}, for $i = 1,\dots, m$,
we derive the clause $D_i:$
$(H \leftarrow G_1, bd(C_i), G_2)\vartheta_i$.

\noindent
Program $P_{k+1}$ is the program $(P_k - \{C\}) \cup
\{D_1,\ldots,D_m\}$.

\medskip
\noindent
{\bf Rule PD4 (Atomic Folding)}.
Let $C$ be a renamed apart clause of $P_k$ of the form:
$H \leftarrow G_1, A\vartheta, G_2$, where:
(i)~$A$ is a non-basic atom, and (ii)~$\vartheta$
is a substitution, and
let $D$ be an atomic definition clause in $\mathit{Defs_k}$ of the form: $N
\leftarrow A$. By {\em folding $C$ \wrt $A\vartheta$ using $D$} we
derive the non-basic atom $N\vartheta$ and we derive
the clause $E:$ $H \leftarrow G_1, N\vartheta, G_2$.

\noindent
Program $P_{k+1}$ is the program $(P_k - \{C\}) \cup
\{E\}$.

\medskip

The partial deduction of a program $P$ may be realized by applying
the atomic definition introduction, definition elimination, unfolding, and
atomic folding rules, according to the so called {\em partial
deduction strategy} which we will describe below. Our partial deduction
strategy uses two subsidiary strategies: (1) an {\em Unfold}  strategy,
which derives new sets of clauses by repeatedly applying the unfolding
rule, and (2) a {\em Define-Fold} strategy, which introduces new atomic
definition clauses and it folds the clauses derived
by the {\em Unfold} strategy.
These subsidiary strategies use an {\em unfolding selection function} and a
{\em generalization function}, which we now define. Let us first
introduce the following notation: (i)~{\em NBAtoms} is the
set of all non-basic atoms, (ii)~{\em Clauses} is the set of all
clauses, (iii)~$\mathit{Clauses}^*$ is the set of all finite
sequences of clauses, (iv)~$\mathcal{P}(\mathit{Clauses})$ is the powerset
of {\it Clauses}, (v)~a sequence of clauses is denoted by $C_1,\ldots,C_n$,
and (vi)~the empty sequence of clauses is denoted by $()$.

An {\em unfolding
selection function} is a total
function $\mathit{Select}:\mathit{Clauses}^* \times \mathit{Clauses}
\rightarrow \mathit{NBAtoms} \cup \{\mathit{halt}\}$, where {\em halt}
is a symbol not occurring in $\mathit{NBAtoms}$. We assume that, for
$C_1,\ldots,C_n \in \mathit{Clauses}^*$ and $C \in \mathit{Clauses}$,
$\mathit{Select}((C_1,\ldots,C_n),C)$ is a non-basic
atom in the body of $C$.

When applying the {\it Unfold}
strategy the $\mathit{Select}$ function is used as follows.
During the unfolding process starting from a set {\it Cls} of clauses, we
consider a clause, say $C$, to be unfolded, and the sequence of its
{\em ancestor clauses}, that is, the sequence $C_1,\ldots,C_n$ of
clauses such that: (i)~$C_1\in{\it Cls}$, (ii)~for $k=1,\ldots,n\!-\!1$,
$C_{k+1}$ is derived by unfolding $C_k$, and (iii)~$C$ is derived by 
unfolding $C_n$. Now, (i)~if $\mathit{Select}((C_1,\ldots,C_n),C) = A$,
where $A$ is a non-basic atom in the body of $C$, then $C$ is unfolded \wrt
$A$, and (ii)~if $\mathit{Select}((C_1,\ldots,C_n),C)=\mathit{halt}$ then
$C$ is not unfolded.

A {\em generalization function}
$\mathit{Gen} : \mathcal{P}(\mathit{Clauses}) \times \mathit{NBAtoms}
\rightarrow \mathit{Clauses}$ is defined
for any set $\mathit{Defs}$ of atomic definition clauses and for any
non-basic atom $A$. $\mathit{Gen}(\mathit{Defs}, A)$ is
either a clause in {\it Defs} or a clause of the
form $g(X_1,\dots,X_h) \leftarrow \mathit{GenA}$,
where:
(i) $\{X_1,\dots,X_h\} = \mathit{vars}(\mathit{GenA})$,
(ii) $A$ is an instance of {\it GenA}, and
(iii) {\it g} is a new  predicate, that is, it occurs neither in
$P$ nor in {\it Defs}.

When applying the {\em Define-Fold} strategy  the generalization
function {\it Gen} is used as follows: when we want to fold a clause
$C$ \wrt a non-basic atom $A$ in its body, we consider the set {\it Defs}
of all atomic definition clauses introduced so far and
we apply the folding rule using  $\mathit{Gen}(\mathit{Defs}, A)$.
This application of the folding rule is indeed possible
because, by construction, $A$ is an instance of the body of
$\mathit{Gen}(\mathit{Defs}, A)$.

\noindent \hrulefill

\vspace*{-4mm}
\subsection*{Partial Deduction Strategy}

\vspace*{-2mm}
\noindent
{\bf Input}: A program $P$ and a non-basic atom
$p(t_1,\ldots,t_h)$ \wrt which we want to
specialize $P$.

\noindent
{\bf Output}: A program $P_\mathit{pd}$ and a non-basic
atom $p_\mathit{pd}(X_1,\ldots,X_r)$,
such that: (i)~$\{X_1,\ldots,X_r\} = \mathit{vars}(p(t_1,\ldots,t_h))$, and
(ii)~for every ground substitution $\vartheta =
\{X_1/u_1,\ldots,X_r/u_r\}$,

\smallskip
$M(P)\models p(t_1,\ldots,t_h)\vartheta$ iff
$M(P_\mathit{pd})\models p_{\mathit{pd}}(X_1,\ldots,X_r)\vartheta$.

\medskip
\noindent  {\it Initialize}:~~Let $S$ be the clause
$p_{\mathit{pd}}(X_1,\ldots,X_r)\If p(t_1,\ldots,t_h)$.
Let $\mathit{Ancestors}(S)$ be the empty sequence of clauses.
\\
\( {\it TransfP}:=P  \); ~\({\it Defs}:=\{S\}\);
~\( \mathit {Cls}:= \{S\} \);

\noindent
{\bf while} ~\( \mathit {Cls}\neq \emptyset  \)~ {\bf do}

\vspace{-3mm}

\begin{description}

\item {\rm (1)} {\em Unfold\/}:

\vspace{-3mm}

\begin{description}

\item[{while}]
there exists a clause $C\in\mathit {Cls}$
with \(\mathit{Select}(\mathit{Ancestors}(C),C)\neq\mathit{halt}\)
{\bf do}\\
Let $\mathit{Unf}(C) = \{ E~|~E$ is derived
by unfolding \(C\) w.r.t.~$\mathit{Select}(\mathit{Ancestors}(C),C)\}$.
\\
$\mathit {Cls}:= (\mathit{Cls} - \{C\}) \cup \mathit{Unf}(C)$;
\\
for each $E\in \mathit{Unf}(C)$ let
$\mathit{Ancestors}(E)$  be the sequence $\mathit{Ancestors}(C)$ followed
by $C$

\vspace{-1mm}

\item[{end-while{\rm ;}}]
\end{description}

\vspace{-4mm}
\item {\rm (2)} {\em Define-Fold\/}:

$\mathit{NewDefs} := \emptyset$;

\vspace{-3mm}

\begin{description}

\item[while]
there exists a clause $C\in\mathit{Cls}$ and there exists a non-basic atom
$A\in bd(C)$
which has not been derived by folding
{\bf do}
\\
Let $G$ be the atomic definition clause $\mathit{Gen}(\mathit{Defs},A)$
and $F$ be the clause derived by folding $C$ \wrt $A$ using $G$.
\\
$\mathit{Cls}:= (\mathit{Cls} - \{C\}) \cup \{F\}$;
\\
~{\bf if} $G\not\in \mathit{Defs}$ {\bf then}
($\mathit{Defs} := \mathit{Defs}\cup \{G\}$;
~$\mathit{NewDefs} := \mathit{NewDefs} \cup \{G\}$)

\vspace{-1mm}

\item[{end-while{\rm ;}}]

\end{description}

\vspace{-3mm}

\( {\it TransfP} := {\it TransfP} \cup \mathit {Cls} \);
~\( \mathit {Cls}:=\mathit {NewDefs} \)

\end{description}

\vspace{-3mm}
\noindent
{\bf end-while};

\smallskip
\noindent
We derive the final program $P_{pd}$ by applying
the definition elimination rule and keeping only the clauses of
{\it TransfP} on which $p_{pd}$ depends.

\noindent \hrulefill

\medskip
\noindent
A given unfolding selection function {\it Select} is said to be {\em
progressive} iff  for the empty sequence $()$ of clauses and for any
clause $C$ whose body contains at least one non-basic atom, we have that
$\mathit{Select}((),C)\neq \mathit{halt}$.

We have the following correctness result which is a straightforward
corollary of Theorem~\ref{corr_th_lhm} of Section~\ref{corr}.

\begin{theorem}
[Correctness of Partial Deduction \wrt the Declarative Semantics]
\label{pc_pe}
{\rm

~\\
Let ${\it Select}$ be a progressive unfolding selection
function. Given a program $P$ and a non-basic atom
$p(t_1,\ldots,t_h)$, if the partial deduction strategy
using {\it Select} terminates
with output program $P_\mathit{pd}$ and output atom
$p_\mathit{pd}(X_1,\ldots,X_r)$, then
 for every ground substitution $\vartheta = \{X_1/u_1,\ldots,X_r/u_r\}$,

$M(P)\models p(t_1,\ldots,t_h)\vartheta$ iff
$M(P_\mathit{pd})\models p_{\mathit{pd}}(X_1,\ldots,X_r)\vartheta$.

}
\end{theorem}

We say that an unfolding selection function {\it Select} is
{\em halting} iff
for any infinite sequence $C_1,C_2,\ldots$ of clauses,
there exists $n \geq 0$ such that
$\mathit{Select}((C_1,C_2,\ldots,C_{n}),C_{n+1})=\mathit{halt}$.

Given an infinite sequence $A_1,A_2,\ldots$ of non-basic atoms,
its {\em image} under the generalization function {\it Gen},
is the sequence of sets of clauses defined as follows:

\smallskip

~~~~$G_1 = \{\mathit{newp}(X_1,\ldots,X_n) \leftarrow A_1\}$, where
$\{X_1,\ldots,X_n\} = \mathit{vars}(A_1)$

~~~~$G_{i+1} = G_i \cup \{\mathit{Gen}(G_i,A_{i+1})\}$ ~~~for
$i\geq 1$.

\smallskip

We say that {\it Gen} is {\em stabilizing} iff for any
infinite sequence $A_1,A_2,\ldots$ of non-basic atoms
whose image under {\it Gen} is $G_1,G_2,\ldots,$
there exists $n>0$ such that $G_k = G_n$ for all
$k\geq n$.

We have the following theorem whose proof is similar to the
one in~\cite{Le&98a}.

\begin{theorem}[Termination of Partial Deduction] \label{term_pe}
{\rm
Let {\it Select} be a halting unfolding selection function
and {\it Gen}  be a stabilizing generalization function. Then
for any input program $P$ and non-basic atom
$p(t_1,\ldots,t_h)$, the partial deduction strategy using
{\it Select} and {\it Gen} terminates.
}
\end{theorem}

The following example shows that the unfolding rule
(and thus, the partial deduction strategy) is not
correct w.r.t.~the operational semantics.

\begin{example} \label{incorr_ex_unf}
{\rm Let us consider the following program
$P_1:$

\bpr
\Feq{1.} p \If X\!\neq\!a,\ q(X)
\Nex{2.} q(b) \If
\eeq
\epr

\noindent
By unfolding clause 1 w.r.t.~q(X) we derive the following
program $P_2$:

\bpr
\Feq{3.} p \If b\!\neq\!a
\Nex{2.} q(b) \If
\eeq
\epr

\noindent
We have that the goal $p$ does not succeed in $P_1$,
while it succeeds in $P_2$.
}% end rm
\end{example}

We will address this correctness issue in detail
in Section~\ref{modes}, where we will present a set of transformation rules
which are correct w.r.t.~the
operational semantics for the class of {\em safe} programs (see
Theorem~\ref{corr_th_op}).

\subsection{An Example of Partial Deduction: String Matching}
\label{string_mat}

In this section we illustrate the partial deduction strategy
by means of a well-known program specialization example which
consists in specializing a general string matching program w.r.t.~a
given pattern (see~\cite{Fuj87,Gall93,Smi91} for a similar example). Given
a program for searching a pattern in a string, and a fixed ground pattern
$p$, we want to derive a new, specialized program for searching the pattern
$p$ in a given string. Now we present a general program,
called {\it Match}, for searching a pattern $P$ in a string $S$ in
$\{a,b\}^*$. Strings in $\{a,b\}^*$ are denoted by lists of $a$'s and
$b$'s.
This program is deterministic for atoms of the
form ${\it match}(P,S)$, where $P$ and $S$ are ground lists.

\medskip

\fpr{{\it Match} \hfill (initial, deterministic)} {
\Feq{1.}{\it match}(P,S) \leftarrow {\it match}1(P,S,P,S)
\Nex{2.}{\it match}1([~],S,Y,Z)\leftarrow
\Nex{3.}{\it match}1([C|{\it P}],[C|{\it S}],Y,Z) \leftarrow
{\it match}1({\it P},{\it S},Y,Z)
\Nex{4.}{\it match}1([a|{\it P}],[b|{\it S}],Y,[C|Z]) \leftarrow
{\it match}1(Y,Z,Y,Z)
\Nex{5.}{\it match}1([b|{\it P}],[a|{\it S}],Y,[C|Z]) \leftarrow
{\it match}1(Y,Z,Y,Z)
\eeq}

\medskip

\noindent
Let us assume that we want to specialize this
program {\it Match} \wrt the goal ${\it match}([a,a,b],S)$, that is, we
want to derive a program which tells us whether or not the pattern
$[a,a,b]$ occurs in the string $S$.

We apply our partial deduction strategy using
the following unfolding selection function {\it DetU}
and generalization function {\it Variant}.

\medskip \noindent
(1) The function ${\it DetU} :\mathit{Clauses}^*
\times \mathit{Clauses} \rightarrow \mathit{NBAtoms} \cup
\{\mathit{halt}\}$ is defined as follows:

\smallskip

%\begin{itemize}
%\item[(i)]
\noindent (i) $\mathit{DetU}((),C)=A$ if $A$ is the leftmost non-basic atom
in the body of clause $C$,

%\item[(ii)]
\noindent (ii) $\mathit{DetU}((C_1,C_2,\ldots,C_n),C)=A$
if $n \geq 1$ and $A$ is the leftmost non-basic atom the body of $C$ such
that $A$  is unifiable with at most one clause head in the program to be
partially evaluated, and

%\item[(iii)]
\noindent (iii) $\mathit{DetU}((C_1,C_2,\ldots,C_n),C)=\mathit{halt}$ if
there exists no non-basic atom in the body of $C$ which
is unifiable with at most one clause head in the program to be
partially evaluated.

%\end{itemize}

\medskip \noindent
(2) The  function
$\mathit{Variant} : \mathcal{P}(\mathit{Clauses}) \times
\mathit{NBAtoms} \rightarrow \mathit{Clauses}$ is defined as
follows:

\smallskip

%\begin{itemize}
%\item[(i)]
\noindent (i) $\mathit{Variant}(\mathit{Defs}, A)$ is a
clause $C$ such that
$bd(C)$ is a variant of $A$, if in $\mathit{Defs}$ there
exists any
such clause $C$, and

%\item[(ii)]
\noindent (ii) $\mathit{Variant}(\mathit{Defs}, A)$ is the clause
${\it newp}(X_1,\dots,X_h) \leftarrow \mathit{A}$, where {\it newp} is a
new predicate symbol and $\{X_1,\dots,X_h\}=\mathit{vars}(A)$, otherwise.

%\end{itemize}

\medskip \noindent
The function ${\it DetU}$ corresponds to the
{\em determinate unfolding rule} considered
in~\cite{Gall93}. We have that ${\it DetU}$ is not halting
and {\it Variant} is not stabilizing. Nevertheless, in
our example, as the reader may verify, the partial deduction strategy
using ${\it DetU}$ and {\it Variant} terminates and generates the
following specialized program:

\medskip

\fpr { ${\it Match}_{pd}$ \hfill (specialized by partial deduction,
deterministic)}
{ \Feq{6.} \mathit{match_{pd}}(S) \If \mathit{new}1(S)
\Nex{7.} \mathit{new}1([a|S]) \If \mathit{new}2(S)
\Nex{8.} \mathit{new}1([b|S]) \If \mathit{new}1(S)
\Nex{9.} \mathit{new}2([a|S]) \If \mathit{new}3(S)
\Nex{10.} \mathit{new}2([b|S]) \If \mathit{new}1(S)
\Nex{11.} \mathit{new}3([b|S]) \If
\Nex{12.} \mathit{new}3([a|S]) \If \mathit{new}3(S)
\eeq}

\medskip
\noindent
The program $\mathit {Match_{pd}}$ is deterministic
for atoms of the form $\mathit{match_{pd}}(S)$, where
$S$ is a ground list, and it
corresponds to a DFA in the sense that:
(i) each predicate corresponds to a state,
(ii) each clause, except for clause 6 and 11,
corresponds to a transition from the
state corresponding to the  predicate of the head
to the state corresponding to the predicate of the body,
(iii) each transition is labelled by the symbol (either $a$ or $b$)
occurring in the head of the corresponding clause,
(iv) by clause 6 we have that {\it new}1 is the initial state for
goals of the form $\mathit{match_{pd}}(w)$, where $w$ is any
ground list representing a word in $\{a,b\}^*$, and
(v) clause 11 corresponds to
a transition, labeled by $b$, to an unnamed final state where any remaining
portion of the input word is accepted.

Thus, via partial deduction we may derive a DFA  from a deterministic
string matching program, and the derived program corresponds to the
Knuth-Morris-Pratt string matching
algorithm~\cite{Kn&77}.

\subsection{Some Limitations of Partial Deduction}
\label{limitations_pd}

The fact that the partial deduction strategy derives a DFA is a
consequence of the fact that the initial string matching program {\it
Match} is rather sophisticated and, indeed, the correctness proof of the
program {\it Match} is not straightforward. Actually, the partial deduction
strategy does {\it not\/} derive a DFA if we consider, instead of the
program {\it Match}, the following naive initial program
for string matching:

\medskip

\fpr{$\mathit{Naive\_Match}$ \hfill (initial, nondeterministic)} {
\Feq{1.}{\it naive\_match}(P,S) \nc \leftarrow \nc
            {\it append}(X,{\it R},S), \ {\it append}({\it L},P,X)
\Nex{2.}{\it append}([~],Y,Y) \leftarrow
\Nex{3.}{\it append}([A|X],Y,[A|Z]) \leftarrow  {\it append}(X,Y,Z)
\eeq
}

\medskip

\noindent This program is nondeterministic for atoms of the form
${\it naive\_match}(P,S)$, where $P$ and $S$ are ground lists.
The correctness of this naive program is straightforward because
for a given pattern $P$ and a string $S$, $\mathit{Naive\_Match}$
tests whether or not $P$ occurs in $S$ by  looking in a
nondeterministic way for two strings $L$ and $R$ such that $S$ is the
concatenation of $L$, $P$, and $R$ in this order.

The reader may verify that the partial deduction strategy does not derive a
DFA when starting from the program $\mathit{Naive\_Match}$. Indeed, if we
specialize $\mathit{Naive\_Match}$ \wrt the goal ${\it
naive\_match}([a,a,b],S)$ by applying the partial deduction strategy
using the unfolding selection function $\mathit{DetU}$ and the
generalization function {\it Variant}, then we derive the following
program $\mathit{Naive\_Match}_{pd}$ which  does {\em not} correspond to a
DFA and it is nondeterministic:

\medskip

\fpr{$\mathit{Naive\_Match}_{pd}$  \hfill
(specialized by partial deduction, nondeterministic)}{
\Feq{4.}{\it naive\_match_{pd}}(S) \nc
\leftarrow \nc           {\it new}1(X,{\it R},S),\
              {\it new}2({\it L},X)
\Nex{5.}{\it new}1([~],Y,Y) \leftarrow

\Nex{6.}{\it new}1([A|X],Y,[A|Z]) \leftarrow  {\it
new}1(X,Y,Z)
\Nex{7.}{\it new}2([~],[a,a,b]) \leftarrow
\Nex{8.}{\it
new}2([A|X],[A|Z]) \leftarrow  {\it new}2(X,Z) \eeq}

\medskip

\noindent
Indeed, this $\mathit{Naive\_Match}_{pd}$ program looks in a
nondeterministic way for two strings $L$ and $R$ such that $S$ is the
concatenation of $L$, $[a,a,b]$, and $R$. If the pattern $[a,a,b]$ is
not found within the string $S$ at a given position, then the search
for $[a,a,b]$ is restarted after a shift of one character to the
right of that position.

{F}rom the program $\mathit{Naive\_Match}$ we can derive a specialized
program which is much more efficient than $\mathit{Naive\_Match}_{pd}$
by applying {\em conjunctive} partial deduction,
instead of partial deduction. Conjunctive partial deduction,
viewed as a sequence of applications of transformation rules, enhances
partial deduction because: (i)~one may introduce a definition clause whose
body is a conjunction of atoms, instead of one atom only (see rule PD1),
and (ii)~one may fold a clause w.r.t.~a conjunction of
atoms in its body, instead of one atom only (see rule PD4). By applying
conjunctive partial deduction one may avoid
intermediate data structures, such as the list $X$ constructed by using
clause 1 of program $\mathit{Naive\_Match}$. Indeed, by using
the ECCE system for conjunctive partial 
deduction~\cite{Leu00}, from the $\mathit{Naive\_Match}$ program we derive
the following specialized program:

\medskip

\fpr{$\mathit{Naive\_Match}_{cpd}$  \hfill
(specialized by conjunctive partial deduction, nondeterministic)}{
\Feq{9.}{\it naive\_match_{cpd}}([X,Y,Z|S]) \nc\leftarrow \nc
        {\it new}1(X,Y,Z,S)
\Nex{10.}{\it new}1(a,a,b,S) \leftarrow
\Nex{11.}{\it new}1(X,Y,Z,[C|S]) \leftarrow {\it new}1(Y,Z,C,S)
\eeq}

\medskip

\noindent
This $\mathit{Naive\_Match}_{cpd}$ program searches for the pattern
$[a,a,b]$ in the input string by looking at the first three elements of
that string. If they are $a$, $a$, and $b$, in this order, then the
search succeeds, otherwise the search for the pattern continues
in the tail of the string. Although this
$\mathit{Naive\_Match}_{cpd}$ program is much more efficient than the
initial $\mathit{Naive\_Match}$ program, it does not correspond to a DFA
because, when searching for the pattern $[a,a,b]$, it looks at a prefix of
length 3 of the input string, instead of one symbol only.

The failure of partial
deduction and conjunctive partial deduction to derive a DFA when starting
from the  $\mathit{Naive\_Match}$ program, is due to some limitations which
can be overcome by using the enhanced transformation rules we will present
in the next section.  By applying these enhanced rules we can define a new
predicate by introducing {\it several} clauses whose bodies are non-atomic
goals, while by applying the rules for partial deduction or conjunctive
partial deduction, a new predicate can be defined by introducing {\it one}
clause only. By folding using definition clauses of the enhanced form, we
can derive specialized programs where nondeterminism is reduced and
intermediate data structures are avoided. Among our enhanced rules we also
have the so called {\it case split rule} which, given a clause, produces
two mutually exclusive instances of that clause by introducing negated
equations. The application of this rule allows subsequent folding steps
which reduce nondeterminism.

By applying the enhanced transformation rules according to the
{\em Determinization Strategy} we will present in Section~\ref{strategy},
one can automatically specialize the nondeterministic program
$\mathit{Naive\_Match}$ \wrt the goal ${\it naive\_match}([a,a,b],S)$
thereby deriving the following deterministic program (this derivation is
not presented here and it is similar to the one presented in
Section~\ref{match_pos_ex}):

\medskip

\fpr{$\mathit{Naive\_Match}_{s}$ \hfill (specialized by Determinization,
deterministic)}
{
\Feq{12.}{\it naive\_match}_{s}(S) \leftarrow {\it new}1(S)
\Nex{13.}{\it new}1([a|S]) \leftarrow {\it new}2(S)
\Nex{14.}{\it new}1([C|S]) \leftarrow C\!\neq\! a,{\it new}1(S)
\Nex{15.}{\it new}2([a|S]) \leftarrow {\it new}3(S)
\Nex{16.}{\it new}2([C|S]) \leftarrow C\!\neq\! a,{\it new}1(S)
\Nex{17.}{\it new}3([b|S]) \leftarrow {\it new}4(S)
\Nex{18.}{\it new}3([a|S]) \leftarrow {\it new}3(S)
\Nex{19.}{\it new}3([C|S]) \leftarrow C\!\neq\! b, C\!\neq\! a,
         {\it new}1(S)
\Nex{20.}{\it new}4(S) \leftarrow
\eeq}

\medskip

\noindent The program $\mathit{Naive\_Match}_{s}$ corresponds in a
straightforward way to a DFA. Moreover, since
the clauses of $\mathit{Naive\_Match}_{s}$ are pairwise
mutually exclusive, the disequations in their bodies can be dropped in
favor of  {\em cuts} (or equivalently, {\em if-then-else} constructs) as
follows:

\medskip

\fpr{$\mathit{Naive\_Match}_{cut}$ \hfill (specialized, with cuts)}
{
\Feq{21.}{\it naive\_match}_{s}(S) \leftarrow {\it new}1(S)
\Nex{22.}{\it new}1([a|S]) \leftarrow !, \ {\it new}2(S)
\Nex{23.}{\it new}1([C|S]) \leftarrow {\it new}1(S)
\Nex{24.}{\it new}2([a|S]) \leftarrow !, \ {\it new}3(S)
\Nex{25.}{\it new}2([C|S]) \leftarrow {\it new}1(S)
\Nex{26.}{\it new}3([b|S]) \leftarrow !, \ {\it new}4(S)
\Nex{27.}{\it new}3([a|S]) \leftarrow !, \ {\it new}3(S)
\Nex{28.}{\it new}3([C|S]) \leftarrow {\it new}1(S)
\Nex{29.}{\it new}4(S) \leftarrow
\eeq}

\medskip

\noindent Computer experiments confirm that the final
$\mathit{Naive\_Match}_{cut}$ program is indeed more efficient than the
$\mathit{Naive\_Match}$, $\mathit{Naive\_Match}_{pd}$, and
$\mathit{Naive\_Match}_{cpd}$ programs. In Section~\ref{examples} we will
present more experimental results which demonstrate that the specialized
programs derived by our technique are more efficient than those derived by
partial deduction or conjunctive partial deduction.

\section{Transformation Rules for Logic Programs
with Equations and Disequations between Terms}
\label{rules_sect}

In this section we present the program transformation rules
which we use for program specialization. These rules extend the
unfold/fold rules considered in~\cite{GeK94,Ro&99a,TaS84} to
logic programs with atoms which denote equations and disequations between
terms. The transformation rules we present in this section enhance in
several respects the rules PD1-PD4 for partial deduction which we have
considered in Section~\ref{rules_stra_pe_sect}.
In particular, we consider a definition introduction rule (see
Rule~\ref{def_intro_r}) which allows the introduction of new predicates
defined by {\it several} clauses whose bodies are {\em non-atomic} goals,
while by rule PD1 a new predicate can be defined by introducing {\it one}
clause whose body is an {\em atomic} goal.
We also consider a folding rule (see Rule~\ref{fold_r}) by which we can fold
several clauses at a time, while by rule PD4 we can fold one clause only.
In addition, we consider the subsumption rule and the following
transformation rules for introducing and eliminating equations and
disequations: (i)~head generalization, (ii)~case split, (iii)~equation
elimination, and (iv)~disequation replacement. Our rules preserve the least
Herbrand model as indicated in Theorem~\ref{corr_th_lhm} below.

\subsection{Transformation Rules}
\label{rules}

Similarly to Section~\ref{rules_stra_pe_sect}, the process of program
transformation is viewed as a transformation
sequence constructed by applying some transformation rules. However, as
already mentioned, in this section we consider an enhanced set of
transformation rules. A transformation sequence  $P_0, \ldots, P_n$ is
constructed from a given initial program $P_0$
by applications of the transformation
rules~\ref{def_intro_r}--\ref{diseq_repl_r}
given below, as follows. For $k=0,\ldots,n-1$,
program $P_{k+1}$ is derived from program $P_k$ by:
(i) selecting a (possibly empty) subset
$\gamma_1$ of clauses of $P_{k}$,
(ii) deriving a set $\gamma_2$ of clauses by
applying a transformation rule
to $\gamma_1$,
and (iii) replacing $\gamma_1$ by $\gamma_2$ in $P_{k}$.

Notice that rules~\ref{def_elim_r} and~\ref{unf_r} are in fact equal
to rules PD2 and PD3, respectively. However, we rewrite them below for the
reader's convenience.

\begin{trule} [Definition Introduction] \label{def_intro_r}
{\rm
We introduce $m$ $(\geq \!1)$ new clauses, called {\it definition
clauses}, of the form:

\smallskip

$\left\{
\begin{array}{ll}
D_1. \ \ \mathit{newp}(X_1,\dots,X_h)  \leftarrow \mathit{Body}_1\\
  \hspace{1cm} \cdots & \\
D_m. \ {\it newp}(X_1,\dots,X_h)  \leftarrow \mathit{Body_m}&
\end{array}
\right.$

\smallskip

\noindent
where:
(i) {\it newp} is a non-basic predicate symbol not occurring in
$P_0, \ldots, P_k$,  (ii) the variables $X_1,\dots,X_h$ are all
distinct and for all $i \in \{1,\ldots,h\}$  there exists
$j\in\{1,\ldots, m\}$  such that $X_i$ occurs in the goal
$\mathit{Body_j}$, (iii) for all $j \in \{1,\ldots,m\}$, every
non-basic predicate occurring in $\mathit{Body_j}$ also
occurs in $P_0$, and (iv) for all $j \in \{1,\ldots,m\}$, there
exists at least one non-basic atom in $\mathit{Body_j}$.

\noindent
Program $P_{k+1}$ is the program $P_k \cup \{D_1,\ldots,D_m\}$.
} %end rm
\end{trule}

As in Section~\ref{rules_stra_pe_sect},  we denote by
$\mathit{Defs_k}$ the set of definition clauses introduced by
the definition introduction rule during the construction of the
transformation sequence $P_0, \ldots, P_k$. In particular,
we have that $\mathit{Defs}_0 = \emptyset$.

\begin{trule} [Definition Elimination] \label{def_elim_r}
{\rm
Let $p$ be a predicate symbol. By {\em definition elimination
\wrt}$\!p$ we derive the program $P_{k+1}= \{C \in P_k ~|~ p$
depends on $C\}$.
} %end rm
\end{trule}

\begin{trule} [Unfolding] \label{unf_r}
{\rm
Let $C$ be a renamed apart clause of $P_k$ of the
form: $H \leftarrow G_1, A, G_2$,
where~$A$ is a non-basic atom.  Let
$C_1,\dots, C_m$, with $m\geq 0$, be the clauses of
$P_k$ such that, for $i = 1,\dots, m$, $A$ is unifiable with the head of
$C_i$ via  the mgu\ $\vartheta_i$.
By {\em unfolding $C$ \wrt $A$}, for $i = 1,\dots, m$,
we derive the clause $D_i:$
$(H \leftarrow G_1, bd(C_i), G_2)\vartheta_i$.

\noindent
Program $P_{k+1}$ is the program $(P_k - \{C\}) \cup
\{D_1,\ldots,D_m\}$.
}
\end{trule}

Notice that an application of the unfolding rule
to clause $C$
amounts to the deletion of $C$  iff $m\!=\!0$. Sometimes in the
literature this particular instance of the unfolding rule is treated as an
extra rule.

\begin{trule}[Folding] \label{fold_r}
{\rm
Let
\smallskip

$\left\{
\begin{array}{ll}
C_1. \ \ H  \leftarrow \ G_1,  \mathit{Body}_1\vartheta, G_2\\
 \cdots & \\
C_m. \ H  \leftarrow \ G_1,  \mathit{Body}_m\vartheta, G_2 &
\end{array}
\right.$

\smallskip

\noindent
be renamed clauses of $P_k$, for a suitable substitution
$\vartheta$, and let

\smallskip

$\left\{
\begin{array}{ll}
{D_1. \ \ \it newp}(X_1,\dots,X_h)  \leftarrow \mathit{Body}_1\\
  \cdots & \\
{D_m. \ \ \it newp}(X_1,\dots,X_h)  \leftarrow \mathit{Body}_m &
\end{array}
\right.$

\smallskip
\noindent
be all clauses in $\mathit{Defs_k}$
which have {\it newp} as head predicate. Suppose that for
$i = 1,\dots, m,$ the following condition holds:
for every variable $X$ occurring in the goal
        ${\it Body}_i$ and not in $\{X_1,\dots,X_h\}$,
        we have that: (i)
           $X\vartheta$ is a variable which does not occur
                 in $(H, G_1, G_2)$,  and  (ii) $X\vartheta$
          does not occur in
                 $Y\vartheta$, for any variable $Y$ occurring in
                 $\mathit{Body_i}$ and different from $X$.
By {\em folding $C_1,\dots,C_m$ using} $D_1,\dots,D_m$ we derive
the single clause $E$:
$H\leftarrow G_1, {\it newp}(X_1,\dots,X_h)\vartheta, G_2$.

\noindent
Program $P_{k+1}$ is the program $(P_k - \{C_1,\dots,C_m\}) \cup \{E\}$.
} %end rm
\end{trule}

{F}or instance, the clauses
$C_1$: $p(X) $ $\leftarrow q(t(X),Y), r(Y)$ and
$C_2$: $p(X) $ $\leftarrow s(X), r(Y)$
can be folded (by considering the substitution $\vartheta = \{U/X, V/Y\}$)
using the two definition clauses $D_1$: $a(U,V) $ $\leftarrow q(t(U),V)$
and $D_2$: $a(U,V) $ $\leftarrow s(U)$, and we
replace $C_1$ and $C_2$ by the
clause $E$: $p(X) $ $\leftarrow a(X,Y), r(Y)$.

\begin{trule} [Subsumption] \label{subs_r}
{\rm (i) Given a substitution $\vartheta$, we say that a clause $H
\leftarrow G_1$ {\em subsumes} a clause $(H \leftarrow G_1, G_2)\vartheta$.

\noindent
Program
$P_{k+1}$  is derived from program $P_k$ by deleting
 a clause which is subsumed by another clause in
$P_k$.
} %end rm
\end{trule}

\begin{trule} [Head Generalization]   \label{head_gen_r}
{\rm
Let $C$ be a clause of the form: $H\{X/t\} \leftarrow {\it Body}$
in $P_k$, where $\{X/t\}$ is a substitution such that $X$ occurs in
$H$ and $X$ does not occur in $C$.
By {\em head generalization}, we derive the clause {\it GenC\/}: $H
\leftarrow X\!=\!t, {\it Body}$.

\noindent
Program $P_{k+1}$ is the program $(P_k - \{C\}) \cup \{\mathit{GenC}\}$.
} %end rm
\end{trule}

Rule~\ref{head_gen_r} is a particular case of the rule
of {\em generalization} + {\em equality introduction} considered, for
instance, in~\cite{PrP95a}.

\begin{trule}  [Case Split] \label{case_split_r}
{\rm
Let $C$ be a clause in $P_k$
of the form: $H \leftarrow {\it Body}$. By {\em case
split of  $C$ \wrt{\em the binding}} $X/t$ where $X$ does not occur in
$t$,  we derive the following two clauses:

\smallskip

$C_1$. $(H  \leftarrow $ ${\it Body}) \{X/t\}$

$C_2$. $H  \leftarrow $ $X\! \neq\! t, {\it Body}$.

\smallskip
\noindent
Program $P_{k+1}$ is the program $(P_k - \{C\}) \cup \{C_1,C_2\}$.
} %end rm
\end{trule}

In this Rule~\ref{case_split_r} we do not assume that $X$ occurs in
$C$. However, in the Determinization Strategy of Section~\ref{strategy}, we
will always apply the case split rule to a clause $C:$
$H \leftarrow {\it Body}$ w.r.t.~a binding
$X/t$ where $X$ occurs in $H$. This use of the case
split rule will be sufficient to derive mutually exclusive clauses. Indeed,
according to our operational semantics, if  $G \longmapsto_{P_{k+1}} G_1$
using clause $C_1$ and $X$ occurs in $H$, then no $G_2$ exists such that $G
\longmapsto_{P_{k+1}} G_2$ using clause $C_2$. The same holds by
interchanging $C_1$ and $C_2$. We will return to this property in
Definitions~\ref{semideter_def} (Semideterminism) and \ref{mutual_excl_def}
(Mutual Exclusion) below.

\begin{trule} [Equation Elimination] \label{eq_elim_r}
{\rm
Let $C_1$ be a clause in $P_k$ of the form:

$C_1$. $H\leftarrow G_1,\ t_1 \!=\!t_2,\  G_2$

\noindent
If $t_1$ and $t_2$
are unifiable via the most general unifier $\vartheta$,
then by equation elimination we derive the following clause:

$C_2$. $(H\leftarrow G_1, G_2)\vartheta$

\noindent
Program $P_{k+1}$ is the program $(P_k - \{C_1\}) \cup \{C_2\}$.

\noindent
If $t_1$ and $t_2$ are not unifiable
then by equation elimination we derive program $P_{k+1}$ which is  $P_k -
\{C_1\}$.
}  %end rm
\end{trule}

\begin{trule} [Disequation Replacement] \label{diseq_repl_r}
{\rm
Let $C$ be a clause in program $P_k$.
Program $P_{k+1}$ is derived from $P_k$ by
either removing $C$ or replacing
$C$ as we now indicate:

\begin{enumerate}
\item[\ref{diseq_repl_r}.1] if $C$ is of the form:
$H\leftarrow G_1, t_1 \!\neq\!t_2,  G_2$
\ and $t_1$ and $t_2$ are not unifiable,
then $C$ is replaced by
$H\leftarrow G_1, G_2$

\item[\ref{diseq_repl_r}.2] if $C$ is of the form:
$H\leftarrow  G_1,\, f(t_1,\ldots,t_m)\!\neq\!f(u_1,\ldots,u_m), \, G_2$,
then $C$ is replaced by the following $m\ (\geq0)$ clauses:
$H\leftarrow G_1, t_1\!\neq\! u_1, G_2$, \ \ $\ldots,$\ \ \
$H\leftarrow G_1, t_m\!\neq\! u_m, G_2$

\item[\ref{diseq_repl_r}.3] if $C$ is of the form:
$H\leftarrow G_1, X \!\neq\!X, G_2$,
then $C$ is removed from $P_k$

\item[\ref{diseq_repl_r}.4] if $C$ is of the form:
$H\leftarrow G_1, t \!\neq\!X, G_2$,
then $C$ is replaced by
$H\leftarrow G_1, X \!\neq\!t, G_2$

\item[\ref{diseq_repl_r}.5] if $C$ is of the form:
$H\leftarrow G_1, X \!\neq\!t_1, G_2, X \!\neq\!t_2, G_3$
and there exists a substitution $\rho$ which is a bijective
mapping from the set of the local variables of $X\deq t_1$ in $C$
onto the set of the local variables of $X\deq t_2$ in $C$ such
that $t_1\rho = t_2$,
then $C$ is replaced by
$H\leftarrow G_1, X \!\neq\!t_1,  G_2, G_3$.
\end{enumerate}
}  %end rm
\end{trule}

In particular, by Rule~\ref{diseq_repl_r}.5, if a disequation has occurs
twice in the body of a clause, then we
can remove the rightmost occurrence.

\subsection{Correctness of the Transformation Rules \wrt
the Declarative Semantics}\label{corr}

In this section we show that, under suitable hypotheses,
our transformation rules preserve the declarative semantics
presented in Section~\ref{decl_sem}. In that sense we also say that
our transformation rules are {\em correct}) \wrt the given declarative
semantics. The following correctness theorem extends similar results
holding for logic programs~\cite{GeK94,Ro&99a,TaS84} to
the case of logic programs with equations and disequations.

\begin{theorem}[Correctness of the Rules \wrt the Declarative Semantics]
\label{corr_th_lhm}
{\rm
Let $P_0,\dots,P_n$  be a transformation sequence constructed
by using the transformation rules~\ref{def_intro_r}--\ref{diseq_repl_r}
and let $p$ be a non-basic predicate in $P_n$. Let us assume that:

\begin{enumerate}
\item \label{fold_cond} {\em if} the folding rule is applied
for the derivation of
a clause $C$ in program $P_{k+1}$ from clauses $C_1,\dots, C_m$ in
program  $P_{k}$ using clauses $D_1,\dots, D_m$ in $\mathit{Defs_k}$,
with $0\!\leq\!k\!<\!n$,\\
{\em then}  for every $i \in \{1,\dots,m\}$ there exists
$j\in\{1,\dots,n\!-\!1\}$ such that $D_i$ occurs in
$P_j$ and $P_{j+1}$ is derived from $P_j$ by unfolding $D_i$.

\item \label{elim_cond} during the transformation
sequence $P_0,\dots,P_n$ the definition elimination rule {\em
either} is never applied {\em or} it is applied \wrt predicate $p$
once only, in the last step, that is, when deriving $P_n$ from $P_{n-1}$.
\end{enumerate}

\noindent
Then, for every ground atom $A$ with predicate $p$, we have that
$M(P_0 \cup {\it Defs}_n) \models A$ iff $M(P_n)\models A$.
} %end rm
\end{theorem}

\begin{proof} It is a simple extension of a similar result
presented in \cite{GeK94} for the case where we use the unfolding, folding,
and {\em generalization} + {\em equality introduction} rules.
The proof technique used in \cite{GeK94} can be adapted
to prove also the correctness of our extended set of rules.
\end{proof}

\noindent
In Example~\ref{incorr_ex_unf} of Section~\ref{rules_stra_pe_sect} we have
shown that the unfolding rule may not preserve the operational semantics.
The following examples show that also other transformation rules may not
preserve the operational semantics.

\begin{example} \label{incorr_ex2}
{\rm
Let us consider the following program
$P_1:$

\bpr
\Feq{1.} p(X) \If q(X),\ X\!\neq\!a
\Nex{2.} q(X) \If
\Nex{3.} q(X) \If X\!=\!b
\eeq
\epr

\noindent
By rule~\ref{subs_r} we may delete clause 3 which is subsumed
by clause 2 and we derive a new program
$P_2$. Now, we have that $p(X)$ succeeds in $P_1$,
while it does not succeed in $P_2$.
}% end rm
\end{example}

\begin{example} \label{incorr_ex3}
{\rm Let us consider the following program
$P_3:$

\bpr
\Feq{1.} p(X) \If
\eeq
\epr

\noindent
By the case split rule we may replace clause 1
by the two clauses:

\bpr
\Feq{2.} p(a) \If
\Nex{3.} p(X) \If X\!\neq\!a
\eeq
\epr

\noindent
and we derive a new program
$P_4$. The goal $p(X), X\!=\!b$ succeeds in $P_3$,
while it does not succeed in $P_4$.

}% end rm
\end{example}

\begin{example} \label{incorr_ex4}
{\rm Let us consider the following program
$P_5:$

\bpr
\Feq{1.} p \If X\!\neq\!a,\ X\!=\!b
\eeq
\epr

\noindent
By rule~\ref{eq_elim_r}  we may replace clause 1
by:

\bpr
\Feq{2.} p \If b\!\neq\!a
\eeq
\epr

\noindent
and we derive a new program
$P_6$. The goal $p$ does not succeed in $P_5$,
while it succeeds in $P_6$.
}% end rm
\end{example}

\noindent
{F}inally, let us consider the following two operations on the body of a
clause: (i)~removal of a duplicate atom, and (ii)~reordering
of atoms. The following examples show that these two operations, which
preserve the declarative semantics,
may not preserve the operational semantics.
Notice, however, that the removal of a duplicate atom and the reordering
of atoms cannot be accomplished by the transformation rules listed in
Section~\ref{rules_sect}, except for special case considered at
Point~\ref{diseq_repl_r}.5 of the disequation replacement rule.

\begin{example} \label{incorr_ex_dupl}
{\rm
Let us
consider the program $P_7$:

\bpr
\Feq{1.} p \If q(X,Y),\ q(X,Y),\ X\!\neq\!Y
\Nex{2.} q(X,b) \If
\Nex{3.} q(a,Y) \If
\eeq
\epr

\noindent
and the program $P_8$ obtained from $P_7$ by replacing clause 1 by
the following clause:

\bpr
\Feq{$4$.} p \If q(X,Y),\ X\!\neq\!Y
\eeq
\epr

\noindent
The goal $p$ succeeds in $P_7$,
while it does not succeed in $P_8$. Indeed, (i) for program $P_7$ we have
that:

\noindent  $p\ \longmapsto_{P_7}\ $
$q(X,Y),q(X,Y),X\!\neq\!Y\  \longmapsto_{P_7}\ $
$q(X,b), X\!\neq\!b\ \longmapsto_{P_7}\  $
$a\!\neq\!b\ \longmapsto_{P_7} {\it true}$,
and (ii) for program $P_8$ we have that:
{\it either} \ $p \longmapsto_{P_8} X\!\neq\!b$
\ {\it or} \ $p \longmapsto_{P_8} a\!\neq\!Y$. In Case~(ii),  since
$X$ and $Y$ are unifiable with $b$ and $a$, respectively, we have that
$p \longmapsto_{P_8}^* {\it true}$ does not hold.
}% end rm
\end{example}

\begin{example} \label{incorr_ex_reord}
{\rm
Let us
consider the program $P_9$:

\bpr
\Feq{1.} p \If q(X),\ r(X)
\Nex{2.} q(a) \If
\Nex{3.} r(X) \If  X\!\neq\!b
\eeq
\epr

\noindent
and the program $P_{10}$ obtained from $P_9$ by replacing clause 1 by
the following clause:

\bpr
\Feq{$4$.} p \If r(X),\ q(X)
\eeq
\epr

\noindent
The goal $p$ succeeds in $P_9$,
while it does not succeed in $P_{10}$.
}% end rm
\end{example}

\noindent In the next section we will introduce a class of
programs and a class of goals for which our transformation rules
preserve both the declarative semantics
 and the operational semantics. In order to do so,
 we associate a {\em mode}
with every predicate. A mode of a predicate specifies the {\em input}
arguments of that predicate, and we assume that whenever the
predicate is called, its input arguments are bound to ground terms.
We will see that, if some suitable conditions are satisfied,
compliance to modes guarantees the preservation of the operational
semantics. This fact is illustrated by the above
Examples~\ref{incorr_ex2} and \ref{incorr_ex3}, and indeed, in each
of them, if we restrict ourselves to calls of the predicate $p$ with
ground arguments, then the initial program and the derived program
have the same operational semantics.

Notice, however, that the incorrectness of the transformation of
Example~\ref{incorr_ex4} does not depend on the
modes. % associated with the predicates.
Thus, in order to ensure correctness
\wrt the operational semantics we have to rule out clauses such as
clause 1 of program $P_5$. Indeed, as we will see in the next section,
the clauses we will consider satisfy the following condition:
each variable which occurs in a disequation {\em either}
occurs in an input
argument of the head predicate {\em or} it is a local variable of the
disequation.

\section{Program Transformations based on Modes}
\label{modes}

Modes provide information about the directionality of predicates,
by specifying whether an argument should be used as input or output
(see, for instance,~\cite{Mel87,War77}).
Mode information is very useful for specifying and verifying
logic programs~\cite{Apt97,Dev90} and it is used
in existing compilers, such as Ciao and Mercury, to generate very
efficient code~\cite{He&99,So&96a}.
Mode information has also been used in the context of program
transformation to provide sufficient conditions which ensure that
reorderings of atoms in the body of a clause preserve
program termination~\cite{Bo&99}.

In this paper we use mode information for: (i)~specifying classes of
programs and goals \wrt which the transformation rules we have
presented in Section~\ref{rules} preserve
the operational semantics (see Section~\ref{op_sem}), and
(ii)~designing our  strategy for
specializing programs and reducing nondeterminism.

\subsection{Modes}

A {\it mode for a non-basic predicate} $p$ of arity
$h$ $(\geq 0)$ is an expression of
the form $p(m_1,\dots,m_h)$, where for $i=1,\ldots,h$,
$m_i$ is either $+$ (denoting {\em any ground} term) or
$?$ (denoting {\em any} term). In particular, if $h\!=\! 0$, then $p$ has a
unique mode which is $p$ itself.
Given an
atom $p(t_1,\dots,t_h)$ and a mode $p(m_1,\dots,m_h)$,\\
(1) for $i=1,\ldots,h$, the term $t_i$ is said to be an {\em input
argument} of $p$ iff $m_i$ is $+$, and\\
(2) a variable of $p(t_1,\dots,t_h)$ with an occurrence in an
input argument of $p$, is said to be an {\em input
variable} of $p(t_1,\dots,t_h)$.

A {\em mode for a program\/} $P$ is a set of modes for non-basic
predicates containing exactly one mode
for every distinct, non-basic predicate $p$ occurring in $P$.

Notice that a mode for a program $P$ may or may not contain modes for
non-basic predicates which do {\em not} occur in $P$.
Thus, if $M$ is a mode for a program $P_1$ and, by applying a
transformation rule, from $P_1$ we derive a new program $P_2$ where
all occurrences of a predicate have been eliminated, then $M$ is a
mode also for $P_2$. The following rules may eliminate occurrences
of predicates: definition elimination, unfolding, folding, subsumption,
disequation replacement (case \ref{diseq_repl_r}.5).
Clearly, if from $P_1$ we derive $P_2$ by applying
the definition introduction rule, then in order to obtain a mode for
$P_2$ we should add to $M$ a mode for the newly introduced predicate
(unless it is already in $M$).

\begin{example}\label{mode_ex}
{\rm
Given the program $P$:

\bpr
\Feq{} p(0,1) \If
\Nex{} p(0,Y) \If q(Y)
\eeq
\epr

\noindent
the set $M_1=\{p(+,?),q(?)\}$ is a mode for $P$.
$M_2=\{p(+,?), q(+),r(+)\}$ is a different mode for $P$.
} % end of roman
\end{example}

\begin{definition}\label{mode_sat}
{\rm
Let $M$ be a mode for a program $P$ and $p$ a non-basic predicate.
We say that an atom
$p(t_1,\ldots,t_h)$ {\em satisfies} the mode $M$ iff
(1) a mode for $p$ belongs
to $M$ and (2) for $i=1,\ldots,h$, if the argument $t_i$ is an input
argument of $p$ according to $M$, then $t_i$ is a ground term. In
particular, when $h\!=\!0$, we have that $p$ {\em satisfies} $M$ iff $p \in
M$.

\noindent
The program $P$ {\em satisfies} the mode $M$
iff for each non-basic atom $A_0$
which satisfies $M$, and for each non-basic atom $A$ and goal $G$
such that $A_0 \longmapsto^*_P (A,G)$, we have that $A$ satisfies $M$. }
%end rm
\end{definition}

\noindent
With reference to Example~\ref{mode_ex} above, program $P$ satisfies
mode $M_1$, but it does {\em not} satisfy mode $M_2$.

In general, the
property that a program satisfies a mode is undecidable.
Two approaches are usually followed
for verifying this property: (i)~the first one uses
{\em abstract interpretation} methods (see, for
instance,~\cite{DeW88,Mel87}) which always terminate, but may return a
{\em don't know} answer, and (ii)~the second one checks suitable
syntactic properties of the program at hand, such as {\em
well-modedness}~\cite{Apt97}, which imply that the mode is satisfied.

Our technique is
independent of any specific method used for verifying that a program
satisfies a mode. However, as the reader may verify, all programs presented
in the examples of Section~\ref{examples} are well-moded and, thus, they
satisfy the given modes.

\subsection{Correctness of the Transformation Rules
\wrt the Operational Semantics}

Now we introduce a class of programs, called
{\em safe\/} programs, and we
prove that if the transformation rules are
applied to a safe program and suitable restrictions hold, then
the given program and the derived program are equivalent
\wrt the operational semantics.

\begin{definition} [Safe Programs]
{\rm
Let $M$ be a mode for a program $P$. We say that
a clause $C$ in $P$ is
{\em safe\/} \wrt $M$ iff for each disequation
$t_1\!\neq\! t_2$  in the body of $C$, we have that:
for each variable $X$ occurring in $t_1\!\neq \!t_2$
{\em either} $X$ is an input variable of ${\it hd}(C)$ {\em or}
$X$ is a local variable of $t_1\!\neq\! t_2$ in $C$.
Program $P$ is safe \wrt $M$
iff all its clauses are safe \wrt
$M$. }
\end{definition}

\noindent
{F}or instance, let us consider the mode $M = \{p(+), q(?)\}$.
Clause $p(X)\If X\!\neq\! f(Y)$
is safe \wrt $M$ and clause
$p(X)\If X\!\neq\! f(Y),\ q(Y)$ is not safe \wrt $M$ because $Y$
occurs both in $f(Y)$ and in $q(Y)$.

When mentioning the safety property \wrt a given mode $M$, we feel
free to omit the reference to $M$, if it is irrelevant
or understood from the context.

In order to get our desired correctness result (see
Theorem~\ref{corr_th_op} below), we need to restrict the use of
our transformation rules as indicated in
Definitions~\ref{safe_unfold}-\ref{safe_cs} below.
In particular, these restrictions ensure
that, by applying the transformation rules, program safety and
mode satisfaction are preserved (see
Propositions~\ref{safety_preserv} and~\ref{mode_preserv} in Appendix
A).

\begin{definition} [Safe Unfolding]  \label{safe_unfold}
{\rm
Let $P_k$ be a program and $M$ be a mode for $P_k$.
Let us consider an application of the unfolding rule (see
Rule~\ref{unf_r} in Section~\ref{rules}) whereby
from the following clause of $P_k$:

$H \leftarrow G_1, A, G_2$

\noindent
we derive the clauses:

\medskip

$\left\{
\begin{array}{ll}
D_1.\ \ (H \leftarrow G_1, bd(C_1), G_2)\vartheta_1\\
 \cdots & \\
D_m.\ (H \leftarrow G_1, bd(C_m), G_2)\vartheta_m\\
\end{array}
\right.$

\medskip

\noindent
where $C_1,\dots, C_m$ are the clauses in
$P_k$ such that, for
$i \in \{1,\dots, m\}$, $A$ is unifiable with the head of
$C_i$ via  the mgu\ $\vartheta_i$.

\smallskip
\noindent We say that this application of the unfolding rule
is {\em safe\/} \wrt mode $M$ iff  for all $i=1,\ldots,m$,
for all disequations $d$ in ${\it bd}(C_i)$,
and for all variables $X$ occurring in $d\vartheta_i$,
we have that either $X$ is an input variable of $H\vartheta_i$ or
$X$ is a local variable of $d$ in $C_i$.
}
\end{definition}

To see that
unrestricted applications of the unfolding rule may not preserve safety,
let us consider the following program:

\bpr
\Feq{1.} p \If q(X), \ r(X)
\Nex{2.} q(1) \If
\Nex{3.} r(X) \If X\deq 0
\eeq
\epr

\noindent
and the mode $M=\{p,\ q(?),\ r(+)\}$ for it.
By unfolding clause 1 \wrt the atom $r(X)$ we derive
the clause:

\oeq{4.}{p \If q(X), \ X\deq 0}

\noindent
This clause is not safe \wrt $M$ because $X$ does not occur in its head.

\begin{definition} [Safe Folding]  \label{safe_fold}
{\rm
Let us consider a program $P_k$ and a mode $M$ for $P_k$.
Let us also consider an application of the folding rule (see
Rule~\ref{fold_r} in Section~\ref{rules}) whereby
from the following clauses in $P_k$:

\medskip

$\left\{
\begin{array}{ll}
C_1.\ H  \leftarrow \ G_1,  (A_1,K_1)\vartheta, G_2\\
 \cdots & \\
C_m.\ H  \leftarrow \ G_1,  (A_m,K_m)\vartheta, G_2 &
\end{array}
\right.$

\medskip

\noindent
and the following definition clauses in $\mathit{Defs_k}$:

\medskip
$\left\{
\begin{array}{ll}
D_1.\ {\it newp}(X_1,\dots,X_h)  \leftarrow A_1,K_1\\
  \cdots & \\
D_m.\ {\it newp}(X_1,\dots,X_h)  \leftarrow A_m,K_m &
\end{array}
\right.$

\medskip
\noindent we derive the new clause:

\smallskip
$H\leftarrow G_1, {\it newp}(X_1,\dots,X_h)\vartheta, G_2$

\smallskip
\noindent We say that this application of the folding rule
is {\em safe\/} \wrt mode $M$ iff the following Property $\Sigma$
holds:

\smallskip
\noindent
({\em Property} $\Sigma$) \ Each input variable of ${\it
newp}(X_1,\dots,X_h)\vartheta$ is also an input variable
of at least one of the non-basic atoms occurring in $(H,G_1,
A_1\vartheta,\dots,A_m\vartheta)$.
} %end rm
\end{definition}

Unrestricted applications of the folding rule may not preserve modes.
Indeed, let us consider the following initial program:

\bpr
\Feq{1.} p \If q(X)
\Nex{2.} q(1) \If
\eeq
\epr

\noindent
Suppose that first we introduce the definition clause:

\oeq{3.}{{\it new}(X) \If q(X)}

\noindent
and then we apply the clause split rule, thereby deriving:

\bpr
\Feq{4.} {\it new}(0) \If q(0)
\Nex{5.} {\it new}(X) \If X\deq 0,\ q(X)
\eeq
\epr

\noindent
The program made out of clauses 1, 2, 4, and 5 satisfies the mode $M =
\{p,\ q(?),\ {\it new}(+)\}$. By folding clause 1 using
clause 3 we derive:

\oeq{6.}{p \If {\it new}(X)}

\noindent
This application of the folding rule is not safe and the program
we have derived, consisting of clauses 2, 4, 5, and 6, does not
satisfy $M$.

\begin{definition} [Safe Head Generalization] \label{safe_head_gen}
{\rm Let us consider a program $P_k$ and a mode $M$ for $P_k$.
We say that an application of the head generalization rule (see
Rule~\ref{head_gen_r} in Section~\ref{rules}) to a clause of $P_k$ is
{\em safe\/} iff $X$ is not an input variable w.r.t.~$M$.
}
\end{definition}

The restrictions
considered in Definition~\ref{safe_head_gen} are needed to preserve safety.
For instance, the clause $p(t(X)) \If X\deq0$ is safe \wrt the mode
$M=\{p(+)\}$, while $p(Y) \If Y\eq t(X), X\deq0$ is not.

\begin{definition} [Safe Case Split]\label{safe_cs}
{\rm
Let us consider a program $P_k$ and a mode $M$ for $P_k$.
Let us consider also an application of the case split rule
(see Rule~\ref{case_split_r} in Section~\ref{rules})
whereby from a clause $C$ in $P_k$
of the form: $H \leftarrow {\it Body}$ we derive the following
two clauses:

\smallskip
$C_1$. $(H  \leftarrow $ ${\it Body}) \{X/t\}$

$C_2$. $H  \leftarrow $ $X\! \neq\! t, {\it Body}$.

\smallskip
\noindent
We say that this application of the case split rule is
{\em safe\/} \wrt mode $M$ iff
$X$ is an input variable of $H$, $X$ does not occur in $t$,
and for all variables $Y\in {\it vars}(t)$,
{\em either} $Y$ is an input variable of $H$ {\em or}
$Y$ does not occur in $C$.
}
\end{definition}

When applying the safe case split rule, $X$ occurs in
$H$ and thus, given a goal $G$, it is not the case that for some
goals $G_1$ and $G_2$, we have both $G \longmapsto G_1$
using clause $C_1$ and $G \longmapsto G_2$ using clause $C_2$.
In Definition~\ref{mutual_excl_def} below,
we will formalize this property by saying that
the clauses $C_1$ and $C_2$ are {\it mutually exclusive}.

Similarly to the unfolding and head
generalization rules,
the unrestricted use of the case split rule may not preserve safety.
For instance, from the clause $p(X) \If$ which is safe \wrt the
mode $M = \{p(?)\}$, we may derive the two clauses $p(0) \If$
and $p(X) \If X\deq 0$, and this last clause is not safe \wrt $M$.

\medskip

We have shown in Section~\ref{rules} (see Example~\ref{incorr_ex_reord}),
that the reordering of atoms in the body of a clause may
not preserve the operational semantics.
Now we prove that a particular reordering of atoms, called {\em disequation
promotion}, which consists in moving to the left the disequations occurring
in the body of a safe clause, preserves the operational
semantics.
Disequation promotion (not included, for reason of simplicity,
among the transformation rules) allows us to rewrite the body of a safe
clause so that every disequation occurs to the left of every atom
different from a disequation thereby deriving the {\em normal
form} of that clause (see Section~\ref{strategy}).
The use of normal forms will simplify the proof of Theorem~\ref{corr_th_op}
below and the presentation of the
Determinization Strategy in Section~\ref{strategy}.

\begin{proposition}[Correctness of Disequation Promotion]
\label{rearr_prop}
{\rm
Let $M$ be a mode for a program $P_1$. Let us assume that
$P_1$ is safe \wrt  $M$ and $P_1$ satisfies $M$.
Let $C_1$: $H \If G_1,\ G_2,\ t_1\!\neq\! t_2,\ G_3$ be a clause in
$P_1$.
Let $P_2$ be the program derived from $P_1$
by replacing clause $C_1$ by clause $C_2$: $H \If G_1,\  t_1\!\neq\!
t_2,\ G_2,\ G_3$. Then:
(i) $P_2$ is safe \wrt $M$,
(ii) $P_2$ satisfies $M$, and
(iii) for each non-basic  atom $A$ which satisfies mode $M$,
$A$ succeeds in $P_1$  {\rm iff} $A$ succeeds in $P_2$.
} %end \rm
\end{proposition}

\begin{proof} Point~(i) follows from the fact that
safety does not depend on the position of the disequation in a clause.
Moreover, the evaluation of goal $G_2$
in program $P_1$ according our operational semantics, does not bind
any variable in $t_1\!\neq\!t_2$, and thus, we get Point~(ii).
Point~(iii) is a
consequence of Points~(i) and~(ii) and the fact that the evaluation of
$t_1\!\neq\!t_2$ does not bind any variable in the goals
$G_2$ and $G_3$.
\end{proof}

The above proposition does not hold if we interchange clause $C_1$
and $C_2$. Consider, in fact, the following clause which is safe
\wrt mode $M = \{p(+), q(+)\}$:

\smallskip
$C_3$. $p(X) \leftarrow X\!\neq\!Y,\ q(Z)$

\smallskip
\noindent
This clause satisfies $M$ because for all derivations starting
from a ground instance $p(t)$ of $p(X)$ the atom $t\!\neq\!Y$ does not
succeed. In contrast, if we use the clause $C_4$: $p(X) \leftarrow q(Z),
X\!\neq\!Y$, we have that in the derivation starting from $p(t)$, the
variable $Z$ is not bound to a ground term and thus, clause $C_4$
does not satisfy the mode $M$ which has the element $q(+)$.

In Theorem~\ref{corr_th_op} below we will show that if we apply
our transformation rules and their safe versions
in a restricted way, then a program $P$ which satisfies a
mode $M$ and is safe w.r.t.~$M$, is transformed
into a new program, say $Q$, which satisfies $M$ and is safe w.r.t.~$M$.
Moreover, the programs $P$ and $Q$ have the same operational semantics.

\begin{theorem}[Correctness of the Rules \wrt the Operational Semantics]
\label{corr_th_op}
{\rm
Let $P_0,\dots,P_n$  be a transformation sequence constructed
by using the transformation rules
\ref{def_intro_r}--\ref{diseq_repl_r} and let $p$ be a
non-basic predicate in $P_n$.
Let $M$ be a mode for $P_0\cup \mathit{Defs_n}$ such that:
(i) $P_0\cup \mathit{Defs_n}$ is safe \wrt $M$,
(ii) $P_0\cup \mathit{Defs_n}$ satisfies $M$, and
(iii) the applications of the unfolding,
folding, head generalization, and case split rules during
the construction of $P_0,\dots,P_n$ are all safe \wrt $M$.  Suppose also
that Conditions~\ref{fold_cond} and~\ref{elim_cond} of
Theorem~\ref{corr_th_lhm} hold. Then:
(i) $P_n$ is safe \wrt $M$,
(ii) $P_n$ satisfies $M$, and
(iii) for each atom $A$ which has predicate $p$ and
satisfies mode $M$, $A$ succeeds in $P_0\cup \mathit{Defs_n}$
iff $A$ succeeds in $P_n$.
} %end rm
\end{theorem}
\begin{proof} See Appendix A. \end{proof}

\subsection{Semideterministic Programs}
\label{semidet}

In this section we introduce the concept of {\em semi\-de\-ter\-minism}
which characterizes the class of programs which can be obtained by using
the Determinization Strategy of Section~\ref{strategy}.
(The reader should not confuse the notion of semideterminism presented
here with the one considered in~\cite{Fe&96}.)

We have already noticed that if a program $P$ is deterministic for
an atom $A$ according to Definition~\ref{determ_def}, then there is at
most one successful derivation starting from $A$, and $A$ succeeds in $P$
with at most one answer substitution.
Thus, if an atom succeeds in a program with more than one answer
substitution, and none of these substitutions is more general
than another, then there is no chance to transform that program into a
new program which is deterministic for that atom.

{F}or instance, let us consider the following generalization
of the problem of Sections~\ref{string_mat} and~\ref{limitations_pd}:
Given a pattern $P$ and a string $S$ we want to
compute the {\em position}, say $N$, of an occurrence
of $P$ in $S$, that is, we want to find two strings $L$
and $R$ such that: (i)~$S$ is the concatenation of $L$, $P$,
and $R$, and (ii)~the length of $L$ is $N$.
The following program {\it Match\_Pos} computes $N$ for any given $P$ and 
$S$:

\medskip

\fpr{{\it Match\_Pos}\hfill (initial, nondeterministic)}{
\Feq{1.} {\it match\_pos}(P,S,N) \nc \If \nc
        {\it append}(Y,{\it R},S),\
        {\it append}({\it L},P,Y),\  {\it length}({\it L},N)
\Nex{2.} {\it length}([~],0)  \If
\Nex{3.} {\it length}([H|T],s(N))  \If  {\it length}(T,N)
\Nex{4.} {\it append}([~],Y,Y)  \If
\Nex{5.} {\it append}([A|X],Y,[A|Z])  \If  {\it append}(X,Y,Z)
\eeq }

\medskip

\noindent
The ${\it Match\_Pos}$ program is nondeterministic for atoms of the form
${\it match\_pos}(P,S,N)$ where $P$ and $S$ are ground lists, and
it computes one answer substitution for each occurrence of $P$ in $S$.

Suppose that we want to specialize ${\it Match\_Pos}$
w.r.t.~the atom ${\it match\_pos}([a,a,b],S,N)$.
Thus, we want to derive a new, specialized program
${\it Match\_Pos_s}$ and a new binary predicate
${\it match\_pos_s}$. This new program should be able to
compute multiple answer substitutions for a goal. For instance,
for the atom ${\it match\_pos_s}([a,a,b,a,a,b],N)$ the
program ${\it Match\_Pos_s}$ should compute the two substitutions
$\{N/0\}$ and $\{N/s(s(s(0)))\}$ and, thus, ${\it Match\_Pos_s}$
cannot be deterministic for the atom
${\it match\_pos_s}([a,a,b,a,a,b],N)$.

Now, in order to deal with programs which may return multiple answer
substitutions, we
introduce the notion of semideterminism, which is
weaker than that of determinism.
Informally, we may say that a semideterministic program
has the minimum amount of nondeterminism which is needed to compute
multiple answer substitutions.
In Section~\ref{strategy} we will prove that the Determinization Strategy,
if it terminates, derives a semideterministic program.

\begin{definition}  [Semideterminism] \label{semideter_def}
{\rm
A program $P$ is {\em semideterministic} for a non-basic atom  $A$
iff for each goal $G$  such that $A \Rightarrow^*_P G$, there
exists {\em at most one} clause $C$ such that
$G\Rightarrow_C G'$ for some goal $G'$ different from {\it true}.

\noindent
Given a mode $M$ for a program $P$, we say that $P$ is {\em
semideterministic} \wrt $M$ iff $P$ is semideterministic for
each non-basic atom which satisfies $M$.
} %end rm
\end{definition}

We will show in Section~\ref{match_pos_ex} that
by applying the Determinization Strategy, from ${\it Match\_Pos_s}$ we
derive the following specialized program ${\it Match\_Pos_s}$ which
is semideterministic for atoms of the form
${\it match\_pos_s}(S,N)$, where $S$ is a ground list.

\medskip

\fpr{${\it Match\_Pos_s}$\hfill (specialized, semideterministic)}{
\Feq{9.}  {\it match\_pos_s}(S,N) \leftarrow {\it new}1(S,N)
\Nex{20.} {\it new}1([a|S],M) \leftarrow {\it new}2(S,M)
\Nex{21.} {\it new}1([C|S],s(N)) \leftarrow C\!\neq\! a,\ {\it
new}1(S,N) \Nex{32.} {\it new}2([a|S],M) \leftarrow {\it new}3(S,M)
\Nex{33.}  {\it new}2([C|S],s(s(N))) \leftarrow
        C\!\neq\! a,\ {\it new}1(S,N)
\Nex{46.} {\it new}3([a|S],s(M)) \leftarrow {\it new}3(R,S)
\Nex{47.} {\it new}3([b|S],M) \leftarrow {\it new}4(R,S)
\Nex{48.} {\it new}3([C|S],s(s(s(N)))) \leftarrow
        C\!\neq\!a,\ C\!\neq\!b,\
        {\it new}1(S,N)
\Nex{49.} {\it new}4(S,0) \leftarrow
\Nex{55.} {\it new}4([a|S],s(s(s(M)))) \leftarrow {\it new}2(S,M)
\Nex{56.} {\it new}4([C|S],s(s(s(s(N))))) \leftarrow
        C\!\neq\!a,\  {\it new}1(S,N)
\eeq}

\medskip

\noindent
Now we give a simple sufficient condition which ensures
semideterminism. It is based on the concept of {\it mutually
exclusive\/} clauses which we introduce below. We need some
preliminary definitions.

\begin{definition} [Satisfiability of Disequations \wrt
a Set of Variables]
\label{satisfiability}
{\rm
Given a set $V$ of variables, we say that a conjunction  $D$
of disequations, is {\em satisfiable \wrt $V$} iff
there exists a ground substitution
$\sigma$ with domain $V$, such that every ground instance of
$D\sigma$ holds (see Section~\ref{decl_sem}).
In particular,  $D$
is satisfiable \wrt $\emptyset$ iff every ground instance of $D$ holds.
}
%end rm
\end{definition}

The satisfiability of a conjunction $D$ of
disequations \wrt a given set $V$ of variables, can be checked
by using the following algorithm defined by structural induction:

\smallskip
\noindent
(1) ${\it true}$, i.e., the
empty conjunction of disequations, is satisfiable \wrt $V$,

\smallskip\noindent
(2) $(D_1, D_2)$ is satisfiable \wrt $V$ iff both $D_1$ and $D_2$
are satisfiable \wrt $V$,

\smallskip\noindent
(3) $X\!\neq\! t$ is satisfiable \wrt $V$ iff $X$ occurs in $V$
and $t$ is either a non-variable term or a variable occurring in $V$
distinct from $X$,

\smallskip\noindent
(4) $t \neq X$ is satisfiable \wrt $V$ iff
$X\!\neq\! t$ is satisfiable \wrt $V$,

\smallskip\noindent
(5) $f(\ldots) \neq
g(\ldots)$, where $f$ and $g$ are distinct function symbols,
is satisfiable \wrt $V$, and

\smallskip\noindent
(6) $f(t_1,\ldots,t_m) \neq f(u_1,\ldots,u_m)$ is satisfiable \wrt
$V$ iff at least one disequation among $t_1\!\neq\! u_1, \ldots,$
$t_m\!\neq\!u_m$ is satisfiable \wrt $V$.

The correctness of
this algorithm relies on the fact that
the set of function symbols is infinite (see Section \ref{syntax}).

\begin{definition} [Linearity] \label{linear}
{\rm
A program $P$ is said to be {\em linear} iff every clause of $P$
has at most one non-basic atom in its body.
} %end rm
\end{definition}

\begin{definition} [Guard of a Clause]
{\rm
The {\it guard} of a clause $C$, denoted ${\it grd}(C)$, is
${\it bd}(C)$ if all atoms in ${\it bd}(C)$ are disequations,
otherwise ${\it grd}(C)$ is the (possibly empty) conjunction of the
disequations occurring in ${\it bd}(C)$ to the left
of the leftmost atom which is not a disequation.
} %end rm
\end{definition}

\begin{definition}[Mutually Exclusive Clauses]
\label{mutual_excl_def}
{\rm
Let us consider a mode $M$ for the following two,
renamed apart clauses:

\smallskip
$C_1$. $p(t_1,u_1) \leftarrow G_1$

$C_2$. $p(t_2,u_2) \leftarrow G_2$

\smallskip
\noindent
where: (i) $p$ is a predicate of arity $k$ $(\geq\!0)$ whose
first $h$ arguments, with $0\!\leq\! h\!\leq\!
k$, are input arguments according to $M$,
(ii) $t_1$ and $t_2$ are $h$-tuples of
terms denoting the input arguments of $p$, and
(iii) $u_1$ and $u_2$ are $(k\!-\!h)$-tuples of terms.

\noindent
We say that $C_1$ and $C_2$ are {\em mutually exclusive}
\wrt mode $M$ iff
either (i) $t_1$ is not unifiable with $t_2$
or (ii) $t_1$ and $t_2$ are unifiable via an mgu $\vartheta$ and
$({\it grd}(C_1),{\it grd}(C_2))\vartheta$ is not satisfiable
\wrt ${\it vars}(t_1,t_2)$.

\noindent
If $h\!=\!0$ we stipulate that the empty tuples $t_1$ and $t_2$
are unifiable via an mgu which is the identity
substitution. } %end rm
\end{definition}

The following proposition is useful for proving that a
program is semideterministic.

\begin{proposition}[Sufficient Condition for Semideterminism]
\label{mutual_excl}
{\rm
If (i) $P$ is a linear program, (ii) $P$ is safe \wrt a given
mode $M$, (iii) $P$ satisfies $M$, and (iv) the non-unit clauses of
$P$ are pairwise mutually exclusive \wrt $M$, then $P$ is
semideterministic \wrt $M$.} %end rm
\end{proposition}

\begin{proof} See Appendix B. \end{proof}

In Section~\ref{strategy}, we will present a strategy for deriving
specialized programs which satisfies the hypotheses~(i)--(iv) of the above
Proposition~\ref{mutual_excl}, and thus, these derived programs are
semideterministic.

The following examples show that in Proposition~\ref{mutual_excl}
no hypothesis on program $P$ can be discarded.

\begin{example}
{\rm
Consider the following program $P$ and the mode
$M = \{p, q\}$ for $P$:

\bpr
\Feq{1.} p \leftarrow q,\ q
\Nex{2.} q \leftarrow
\Nex{3.} q \leftarrow q
\eeq
\epr

\noindent
$P$ is not linear, but $P$ is safe \wrt $M$ and $P$
satisfies $M$. The non-unit clauses of $P$ which are the clauses 1 and 3,
are pairwise mutually exclusive. However, $P$ is not semideterministic \wrt
$M$, because $p \longmapsto^*_P (q,q)$, and there exist two non-basic goals,
namely $q$ and $(q,q)$, such that $(q,q) \Rightarrow_P q$ and $(q,q)
\Rightarrow_P (q,q)$.
} %end rm
\end{example}

\begin{example}
{\rm
Consider the following program $Q$ and the mode
$M = \{p(?), q_1, q_2\}$ for $Q$:

\bpr
\Feq{1.} p(X) \leftarrow X\!\neq\!0,\ q_1
\Nex{2.} p(1) \leftarrow q_2
\eeq
\epr

\noindent
$Q$ is linear and it
satisfies $M$, but $Q$ is not safe \wrt $M$ because $X$ is not an
input variable of $p$. Clauses 1 and 2 are mutually exclusive \wrt
$M$, because the set of input variables in $p(X)$ is empty and
$X\!\neq\!0$ is not satisfiable \wrt $\emptyset$. However, $Q$ is not
semideterministic \wrt $M$, because $p(1) \longmapsto^*_Q p(1)$, and
there exist two non-basic goals, namely $q_1$ and $q_2$, such that
$p(1) \Rightarrow_Q q_1$ and $p(1) \Rightarrow_Q q_2$.} %end rm
\end{example}

\begin{example}
{\rm
Consider the following program $R$ and the mode
$M = \{p, r(+), r_1, r_2\}$ for $R$:

\bpr
\Feq{1.} p \leftarrow r(X)
\Nex{2.} r(1) \leftarrow r_1
\Nex{3.} r(2) \leftarrow r_2
\eeq
\epr

\noindent
$R$ is linear and safe \wrt $M$, but $R$ does not satisfy $M$,
because $p \longmapsto_R r(X)$ and $X$ is not a ground term. Clauses 1, 2,
and 3 are pairwise mutually exclusive. However, $R$ is not
semideterministic \wrt $M$, because $p \longmapsto^*_R r(X)$ and there
exist two non-basic goals, namely $r_1$ and $r_2$, such that
$r(X) \Rightarrow_R r_1$ and $r(X) \Rightarrow_R r_2$.} %end rm
\end{example}

\begin{example}
{\rm
Consider the following program $S$ and the mode
$M = \{p, r_1, r_2\}$ for $S$:

\bpr
\Feq{1.}  p \leftarrow r_1
\Nex{2.} p \leftarrow r_2
\eeq
\epr

\noindent
$S$ is linear and safe \wrt $M$, and $S$ satisfies $M$.
Clauses 1 and 2 are not pairwise mutually
exclusive. $S$ is not semideterministic \wrt $M$, because $p
\longmapsto^*_S p$, and there exist two non-basic goals, namely $r_1$ and
$r_2$, such that $p \Rightarrow_S r_1$ and
$p \Rightarrow_S r_2$.} %end rm
\end{example}

We conclude this section by observing that
when a program consists of mutually exclusive clauses and, thus,
it is semideterministic, it may be executed very efficiently
on standard Prolog systems by inserting cuts in a suitable
way. We will return to this point in Section~\ref{statistics}
when we discuss the speedups obtained by our specialization
technique.

\section{A Transformation Strategy for Specializing Programs
and Reducing Nondeterminism} \label{strategy}

In this section we present a strategy, called {\em Determinization},
for guiding the application of the transformation rules presented in
Section~\ref{rules}.
Our strategy pursues the following objectives. (1)~The specialization
of a program \wrt a particular goal. This is similar to
what partial deduction does. (2)~The elimination of multiple or
intermediate data structures. This is similar to what the strategies
for eliminating {\em unnecessary variables}~\cite{PrP95a} and conjunctive
partial deduction do. (3)~The reduction of nondeterminism. This is
accomplished by deriving programs whose non-unit clauses
are mutually exclusive \wrt a given mode, that is, by
Proposition~\ref{mutual_excl}, semideterministic programs.

The Determinization Strategy
is based upon three subsidiary strategies:
(i)~the {\em Unfold-Simplify} subsidiary strategy,
which uses the safe unfolding, equation elimination,
disequation replacement, and subsumption rules,
(ii)~the {\em Partition} subsidiary strategy,
which  uses  the safe case split,
equation elimination, disequation replacement,
subsumption, and safe head generalization rules, and
(iii)~the {\em Define-Fold} subsidiary strategy  which
uses the definition introduction and safe folding rules.
For reasons of clarity, during the presentation of the
Determinization Strategy we use high-level
descriptions of the subsidiary strategies.
These descriptions are used to establish the
correctness of Determinization (see Theorem~\ref{determ_corr}).
Full details of the subsidiary strategies will be given
in Sections~\ref{unfolding},
\ref{part_proc}, and~\ref{def_fold_strat}, respectively.

\subsection{The Determinization Strategy}\label{determ_strat}

Given an initial program $P$,
a mode $M$ for $P$, and an atom
$p(t_1,\dots,t_h)$ \wrt which we want to
specialize $P$,  we introduce  by the
definition introduction rule, the clause

\smallskip
 $S$: $p_s(X_1,$ $\dots,
$ $X_r) \leftarrow p(t_1,\dots,t_h)$
\smallskip

\noindent
where $X_1,\dots,X_r$ are
the distinct variables occurring in $p(t_1,\dots,t_h)$.

\noindent
We also define a mode  $p_s(m_1,$ $\dots, $ $m_r)$ for the
predicate $p_s$  by stipulating that, for any $j= 1,\ldots, r,\ $
 $m_j$ is $+$ iff  $X_j$ is an input
variable of $p(t_1,\dots,t_h)$  according to the mode $M$.
We assume that the program $P$ is safe \wrt $M$. Thus, also
program $P\cup \{S\}$ is safe w.r.t.
$M \cup \{p_s(m_1,$ $\dots, $ $m_r)\}$. We also assume that
$P$ satisfies mode $M$ and thus,  program $P \cup \{S\}$
satisfies mode $M \cup \{p_s(m_1,\dots,m_r)\}$.

Our Determinization Strategy is
presented below as an iterative procedure that,
at each iteration,
manipulates the following three sets of clauses:
 (1) {\it TransfP}, which is the set of
clauses from which we will construct the specialized program,
 (2) {\it Defs}, which is the set of clauses
introduced by the definition introduction rule, and
 (3) {\it Cls}, which is the set of clauses to be transformed
during the current iteration.
Initially, {\it Cls} consists of the single clause $S$:
$p_s(X_1,\dots,X_r) \leftarrow p(t_1,\dots,t_h)$ which is constructed
as we have indicated above.

The Determinization Strategy starts off each iteration
by applying the Unfold-Simplify subsidiary strategy to
the set {\it Cls}, thereby deriving a new set
of clauses called {\it UnfoldedCls}.
The Unfold-Simplify strategy first unfolds the
clauses in {\it Cls},
and then it simplifies the derived set of clauses by applying the
equation elimination, disequation replacement, and subsumption rules.

Then the set {\it UnfoldedCls} is divided into two sets:
(i)~{\it UnitCls}, which is the set of unit clauses, and
(ii)~{\it NonunitCls}, which is the set of non-unit clauses.
The Determinization Strategy proceeds by applying the Partition
subsidiary strategy to {\it NonunitCls}, thereby deriving a new set
of clauses called {\it PartitionedCls}.
The Partition strategy consists of suitable applications
of the case split, equation elimination, disequation
replacement, and head generalization rules such that the
set {\it PartitionedCls} has the following property:
it can be partitioned into sets of clauses, called {\em packets},
such that two clauses taken from different packets are
mutually exclusive (\wrt a suitable mode).

The Determinization Strategy continues by applying
the Define-Fold subsidiary strategy to the clauses in
{\it PartitionedCls}, thereby deriving
a new, semideterministic set of
clauses called {\it FoldedCls}.
The Define-Fold subsidiary strategy
introduces a (possibly empty) set {\it NewDefs} of
definition clauses such that
each packet can be folded into a single clause
by using a set of definition
clauses in ${\it Defs} \cup {\it NewDefs}$.
We have that clauses derived by folding different
packets are mutually exclusive and, thus,
$\mathit{UnitCls\cup FoldedCls}$ is semideterministic.

At the end of each iteration,
$\mathit{UnitCls\cup FoldedCls}$ is added to
{\it TransfP}, {\it NewDefs} is
added to {\it Defs}, and the value of the set {\it Cls}
is updated to {\it NewDefs}.

The Determinization Strategy terminates
when ${\it Cls}\!=\!\emptyset$, that is, no new predicate is
introduced during the current iteration.

%\bracetop
\noindent \hrulefill

\noindent
{\large {\bf Determinization Strategy }}

\smallskip
\noindent
{\bf Input\/}: A program $P$, an atom
$p(t_1,\ldots,t_h)$ \wrt which we want to
specialize $P$, and a mode $M$ for $P$ such that $P$ is safe
\wrt $M$  and  $P$ satisfies $M$.

\noindent
{\bf Output}: A specialized program $P_\mathit{s}$, and an
atom $p_s(X_1,\!\ldots,X_r)$,
with $\{X_1,\!\ldots,X_r\}$ $\!=\!$
$\mathit{vars}(p(t_1,\!\ldots,t_h))$ such that: (i) for every ground
substitution $\vartheta = \{X_1/u_1,\ldots,X_r/u_r\}$, $M(P)\models
p(t_1,\ldots,t_h)\vartheta$ iff $M(P_s)\models
p_s(X_1,\ldots,X_r)\vartheta$, and (ii) for every substitution $\sigma
= \{X_1/v_1,\ldots,X_r/v_r\}$ such that the atom
$p(t_1,\ldots,t_h)\sigma$ satisfies mode $M$, we have that: (ii.1)
$p(t_1,\ldots,t_h)\sigma$ succeeds in $P$ iff
$p_s(X_1,\ldots,X_r)\sigma$ succeeds in $P_s$, and (ii.2) $P_s$ is
semideterministic for $p_s(X_1,\ldots,X_r)\sigma$.

\bigskip

\noindent  {\it Initialize}:~~Let $S$ be the clause
$p_s(X_1,\ldots,X_r)\If p(t_1,\ldots,t_h)$.
\\
 ~${\it TransfP} := P$; ~$\mathit{Defs}:=\{S\}$;
~$\mathit {Cls}:= \{S\}$;
~$M_s := M \cup \{p_s(m_1,\dots,m_r)\}$, where for any
$j= 1,\ldots, r,\ $
 $m_j = +$ iff  $X_j$ is an input
 variable of $p(t_1,\dots,t_h)$  according to the mode $M$;

\medskip \noindent
{\bf while} ~\( \mathit {Cls}\neq \emptyset  \)~ {\bf do}

\vspace{-2mm}

\begin{enumerate}

\item[{\rm (1)}] {\em Unfold-Simplify}:\\
We apply the safe unfolding, equation elimination,
disequation replacement, and subsumption rules according
to the Unfold-Simplify Strategy given in Section~\ref{unfolding}
below, and from {\it Cls} we derive a new set of clauses
$\mathit{UnfoldedCls}$.

\item[{\rm (2)}]{\em Partition}:\\
Let $\mathit{UnitCls}$ be the unit clauses occurring in
$\mathit{UnfoldedCls}$, and $\mathit{NonunitCls}$ be the
set of non-unit clauses in $\mathit{UnfoldedCls}$.

We apply the safe case split, equation elimination, disequation
replacement, and safe head generalization rules according to the
Partition Strategy given in Section~\ref{part_proc} below, and from
$\mathit{NonunitCls}$ we derive a set $\mathit{PartitionedCls}$
of clauses which is the union of disjoint
subsets of clauses. Each subset is called a {\it packet}.
The packets of $\mathit{PartitionedCls}$
enjoy the following properties:

\noindent
(2a) each packet is a set of clauses of the form (modulo
renaming of variables):

\smallskip

~~~~~~~~$\left\{
\begin{array}{ll}
H  \leftarrow  \mathit{Diseqs}, G_1\\
  \hspace{.5cm} \cdots & \\
H  \leftarrow \mathit{Diseqs}, G_m &
\end{array}
\right.$

\smallskip

\noindent
where $\mathit{Diseqs}$ is a conjunction of disequations and for
$k=1,\ldots,m$, no disequation occurs in $G_k$, and

\noindent
(2b) for any two clauses $C_1$ and $C_2$, if the packet of
$C_1$ is different from the packet of $C_2$, then $C_1$ and $C_2$ are
mutually exclusive \wrt mode $M_s$.

\item[{\rm (3)}] {\em Define-Fold}:

We apply the definition introduction and the safe folding rules
according to the Define-Fold subsidiary strategy
given in Section~\ref{def_fold_strat} below.
According to that strategy, we introduce a (possibly empty)
set {\it NewDefs} of new definition clauses and a set $M_\mathit{new}$
of modes such that:
\vspace{-.2cm}
\begin{enumerate}
\item[(3a)] in $M_\mathit{new}$ there exists exactly one
mode for each distinct head predicate in {\it NewDefs\/}, and

\item[(3b)] from each packet in $\mathit{PartitionedCls}$
we derive a single clause of the form:

\smallskip

~~~~~~~~$H \leftarrow \mathit{Diseqs}, {\it newp}(\dots)$

\smallskip
by an application of the folding rule, which is safe \wrt $M_\mathit{new}$,
using the clauses in ${\it Defs}\cup {\it NewDefs}$.

\end{enumerate}

Let $\mathit{FoldedCls}$ be the set of clauses derived
by folding the packets in $\mathit{PartitionedCls}$.

\item[{\rm (4)}]
${\it TransfP} := {\it TransfP}\cup \mathit{UnitCls}\cup
\mathit{FoldedCls}$; ~$\mathit{Defs} := \mathit{Defs}\cup
\mathit{NewDefs}$; ~$\mathit{Cls}:= \mathit{NewDefs}$;

~$M_s := M_s\cup M_\mathit{new}$

\end{enumerate}

\vspace{-2mm}

\noindent
{\bf end-while}

\smallskip

\noindent
We derive the specialized program $P_s$ by applying
the definition elimination rule and keeping only the clauses of
{\it TransfP} on which $p_s$ depends.

%\bracebottom
\noindent  \hrulefill

\medskip

\noindent
The Determinization Strategy may fail to terminate for two reasons:
(i)~the Unfold-Simplify subsidiary strategy may not
terminate, because it may perform infinitely many unfolding steps, and
(ii)~the condition $\mathit {Cls}\neq \emptyset$ for exiting the while-do
loop may always be false, because at each iteration the
Define-Fold subsidiary strategy may introduce new definition clauses.
We will discuss these issues in more detail in
Section~\ref{discussion}.

Now we show that, if the Determinization Strategy
terminates, then the least Herbrand model and the operational
semantics are preserved. Moreover, the derived specialized program
$P_s$ is semideterministic for $p_s(X_1,\ldots,X_r)\sigma$ as
indicated by the following theorem.

\sloppy
\begin{theorem}[Correctness of the Determinization Strategy]
\label{determ_corr}
{\rm
Let us consider a program $P$, a non-basic atom $p(t_1,\ldots,t_h)$,
and a mode $M$ for $P$ such that: (1) $P$ is safe \wrt $M$ and (2)
$P$ satisfies $M$. If the Determinization Strategy terminates
with output program $P_s$ and output atom $p_s(X_1,\ldots,X_r)$ where
$\{X_1,\ldots,X_r\} = \mathit{vars}(p(t_1,\ldots,t_h))$, then
\begin{itemize}
\item[(i)] for every ground substitution  $\vartheta =
\{X_1/u_1,\ldots,X_r/u_r\}$,

~~~~$M(P)\models p(t_1,\ldots,t_h)\vartheta$ iff
$M(P_s)\models p_s(X_1,\ldots,X_r)\vartheta$ \ \ \ and

\item[(ii)] for every substitution $\sigma = \{X_1/v_1,\ldots,X_r/v_r\}$
such that the atom $p(t_1,\ldots,t_h)\sigma$ satisfies mode $M$,

\begin{itemize}
\vspace{-3mm}
\item[(ii.1)] $p(t_1,\ldots,t_h)\sigma$ succeeds in $P$
iff $p_s(X_1,\ldots,X_r)\sigma$ succeeds in $P_s$, and

\item[(ii.2)] $P_s$ is semideterministic for $p_s(X_1,\ldots,X_r)\sigma$.

\end{itemize}
\end{itemize}
} % end rm
\end{theorem}
\fussy

\begin{proof}
Let {\it Defs} and $P_s$ be
the set of definition clauses and the specialized program
obtained at the end of the Determinization Strategy.

\smallskip
\noindent
(i) Since $p_s(X_1,\ldots,X_r)\If p(t_1,\ldots,t_h)$ is the only
clause for $p_s$ in $P\cup \mathit{Defs}$ and
$\{X_1,\ldots,X_r\} = \mathit{vars}(p(t_1,\ldots,t_h))$,
 for every ground substitution $\vartheta =
\{X_1/u_1,\ldots,X_r/u_r\}$ we have that  $M(P)\models
p(t_1,\ldots,t_h)\vartheta$ iff $M(P\cup \mathit{Defs})\models
p_s(X_1,\ldots,X_r)\vartheta$. By the correctness of the
transformation rules \wrt the least Herbrand model (see
Theorem~\ref{corr_th_lhm}), we have that
$M(P\cup \mathit{Defs})\models p_s(X_1,\ldots,X_r)\vartheta$ iff
$M(P_s) \models p_s(X_1,\ldots,X_r)\vartheta$.

\smallskip
\noindent
Point (ii.1) follows from Theorem~\ref{corr_th_op}
because during the Determinization Strategy,
each application of the unfolding, folding,
head generalization,
and case split rule is safe.

\smallskip

\noindent
(ii.2) We first observe that, by construction,
for every substitution $\sigma$,
the atom $p(t_1,\ldots,t_h)\sigma$ satisfies mode $M$
iff $p_s(X_1,\ldots,X_r)\sigma$ satisfies mode $M_s$, where $M_s$ is
the mode obtained  from $M$ at the end of the Determinization
Strategy. Thus, Point (ii.2) can be shown by
proving that $P_s$ is semideterministic \wrt $M_s$. In order to
prove this fact, it is enough to prove that ${\it TransfP_w}\! -\! P$
is semideterministic \wrt $M_s$, where ${\it TransfP_w}$
is the set of clauses which is the value of the variable {\it TransfP\/}
at the end of the while-do statement of the Determinization Strategy.
Indeed, $P_s$ is equal to ${\it TransfP_w}\!-\!P$
because, by construction, $p_s$ does not depend on any clause of $P$,
and thus, by the final application of the definition elimination rule,
all clauses of $P$ are removed from ${\it TransfP_w}$.

By Proposition~\ref{mutual_excl},
it is enough to prove that:
(a) ${\it TransfP_w}\! -\! P$ is linear,
(b) ${\it TransfP_w}\! -\! P$ is safe w.r.t.~$M_s$,
(c) ${\it TransfP_w}\! -\! P$ satisfies $M_s$, and
(d) the non-unit clauses of ${\it TransfP_w}\! -\! P$ are pairwise
mutually exclusive \wrt $M_s$.

Property (a) holds because according to the Determinization
Strategy, after every application of the safe folding rule we get a
clause of the form: $H \If  \mathit{Diseqs}, {\it newp}(\ldots)$, where a
single non-basic atom occurs in the body. All other clauses in ${\it
TransfP_w}\! -\! P$ are unit clauses.

Properties (b) and (c) follow from Theorem~\ref{corr_th_op}
recalling that the application of the unfolding, folding, head
generalization, and case split rules are all safe.

Property (d) can be proved by showing that, during the execution of the
Determinization Strategy, the following Property~(I) holds: all the
non-unit clauses of ${\it TransfP} \! -\! P$ are pairwise mutually
exclusive \wrt $M_s$.
Indeed, initially ${\it TransfP} \! -\! P$ is empty and thus, Property~(I)
holds. Furthermore, Property~(I) is an invariant of the while-do
loop. Indeed, at the end of each execution of the body of the while-do (see
Point (4) of the strategy), the non-unit clauses which are added to the
current value of ${\it TransfP\/}$ are the elements of the set {\it
FoldedCls} and those non-unit clauses are derived by applying the Partition
and Define-Fold subsidiary strategies at Points (3) and (4), respectively.
By construction, the clauses in {\it FoldedCls} are pairwise mutually
exclusive \wrt $M_{new}$, and their head predicates do not occur in ${\it
TransfP\/}$. Thus, the clauses of ${\it TransfP\/} \cup {\it UnitCls}
\cup {\it FoldedCls}$  are pairwise mutually exclusive \wrt $M_s \cup
M_{new}$. As a consequence, after the two assignments (see Point (4)
of the strategy) ${\it TransfP\/} := {\it TransfP\/} \cup {\it
UnitCls} \cup {\it FoldedCls}$ and $M_s := M_s \cup M_{new}$, we have
that Property~(I) holds.
\end{proof}

Now we describe the three subsidiary
strategies for realizing the {\em Unfold-Simplify},
{\em Partition}, and {\em Define-Fold} transformations
as specified by the Determinization Strategy.
We will see these subsidiary strategies in action in the examples of
Section~\ref{examples}.

During the application of our subsidiary strategies it will be
convenient to rewrite every safe clause into its {\em normal form}.
The normal form $N$ of a safe
clause can be constructed by performing disequation replacements
 and disequation promotions, so that the
following Properties N1--N5 hold:

\smallskip
\noindent
(N1) every disequation is of the form: $X\!\neq \!t$, with $t$ different
from $X$ and unifiable with $X$,

\noindent
(N2) every disequation occurs in $bd(N)$  to the
left of every atom different from a disequation,

\noindent
(N3) if $X\!\neq\! Y$ occurs in $bd(N)$ and both $X$ and $Y$ are
input variables of $hd(N)$,
then in $\mathit{hd}(N)$ the leftmost occurrence of $X$ is to the left
of the leftmost occurrence of $Y$,

\noindent
(N4) for every disequation of the form $X\!\neq\! Y$ where Y is an input
variable, we have that also $X$ is an input variable, and

\noindent
(N5) for any pair of disequations $d_1$ and $d_2$
in $bd(N)$, it does not exist a substitution $\rho$ which is a
bijective mapping from the set of the local variables of $d_1$ in $N$
onto the set of the local variables of $d_2$ in $N$ such
that $d_1\rho = d_2$.

\smallskip
\noindent
We have that: (i)~the normal form of a safe clause is unique, modulo
renaming of variables and disequation promotion,
(ii)~no two equal disequations occur in the normal form of a safe clause,
and (iii) given a
program $P$ and a mode $M$ for $P$ such that $P$ is safe \wrt $M$ and $P$
satisfies $M$, if we rewrite a clause of $P$ into its normal form, then the
least Herbrand model semantics and the operational semantics are preserved
(this fact is a consequence of Theorem~\ref{corr_th_lhm},
Theorem~\ref{corr_th_op}, and Proposition \ref{rearr_prop}).

A safe clause for which Properties N1--N5 hold, is
said to be {\em in normal form}.  If a clause $C$  is in normal form,
then by Property N2, every disequation in ${\it bd}(C)$ occurs also
in ${\it grd}(C)$.

\subsection{The Unfold-Simplify Subsidiary Strategy}
\label{unfolding}

The Unfold-Simplify strategy first unfolds the clauses in {\it Cls} \wrt
the leftmost atom in their body, and then it keeps unfolding the derived
clauses as long as input variables are not instantiated. Now, in order to
give the formal definition of the Unfold-Simplify strategy we introduce the
following concept.

\begin{definition}[Consumer Atom]
{\rm
Let $P$ be a program and $M$ a mode for $P$.
A non-basic atom $q(t_1,\dots,t_k)$
is said to be a {\it consumer atom} iff for every non-unit clause in $P$
whose head unifies with that non-basic atom via an mgu $\vartheta$, we
have that for $i=1,\dots,k$, if $t_i$ is an input argument of $q$ then
$t_i\vartheta$ is a variant of $t_i$. } %end rm
\end{definition}

The Unfold-Simplify strategy is realized by the following {\em
Unfold-Simplify} procedure, where the expression ${\it Simplify}({\it
S})$ denotes the set of clauses derived from a given set
{\it S} of clauses by: (1) first, applying whenever
possible the equation elimination rule to the clauses in $S$, (2) then,
rewriting the derived clauses into their normal form, and
(3) finally, applying as long as possible the subsumption rule.

%\bracetop
\noindent \hrulefill

\noindent
{\bf Procedure}
{\em Unfold-Simplify}$({\it Cls},\mathit{UnfoldedCls})$.

\noindent
{\bf Input\/}: A set {\it Cls} of clauses in a program $P$
and a mode $M_s$ for $P$. $P$ is safe w.r.t.~$M_s$ and
for each $C\in\mathit{Cls}$, the input
variables of the leftmost non-basic atom in the body of $C$ are input
variables of the head of $C$.

\noindent
{\bf Output\/}: A new set $\mathit{UnfoldedCls}$ of clauses
which are derived from $\mathit{Cls}$ by applying the safe
unfolding, equation elimination, disequation replacement, and
 subsumption rules. The clauses in {\it UnfoldedCls} are safe
w.r.t.~$M_s$.

\medskip

\noindent
(1) {\em Unfold \wrt Leftmost Non-basic Atom\/}:
\vspace*{-.1cm}
\begin{description}
\item[]~~~~~~$\mathit{UnfoldedCls}:=\{E~|~$ there exists a clause
$C\in\mathit{Cls}$ and clause $E$ is derived by unfolding $C$ \wrt
\\
\hspace*{3.6cm}the leftmost non-basic atom in its body$\}$;\\
~~~~~~~~~~~~~~~~~~~~$\mathit{UnfoldedCls}:=
\mathit{Simplify}(\mathit{UnfoldedCls})$
\end{description}
\vspace*{-.2cm}

\noindent
(2) {\em Unfold \wrt Leftmost Consumer Atom\/}:
\vspace*{-.1cm}
\begin{description}
\item[]~~~~~~{\bf while}~there exists a clause $C\in\mathit
{UnfoldedCls}$ whose body has a leftmost consumer atom, say $A$,
such that the unfolding of $C$ w.r.t.~$A$ is safe {\bf do}\\
$\mathit{UnfoldedCls}:= (\mathit{UnfoldedCls} - \{C\})  \cup \{ E~|~E$
is derived by unfolding $C$ w.r.t.~$A\}$;\\
$\mathit{UnfoldedCls}:= \mathit{Simplify}(\mathit{UnfoldedCls})$
\vspace*{-.2cm}
\item[]~~~~~~{\bf end-while}
\end{description}
\vspace*{-.4cm}

%\bracebottom
\noindent \hrulefill

\smallskip
\noindent
Notice that our assumptions on the input
program $P$ and clauses {\it Cls} ensure that the first unfolding step
performed by the {\em Unfold-Simplify} procedure
is safe.

Notice also that our Unfold-Simplify strategy may fail to terminate.
We will briefly return to this issue in Section~\ref{discussion}.

Our Unfold-Simplify strategy differs from usual unfolding strategies for
(conjunctive) partial deduction (see, for
instance,~\cite{De&99,Gall93,Pre93b,Sah93}), because mode information is
used. We have found this strategy very effective on several examples
as shown in the following Section~\ref{examples}.

\subsection{The Partition Subsidiary Strategy}\label{part_proc}

The Partition strategy is realized by the following procedure, where
we will write $p(t,u)$ to denote an atom with non-basic predicate $p$
of arity $k\ (\geq 0)$,  such that: (i) $t$ is
an $h$-tuple of terms, with $0\!\leq \! h\! \leq \! k$, denoting the
$h$ input arguments of $p$, and (ii) $u$ is a $(k\!-\!h)$-tuple of
terms denoting the arguments of $p$ which are {\it not\/} input arguments.

%\bracetop
\noindent \hrulefill

\noindent
{\bf Procedure} {\em Partition}$({\it NonunitCls}, {\it
PartitionedCls})$.

\noindent
{\bf Input\/}: A set {\it NonunitCls} of non-unit clauses
in normal form and without variables in common. A mode $M_s$ for {\it
NonunitCls\/}. The clauses in {\it NonunitCls} are safe w.r.t.~$M_s$.

\noindent
{\bf Output\/}: A set {\it PartitionedCls} of clauses which is the
union of disjoint packets of clauses such that:\\
(2a) each packet is a set of clauses of the form (modulo
renaming of variables):

\smallskip
~~~~~~~~$\left\{
\begin{array}{ll}
H  \leftarrow   \mathit{Diseqs}, G_1\\
  \hspace{.5cm} \cdots & \\
H  \leftarrow   \mathit{Diseqs}, G_m &
\end{array}
\right.$

\smallskip
\noindent
where $\mathit{Diseqs}$ is a conjunction of disequations and for
$k=1,\ldots,m$, no disequation occurs in $G_k$, and

\noindent
(2b) for any two clauses $C_1$ and $C_2$, if the packet of
$C_1$ is different from the packet of $C_2$, then $C_1$ and $C_2$ are
mutually exclusive \wrt mode $M_s$.

\noindent
The clauses in {\it PartitionedCls} are in normal form and they are safe
w.r.t.~$M_s$.

\bigskip

\noindent
{\bf while} there exist in {\it NonunitCls} two clauses of
the form:

\bpr
\Feq{$C_1$.}p(t_1,u_1) \leftarrow {\it Body}_1
\Nex{$C_2$.}p(t_2,u_2) \leftarrow {\it Body}_2
\eeq
\epr

\noindent
such that: (i) $C_1$ and $C_2$ are not mutually exclusive
\wrt mode $M_s$, and {\it either} \\
(ii.1) $t_1$ is not a variant of $t_2$ {\it or} \\ (ii.2) $t_1$ is a
variant of $t_2$ via an mgu $\vartheta$ such that $t_1\vartheta\!=\!t_2$,
and for any substitution $\rho$ which is a bijective mapping from the set
of local variables of ${\it grd}(C_1\vartheta)$ in $C_1\vartheta$ onto the
set of local variables of ${\it grd}(C_2)$ in $C_2$,
${\it grd}(C_1\vartheta\rho)$ cannot be made syntactically equal to ${\it
grd}(C_2)$ by applying disequation promotion {\bf do}

%%%%%%
% above:
%       ${\it grd}(C_1\vartheta)$ in $C_1\vartheta$
% can be replaced by the simpler, but less understandable:
%       ${\it grd}(C_1)$ in $C_1$
%%%%%%%

\vspace{-.2cm}
\begin{description}
%%%%%%
\item [] We take a binding $X/r$ as follows.

(Case 1) Suppose that $t_1$ is {\em not} a variant of
$t_2$. In this case, since $C_1$ and $C_2$ are not mutually exclusive,
we have that $t_1$ and $t_2$ are unifiable and, for some $i,j \in \{1,2\}$,
with $i\!\neq\! j$, there exists an mgu $\vartheta$ of
$t_i$ and $t_j$ and a binding $Y/t_a$ in $\vartheta$ such that
$t_j\{Y/t_a\}$ is not a variant of $t_j$.
Without loss of generality we may assume that
$i\!=\!1$ and $j\!=\!2$. Then we take the binding $X/r$ to be $Y/t_a$.

(Case 2) Suppose that $t_1$ is a variant of $t_2$ via an mgu $\vartheta$.
Now every safe clause whose normal form has a disequation of the form
$X\deq t$, where $X$ is a local variable of that disequation in that
clause, is mutually exclusive \wrt any other safe clause. This is the
case because, for any substitution $\sigma$ which does not bind $X$,
$t\sigma$ is unifiable with $X$ and, thus, $X\deq t\sigma$ is not
satisfiable.   Thus, for some $i,j \in \{1,2\}$, with $i\!\neq\! j$, there
exists a disequation $(Y\!\neq\!t_a)\vartheta$ in ${\it grd}(C_i\vartheta)$
where $Y\vartheta$ is an input variable of ${\it hd}(C_i\vartheta)$, such
that for any substitution $\rho$ which is a bijective mapping from the set
of local variables of ${\it grd}(C_i\vartheta)$ in $C_i\vartheta$  onto the
set of local variables of ${\it grd}(C_j\vartheta)$ in $C_j\vartheta$ and
for every disequation $(Z\!\neq\!t_b)\vartheta$ in ${\it
grd}(C_j\vartheta)$, we have that $(Y\!\neq\!t_a)\vartheta\rho$ is
different from $(Z\!\neq\!t_b)\vartheta$. We also have that $Y\vartheta$ is
an input variable of ${\it hd}(C_j\vartheta)$. Without loss of generality
we may assume that $i\!=\!1$, $j\!=\!2$, $t_1\vartheta\!=\!t_2$, and
$C_2\vartheta\!=\!C_2$. Then we take the binding $X/r$ to be
$(Y/t_a)\vartheta$.

%%%%%%%%%%
\item [] We apply the case split rule to clause $C_2$ \wrt $X/r$,
that is, we derive the two clauses:

\hspace*{.5cm}
  $C_{21}$. $(p(t_2,u_2) \leftarrow {\it Body}_2)\{X/r\}$\\
\hspace*{.5cm}
  $C_{22}$. $p(t_2,u_2) \leftarrow X\!\neq\!r, {\it Body}_2$

\item [] We update the value of {\it NonunitCls} as follows:\\
${\it NonunitCls} := ({\it NonunitCls} - \{C_2\}) \cup
\{C_{21},C_{22}\}$\\
 ${\it NonunitCls} := {\it Simplify}({\it NonunitCls})$.

\noindent
\hspace{-1cm} {\bf end-while}
\end{description}

\noindent
Now the set {\it NonunitCls\/} is partitioned into
subsets of clauses and after suitable renaming of variables and
disequation promotion, each subset is of the form:

\smallskip

~~~~~~~~$\left\{
\begin{array}{ll}
p(t,u_1)  \leftarrow  \mathit{Diseqs}, {\it Goal}_1\\
  \hspace{.5cm} \cdots & \\
p(t,u_m)  \leftarrow  \mathit{Diseqs}, {\it Goal}_m &
\end{array}
\right.$

\smallskip
\noindent
where $\mathit{Diseqs}$ is a conjunction of disequations and for
$k=1,\ldots,m$, no disequation occurs in ${\it Goal}_k$,
and any two clauses in different subsets are
mutually exclusive \wrt mode $M_s$.

\smallskip \noindent
Then we process every subset of clauses we have derived, by applying
the safe head generalization rule so to replace the non-input
arguments in the heads of the clauses belonging to the same subset by
their most specific common generalization. Thus, every subset of
clauses will eventually take the form:

\smallskip
~~~~~~~~$\left\{
\begin{array}{ll}
p(t,u)  \leftarrow  {\it Eqs}_1, \mathit{Diseqs}, {\it Goal}_1\\
  \hspace{.5cm} \cdots & \\
p(t,u)  \leftarrow  {\it Eqs}_m, \mathit{Diseqs}, {\it Goal}_m &
\end{array}
\right.$

\smallskip \noindent
where $u$ is the most specific common generalization
of the terms $u_1,\dots,u_m$ and,
for $k=1,\ldots,m$, the goal ${\it Eqs}_k$ is a conjunction
of the equations $V_1\!=\!v_1,\dots,V_r\!=\!v_r$ such that
$u\{V_1/v_1,\dots,V_r/v_r\} = u_k$.

\smallskip
\noindent
{F}inally, we move all disequations to the leftmost positions of the body
of every clause whereby getting the set {\it PartitionedCls\/}.
\nopagebreak

%\bracebottom
\noindent \hrulefill

\medskip \noindent
Notice that in the above procedure the application of the case split rule
to clause $C_2$ \wrt $X/r$ is safe because: (i)~clauses $C_1$ and $C_2$ are
safe w.r.t.~$M_s$, (ii)~$X$ is an input variable of ${\it hd}(C_{22})$
(recall that our choice of $X/r$ in Case 2 ensures that $X$ is an input
variable of ${\it hd}(C_2)$), and (iii)~each variable in $r$ is either an
input variable of ${\it hd}(C_{22})$ or a local variable of $X\deq r$ in
$C_{22}$. Thus, clauses $C_{21}$ and $C_{22}$ are safe w.r.t. mode $M_s$
and they are also mutually exclusive \wrt $M_s$.

The following property is particularly important for
the mechanization of our Determinization Strategy.

\begin{theorem} \label{thm:termination_of_partition}
{\rm The Partition procedure terminates.} \end{theorem}
\begin{proof} See Appendix C. \end{proof}

When the Partition procedure terminates, it returns a set {\it
PartitionedCls} of clauses which is the union of packets of clauses
enjoying Properties (2a) and (2b) indicated in the Output
specification of that procedure. These properties are a
straightforward consequence of the termination condition of the
while-do statement of that same procedure.

\subsection{The Define-Fold Subsidiary Strategy}
\label{def_fold_strat}

The Define-Fold strategy is realized by the following procedure.

%\bracetop
\noindent  \hrulefill

\noindent
{\bf Procedure}
{\em Define-Fold}$({\it PartitionedCls},{\it Defs},
{\it NewDefs},\mathit{FoldedCls})$.

\noindent
{\bf Input\/}: (i) A mode $M_s$,
(ii) a set {\it PartitionedCls} of clauses which are safe
w.r.t.~$M_s$, and (iii) a set  {\it Defs} of definition clauses.
{\it PartitionedCls} is the  union of the disjoint packets of clauses
computed by the Partition subsidiary strategy.

\noindent
{\bf Output\/}:
(i) A (possibly empty) set {\it NewDefs} of definition
clauses, together with a mode $M_{\it new}$ consisting of exactly
one mode for each
distinct head predicate in {\it NewDefs}.
For each $C\in\mathit{NewDefs}$, the input
variables of the leftmost non-basic
atom in the body of $C$ are input
variables of the head of $C$.
(ii) A set $\mathit{FoldedCls}$ of
folded clauses.

\bigskip

\noindent
${\it NewDefs}:= \emptyset$;
\ $M_{\it new}:= \emptyset$;
\ $\mathit{FoldedCls}:= \emptyset$;

\noindent
{\bf while} there exists in {\it PartitionedCls}
a packet {\it Q} of the form:

\smallskip

$\left\{
\begin{array}{ll}
H  \leftarrow \ {\it Diseqs},\ G_1\\
  \hspace{.5cm} \cdots \\
H \leftarrow \ {\it Diseqs},\ G_m
\end{array}
\right.$

\smallskip
\noindent
where $\mathit{Diseqs}$ is a conjunction of disequations and for
$k=1,\ldots,m$, no disequation occurs in $G_k$,

\noindent
{\bf do} \ ${\it PartitionedCls} := {\it PartitionedCls} - {\it Q}$
and apply the definition and safe folding rules as follows.

\noindent
\begin{description}

\item {\rm (Case $\alpha$)} Let us suppose that the set ${\it Defs}$
of the available definition clauses contains a subset of
clauses of the form:

$\left\{
\begin{array}{ll}
 {\it newq}(X_1,\dots, X_h)\  \leftarrow \ G_1\\
\hspace{.5cm} \cdots & \\
{\it newq}(X_1,\dots, X_h)\  \leftarrow \ G_m &
\end{array}
\right.$

\noindent
such that: (i) they are all the clauses in ${\it Defs}$ for
predicate {\it newq}, (ii) $X_1,\dots, X_h$ include every variable
which occurs in one of the goals $G_1,\dots, G_m$
and also occurs in one of the goals $H, \mathit{Diseqs}$
(this property is needed for the correctness of folding, see
Section~\ref{rules}), and
(iii) for $i=1,\ldots,h$, if $X_i$ is an input argument of
${\it newq}$ then $X_i$ is either an input variable of $H$
(according to the given mode $M_s$) or an input variable of the
leftmost non-basic atom of one of the goals $G_1,\ldots,G_m$. Then we
fold the given packet and we get:

${\it FoldedCls} := {\it FoldedCls} \cup
        \{H  \leftarrow \mathit{Diseqs}, {\it newq}(X_1,\dots, X_h)\}$

\item {\rm (Case~$\beta$)}
If in ${\it Defs}$ there is no set of
definition clauses satisfying the conditions described
in Case~($\alpha$), then we add to {\it NewDefs}
the following clauses for a new predicate {\it newr\/}:

$\left\{
\begin{array}{ll}
{\it newr}(X_1,\dots, X_h)\  \leftarrow \  G_1\\
\hspace{.5cm}\cdots \\
{\it newr}(X_1,\dots, X_h)\  \leftarrow \  G_m
\end{array}
\right.$

\noindent
where, for $i=1,\ldots,h$,
either (i) $X_i$ occurs in one
of the goals $G_1, \dots, G_m$
and also occurs in one of the goals $H, \mathit{Diseqs}$,
or (ii) $X_i$ is an input variable of the leftmost non-basic
atom of one of the goals  $G_1, \dots, G_m$. We
add to $M_{new}$ the mode ${\it newr}(m_1,$ $\dots, $ $m_h)$
such that for $i= 1,\dots,h$, $m_i\eq+$ iff
$X_i$ is either an input variable of $H$
or an input variable of
the leftmost non-basic
atom of one of the goals  $G_1, \dots, G_m$.
We then fold the packet
under consideration and we get:

${\it FoldedCls} := {\it FoldedCls} \cup
        \{H \leftarrow \mathit{Diseqs}, {\it newr}(X_1,\dots, X_h)\}$

\end{description}

\vspace*{-.2cm}
\noindent
{\bf end-while}

%\bracebottom
\noindent \hrulefill

\medskip

\noindent
Notice that the post-conditions on the set {\it NewDefs}
which is derived by the Define-Fold procedure
(see Point (i) of the Output
of the procedure), ensure the satisfaction of
the pre-conditions on the set {\it Cls}
which is an input of the Unfold-Simplify procedure.
Indeed, recall that
the set {\it Cls} is constructed during
the Determinization Strategy by the assignment
${\it Cls}:={\it NewDefs}$. Recall also that these
pre-conditions are needed to ensure that
the first unfolding step performed by the
Unfold-Simplify procedure is safe.

Notice also that each application of the folding
rule is safe (see Definition~\ref{safe_fold}).
This fact is implied in Case~($\alpha$) by Condition~(iii),
and in Case~($\beta$) by the definition of the mode for ${\it
newr}$.

{F}inally, notice that the Define-Fold procedure
terminates. However, this procedure does not guarantee the termination
of the specialization process, because at each iteration of the
while-do loop of the Determinization Strategy, the Define-Fold
procedure may introduce a nonempty set of new definition clauses. We will
briefly discuss this issue in Section~\ref{discussion}.

\section{Examples of Application of the Determinization Strategy}
\label{examples}

In this section we will present some examples of program
specialization where we will see in action our Determinization
Strategy together with the Unfold-Simplify, Partition, and Define-Fold
subsidiary strategies.

\subsection{A Complete Derivation: Computing
the Occurrences of a Pattern in a String}
\label{match_pos_ex}
We consider again the program ${\it Match\_Pos}$
of Section~\ref{semidet}.
The mode $M$ for the program ${\it Match\_Pos}$ is
$\{{\it match\_pos}(+,+,?)$, ${\it append}(?,?,+)$, ${\it length}(+,?)\}$.
We leave it to the reader to verify that ${\it Match\_Pos}$ satisfies
$M$.

The derivation we will perform using the Determinization Strategy is
more challenging than the ones  presented in the literature
(see, for instance,~\cite{Fuj87,Fu&91,Gall93,GlK93,Smi91}) because  an
occurrence of the pattern $P$ in the string $S$ is specified in the initial
program (see clause 1) in a nondeterministic way by stipulating the
existence of two substrings $L$ and $R$ such that $S$ is the
concatenation of $L$, $P$, and $R$.

We want to specialize the ${\it Match\_Pos}$ program
\wrt the atom  $ {\it match\_pos}([a,a,b],S,N)$.
Thus, we first introduce the definition clause:

\oeq{6.}{{\it match\_pos_s}(S,N)  \leftarrow
{\it match\_pos}([a,a,b],S,N)}

\noindent
The mode of the new predicate is
${\it match\_pos_s}(+,?)$ because $S$
is an input argument of ${\it match\_pos}$
and $N$ is not an input argument. Our transformation strategy starts off
with the following initial values:
${\it Defs} = {\it Cls} =
\{6\}$, ${\it TransfP} = {\it Match\_Pos}$, and
$M_s = M \cup \{{\it match\_pos_s}(+,?)\}$.

\subsubsection*{First iteration}

{\em Unfold-Simplify}.
By unfolding clause 6 \wrt the leftmost atom in its body
we derive:

\oeq{7.} {{\it match\_pos_s}(S,N)  \leftarrow
        {\it append}(Y,R,S),\
        {\it append}(L, [a,a,b],Y),\
        {\it length}(L,N)}

\noindent
The body of clause 7 has no consumer atoms
(notice that, for instance, the mgu of
${\it append}(Y,R,S)$ and the head of clause
5 has the binding $S/[A|Z]$ where $S$ is an input
variable).
Thus, the Unfold-Simplify subsidiary
strategy terminates. We have: ${\it UnfoldedCls} = \{7\}$.

\smallskip \noindent {\em Partition}. {\it NonunitCls} is made out of
clause 7 only, and thus, the Partition subsidiary strategy
immediately terminates and produces a set {\it PartitionedCls} which
consists of a single packet made out of clause 7.

\smallskip \noindent {\em Define-Fold}.
In order to fold clause 7 in {\it PartitionedCls}, the Define-Fold
subsidiary strategy introduces the following definition clause:

\oeq{8.} {{\it new}1(S,N)  \leftarrow
        {\it append}(Y,R,S),\
        {\it append}(L, [a,a,b],Y),\
        {\it length}(L,N)}

\noindent
The mode of {\it new}1 is ${\it new}1(+,?)$.
By folding clause 7 using clause 8 we derive:

\oeq{9.} {{\it match\_pos_s}(S,N) \leftarrow {\it new}1(S,N)}

\noindent
Thus, the first iteration of the Determinization Strategy terminates
with ${\it Defs} = \{6,8\}$, ${\it Cls} = \{8\}$,
${\it TransfP} = {\it Match\_Pos}\cup \{9\}$, and
$M_s = M \cup \{{\it match\_pos_s}(+,?),\ {\it new}1(+,?)\}$.

\medskip

\subsubsection*{Second iteration}

{\em Unfold-Simplify}.
We follow the
subsidiary strategy described in Section~\ref{unfolding} and
we first unfold clause 8 in {\it Cls} \wrt the
leftmost atom in its body. We get:

\bpr
\Feq{10.} {\it new}1(S,N) \leftarrow
        {\underline {{\it append}(L,[a,a,b],[~])}},\
        length(L,N)
\Nex{11.}  {\it new}1([C|S],N)  \leftarrow
        {\it append}(Y,R,S),\
        {\underline {{\it append}(L,[a,a,b],[C|Y])}}, \
        length(L,N)
\eeq
\epr

\noindent
Now we unfold clauses 10 and 11
\wrt the leftmost consumer atom of their bodies
(see the underlined atoms). The
unfolding of clause 10 amounts to its deletion because the
atom ${\it append}(L,[a,a,b],[~])$ is not unifiable with any head in
program {\it Match\_Pos}. The unfolding of clause 11 yields
two new clauses that are further unfolded according to
the Unfold-Simplify subsidiary strategy. After some unfolding steps, we
derive the following clauses:

\bpr
\Feq{12.}  {\it new}1([a|S],0) \leftarrow {\it append}([a,b],R,S)
\Nex{13.}  {\it new}1([C|S],s(N)) \nc \leftarrow
        {\it append}(Y,R,S), \
        {\it append}(L,[a,a,b],Y),\
        length(L,N)
\eeq
\epr

\smallskip
\noindent
{\it Partition}.
We apply the safe case split rule to clause 13 \wrt to the
binding $C/a$, because the input argument in the head of this clause
is unifiable with the input argument in the head of clause 12
via the mgu\
$\{C/a\}$. We derive the following two clauses:

\bpr
\Feq{14.}  {\it new}1([a|S],s(N)) \nc \leftarrow
        {\it append}(Y,R,S),\
        {\it append}(L,[a,a,b],Y),\
        length(L,N)
\Nex{15.}  {\it new}1([C|S],s(N)) \nc \leftarrow C\!\neq\! a,\
        {\it append}(Y,R,S),\
        {\it append}(L,[a,a,b],Y),\
        length(L,N)
\eeq
\epr

\noindent
Now, the set of clauses derived so far by the Partition subsidiary
strategy can be partitioned into two
packets: the first one is made out of clauses 12 and 14, where
the input argument of the head predicate
is of the form $[a|S]$, and the second one is made
out of clause 15 only, where the input argument of the head
predicate is of the form
$[C|S]$ with $C\!\neq\! a$.

The Partition subsidiary strategy terminates
by applying the safe head generalization rule to clauses
12 and 14, so to replace the second arguments
in their heads by the
most specific common
generalization of those arguments, that is, a variable.
We get the packet:

\bpr
\Feq{16.}  {\it new}1([a|S],M) \leftarrow M\! =\! 0,\
        {\it append}([a,b],R,S)
\Nex{17.}  {\it new}1([a|S],M) \nc\nc \leftarrow M\!=\!s(N),\
        {\it append}(Y,R,S), \
        {\it append}(L,[a,a,b],Y),\
        length(L,N)
\eeq
\epr

\noindent
{F}or the packet made out of clause 15 only, no application of
the safe head generalization rule is performed. Thus, we have derived
the set of clauses {\it PartitionCls} which is the union of the
two packets $\{ 16, 17\}$ and $\{ 15 \}$.

\smallskip
\noindent
{\it Define-Fold}.
Since there is no set of definition clauses which can be used to
fold the packet $\{16, 17\}$, we are in Case ($\alpha$) of
the Define-Fold subsidiary strategy. Thus, we introduce
a new predicate ${\it new}2$ as follows:

\bpr
\Feq{18.}{\it new}2(S,M) \leftarrow M\! =\! 0,\ {\it
append}([a,b],R,S) \Nex{19.}{\it new}2(S,M) \nc\nc \leftarrow
M\!=\!s(N),\         {\it append}(Y,R,S),\
        {\it append}(L,[a,a,b],Y),\
        length(L,N)
\eeq
\epr

\noindent
The mode of {\it new}2 is ${\it new}2(+,?)$  because
$S$ is an input variable of the head of each clause
of the corresponding packet.
By folding clauses 16 and 17 using clauses 18 and 19
we derive the following clause:

\oeq{20.}{{\it new}1([a|S],M) \leftarrow {\it new}2(S,M)}

\noindent
We then consider the packet made out of clause 15 only.
This packet can be folded
using clause 8 in ${\it Defs}$. Thus, we are in Case ($\beta$)
of the Define-Fold subsidiary strategy. By folding clause 15
we derive the following clause:

\oeq{21.}{{\it new}1([C|S],s(N)) \leftarrow C\!\neq\! a,\ {\it
new}1(S,N)}

\noindent
Thus, {\it FoldedCls} is the set $\{20,21\}$.

After these folding steps we conclude the
second iteration of the Determinization Strategy
with the following assignments:
${\it Defs} := {\it Defs} \cup \{18,19\}$;
${\it Cls} := \{18,19\}$;
${\it TransfP} := {\it TransfP} \cup \{20,21\}$;
$M_s := M_s \cup \{{\it new}2(+,?)\}$.

\subsubsection*{Third iteration}

{\em Unfold-Simplify}. From {\it Cls}, that is,
clauses 18 and 19, we derive the set {\it UnfoldedCls}
made out of the following clauses:

\bpr
\Feq{22.} {\it new}2([a|S],0) \leftarrow {\it append}([b],R,S)
\Nex{23.}  {\it new}2([a|S],s(0)) \leftarrow
        {\it append}([a,b],R,S)
\Nex{24.}  {\it new}2([C|S],s(s(N))) \leftarrow
        {\it append}(Y,R,S),\
        {\it append}(L,[a,a,b],Y),\
        length(L,N)
\eeq
\epr

\smallskip \noindent {\em Partition}.
The set {\it NonunitCls} is identical to {\it UnfoldedCls}. From {\it
NonunitCls} we derive the set {\it PartitionedCls} which is the union of
two packets. The first packet consists of the
following clauses:

\bpr
\Feq{25.} {\it new}2([a|S],M) \leftarrow M\!=\!0,\ {\it append}([b],R,S)
\Nex{26.} {\it new}2([a|S],M) \leftarrow M\!=\!s(0),\
        {\it append}([a,b],R,S)
\Nex{27.}  {\it new}2([a|S],M) \leftarrow M\!=\!s(s(N)),\
        {\it append}(Y,R,S),\
        {\it append}(L,[a,a,b],Y),\
        length(L,N)
\eeq
\epr

\noindent
The second packet consists of the following clause only:

\bpr
\Feq{28.}  {\it new}2([C|S],s(s(N))) \leftarrow
        C\!\neq\! a,\ {\it append}(Y,R,S),\
        {\it append}(L,[a,a,b],Y),\
        length(L,N)
\eeq
\epr

\smallskip \noindent {\em Define-Fold}.
We introduce the following definition clauses:

\bpr
\Feq{29.} {\it new}3(S,M) \leftarrow M\!=\!0,\ {\it append}([b],R,S)
\Nex{30.} {\it new}3(S,M) \leftarrow M\!=\!s(0),\ {\it append}([a,b],R,S)
\Nex{31.} {\it new}3(S,M) \leftarrow M\!=\!s(s(N)),\
        {\it append}(Y,R,S),\
        {\it append}(L,[a,a,b],Y),\
        length(L,N)
\eeq
\epr

\noindent where the mode for ${\it new}3$ is ${\it new}3(+,?)$.
By folding, from  {\it PartitionedCls}
we derive the following two clauses:

\bpr
\Feq{32.} {\it new}2([a|S],M) \leftarrow {\it new}3(S,M)
\Nex{33.}  {\it new}2([C|S],s(s(N))) \leftarrow
        C\!\neq\! a,\ {\it new}1(S,N)
\eeq
\epr

\noindent
which constitute the set {\it FoldedCls}.

The third iteration of the Determinization Strategy
 terminates with the following assignments:
${\it Defs} := {\it Defs}\cup \{29,30,31\}$;
${\it Cls} := \{29,30,31\}$;
${\it TransfP} := {\it TransfP}\cup \{32,33\}$;
$M_s := M_s \cup \{{\it new}3(+,?)\}$.

\subsubsection*{Fourth iteration}

{\em Unfold-Simplify}. From {\it Cls} we derive the new set {\it
UnfoldedCls} made out of the following clauses:

\bpr
\Feq{34.} {\it new}3([b|S],0) \leftarrow {\it append}([~],R,S)
\Nex{35.} {\it new}3([a|S],s(0)) \leftarrow {\it append}([b],R,S)
\Nex{36.} {\it new}3([a|S],s(s(0))) \leftarrow {\it append}([a,b],R,S)
\Nex{37.} {\it new}3([C|S],s(s(s(N)))) \leftarrow
        {\it append}(Y,R,S),\
        {\it append}(L,[a,a,b],Y),\
        length(L,N)
\eeq
\epr

\smallskip \noindent {\em Partition}.
The set {\it NonunitCls} is identical to {\it UnfoldedCls}. From {\it
NonunitCls} we derive the new set {\em PartitionedCls} made out of the
following clauses:

\bpr
\Feq{38.} {\it new}3([a|S],s(M)) \leftarrow M\!=\!0,\
{\it append}([b],R,S)
\Nex{39.} {\it new}3([a|S],s(M)) \leftarrow M\!=\!s(0),\
        {\it append}([a,b],R,S)
\Nex{40.} {\it new}3([a|S],s(M)) \leftarrow
        M\!=\!s(s(N)),\
        {\it append}(Y,R,S),\
        {\it append}(L,[a,a,b],Y),\
        length(L,N)
\Nex{41.} {\it new}3([b|S],M) \leftarrow M\!=\!0,\ {\it append}([~],R,S)
\Nex{42.} {\it new}3([b|S],M) \leftarrow
        M\!=\!s(s(s(N))),\
        {\it append}(Y,R,S),\
        {\it append}(L,[a,a,b],Y),\
        length(L,N)
\Nex{43.} {\it new}3([C|S],s(s(s(N))))\! \leftarrow  C\!\neq\!a,
        C\!\neq\!b, {\it append}(Y,R,S),
        {\it append}(L,[a,a,b],Y),
        length(L,N)
\eeq
\epr

\noindent
{\em PartitionedCls} consists of three packets:
$\{38,39,40\}$, $\{41,42\}$, and $\{43\}$.

\smallskip \noindent {\em Define-Fold}.
We introduce two new predicates by means of the following
definition clauses:

\bpr
\Feq{44.} {\it new}4(S,M) \leftarrow M\!=\!0,\ {\it append}([~],R,S)
\Nex{45.} {\it new}4(S,M) \leftarrow
        M\!=\!s(s(s(N))),\
        {\it append}(Y,R,S),\
        {\it append}(L,[a,a,b],Y),\
        length(L,N)
\eeq
\epr

\noindent
We now fold the clauses in {\it PartitionedCls}
and we derive the set {\it FoldedCls} made out
of the following clauses:

\bpr
\Feq{46.} {\it new}3([a|S],s(M)) \leftarrow {\it new}3(R,S)
\Nex{47.} {\it new}3([b|S],M) \leftarrow {\it new}4(R,S)
\Nex{48.} {\it new}3([C|S],s(s(s(N)))) \leftarrow
        C\!\neq\!a,\ C\!\neq\!b,\
        {\it new}1(S,N)
\eeq
\epr

\noindent
The fourth iteration terminates
with the following assignments:
${\it Defs} := {\it Defs}\cup \{44,45\}$;
${\it Cls} := \{44,45\}$;
${\it TransfP} := {\it TransfP}\cup \{46,47,48\}$;
$M_s := M_s \cup \{{\it new}4(+,?)\}$.

\subsubsection*{Fifth iteration}

{\em Unfold-Simplify}. From {\it Cls} we derive the new set {\it
UnfoldedCls} made out of the following clauses:

\bpr
\Feq{49.} {\it new}4(S,0) \leftarrow
\Nex{50.} {\it new}4([a|S],s(s(s(0)))) \leftarrow {\it append}([a,b],R,S)
\Nex{51.} {\it new}4([C|S],s(s(s(s(N))))) \leftarrow
        {\it append}(Y,R,S),\
        {\it append}(L,[a,a,b],Y),\
        length(L,N)
\eeq
\epr

\smallskip \noindent {\em Partition}.
The set {\it NonunitCls} is made out of clauses 50 and 51. From {\it
NonunitCls} we derive the new set {\em PartitionedCls} made out of the
following clauses:

\bpr
\Feq{52.} {\it new}4([a|S],s(s(s(M)))) \leftarrow
        M\!=\!0,\ {\it append}([a,b],R,S)
\Nex{53.} {\it new}4([a|S],s(s(s(M)))) \leftarrow
        M\!=\!s(N),\
        {\it append}(Y,R,S),\
        {\it append}(L,[a,a,b],Y),\
        length(L,N)
\Nex{54.} {\it new}4([C|S],s(s(s(s(N))))) \leftarrow
        C\!\neq\!a,\
        {\it append}(Y,R,S),\
        {\it append}(L,[a,a,b],Y),\
        length(L,N)
\eeq
\epr

\noindent
{\em PartitionedCls} consists of two packets:
$\{52,53\}$ and $\{54\}$.

\smallskip \noindent {\em Define-Fold}.
We are able to perform all required folding steps without
introducing new definition clauses (see Case ($\alpha$) of the
Define-Fold procedure).
In particular, (i)~we fold clauses 52 and 53 using clauses 18 and 19,
and (ii)~we fold clause 54 using clause 8.
Since no new definition is introduced, the set {\it Cls} is
empty and the transformation strategy terminates.
Our final specialized program is the program ${\it Match\_Pos_s}$
shown in Section~\ref{semidet}.

\begin{figure}
\begin{centering}
\setlength{\unitlength}{0.00083333in}
\begingroup\makeatletter\ifx\SetFigFont\undefined%
\gdef\SetFigFont#1#2#3#4#5{%
  \reset@font\fontsize{#1}{#2pt}%
  \fontfamily{#3}\fontseries{#4}\fontshape{#5}%
  \selectfont}%
\fi\endgroup%
{\renewcommand{\dashlinestretch}{30}
%\begin{picture}(7736,3384)(-90,-10)
\begin{picture}(7920,3384)(450,200)
\thicklines
\put(4800,1854){\ellipse{150}{150}}
\put(3300,1854){\ellipse{150}{150}}
\put(6375,1854){\ellipse{150}{150}}
\put(6375,1854){\ellipse{150}{150}}
\put(1800,1854){\ellipse{150}{150}}
\put(7920,1854){\ellipse{150}{150}}
\put(7920,1854){\ellipse{246}{246}}
\path(1950,1854)(3150,1854)
\blacken\path(3030.000,1824.000)
             (3150.000,1854.000)
             (3030.000,1884.000)
             (3066.000,1854.000)
             (3030.000,1824.000)
\path(3450,1854)(4650,1854)
\blacken\path(4530.000,1824.000)
             (4650.000,1854.000)
             (4530.000,1884.000)
             (4566.000,1854.000)
             (4530.000,1824.000)
\path(4950,1854)(6150,1854)
\blacken\path(6080.000,1824.000)
             (6200.000,1854.000)
             (6080.000,1884.000)
             (6116.000,1854.000)
             (6080.000,1824.000)
\path(650,1854)(1650,1854)
\blacken\path(1530.000,1824.000)
             (1650.000,1854.000)
             (1530.000,1884.000)
             (1566.000,1854.000)
             (1530.000,1824.000)
\path(4729,2008)(4727,2011)(4723,2018)
	(4715,2030)(4705,2046)(4694,2065)
	(4681,2087)(4668,2108)(4656,2129)
	(4645,2149)(4635,2167)(4627,2184)
	(4620,2200)(4614,2216)(4609,2231)
	(4604,2245)(4600,2261)(4596,2276)
	(4593,2293)(4591,2310)(4589,2327)
	(4588,2346)(4587,2364)(4588,2383)
	(4589,2402)(4591,2420)(4593,2439)
	(4596,2456)(4600,2473)(4605,2490)
	(4610,2505)(4617,2520)(4623,2535)
	(4631,2550)(4640,2565)(4649,2580)
	(4660,2594)(4671,2608)(4684,2621)
	(4697,2633)(4710,2645)(4724,2655)
	(4737,2663)(4751,2670)(4765,2676)
	(4778,2680)(4791,2682)(4804,2683)
	(4817,2682)(4830,2680)(4843,2676)
	(4857,2670)(4871,2663)(4884,2655)
	(4898,2645)(4911,2633)(4924,2621)
	(4937,2608)(4948,2594)(4959,2580)
	(4968,2565)(4977,2550)(4985,2535)
	(4992,2520)(4998,2505)(5003,2490)
	(5008,2473)(5012,2456)(5015,2439)
	(5017,2420)(5019,2402)(5020,2383)
	(5021,2364)(5020,2346)(5019,2327)
	(5017,2310)(5015,2293)(5012,2276)
	(5008,2261)(5004,2245)(4999,2231)
	(4994,2216)(4988,2200)(4981,2184)
	(4973,2167)(4963,2149)(4952,2129)
	(4940,2108)(4927,2087)(4914,2065)
	(4903,2046)(4879,2008)
\blacken\path(4917.714,2125.478)
             (4879.000,2008.000)
             (4968.444,2093.439)
             (4923.855,2079.021)
             (4917.714,2125.478)
\path(3300,1704)(3300,1703)(3300,1699)
	(3299,1690)(3298,1674)(3296,1653)
	(3293,1630)(3289,1606)(3285,1583)
	(3280,1562)(3274,1542)(3267,1525)
	(3259,1509)(3249,1494)(3238,1479)
	(3228,1469)(3218,1458)(3207,1448)
	(3195,1437)(3181,1426)(3166,1415)
	(3150,1404)(3133,1393)(3114,1382)
	(3093,1372)(3072,1361)(3050,1351)
	(3027,1342)(3003,1333)(2979,1324)
	(2954,1317)(2929,1309)(2903,1303)
	(2877,1297)(2850,1291)(2827,1287)
	(2803,1284)(2778,1280)(2752,1277)
	(2725,1275)(2697,1272)(2668,1271)
	(2639,1270)(2608,1269)(2578,1269)
	(2547,1270)(2516,1271)(2485,1273)
	(2455,1276)(2425,1280)(2396,1284)
	(2367,1289)(2340,1294)(2314,1300)
	(2289,1307)(2265,1315)(2242,1323)
	(2221,1332)(2200,1341)(2179,1353)
	(2158,1366)(2139,1380)(2120,1395)
	(2102,1413)(2084,1432)(2065,1454)
	(2047,1479)(2028,1505)(2009,1534)
	(1990,1565)(1971,1598)(1953,1631)
	(1935,1663)(1919,1693)(1905,1720)
	(1894,1742)(1875,1779)
\blacken\path(1956.504,1685.956)
             (1875.000,1779.000)
             (1903.130,1658.548)
             (1913.372,1704.276)
             (1956.504,1685.956)
\path(6375,2004)(6375,2006)(6374,2010)
	(6374,2018)(6372,2030)(6371,2047)
	(6368,2070)(6364,2097)(6360,2130)
	(6355,2167)(6348,2209)(6340,2253)
	(6332,2300)(6322,2348)(6311,2396)
	(6300,2445)(6287,2493)(6273,2539)
	(6258,2583)(6242,2626)(6225,2666)
	(6207,2704)(6188,2740)(6167,2773)
	(6145,2805)(6121,2834)(6096,2861)
	(6069,2886)(6040,2909)(6009,2931)
	(5976,2951)(5941,2970)(5903,2987)
	(5863,3004)(5830,3016)(5796,3028)
	(5760,3039)(5723,3050)(5684,3060)
	(5644,3070)(5602,3079)(5559,3087)
	(5515,3096)(5469,3103)(5421,3110)
	(5373,3117)(5323,3123)(5272,3128)
	(5220,3133)(5167,3137)(5114,3140)
	(5059,3143)(5004,3146)(4949,3147)
	(4893,3148)(4837,3149)(4782,3148)
	(4726,3147)(4671,3146)(4616,3143)
	(4561,3140)(4508,3137)(4455,3133)
	(4403,3128)(4352,3123)(4302,3117)
	(4254,3110)(4206,3103)(4160,3096)
	(4116,3087)(4073,3079)(4031,3070)
	(3991,3060)(3952,3050)(3915,3039)
	(3879,3028)(3845,3016)(3813,3004)
	(3772,2987)(3734,2970)(3699,2951)
	(3666,2931)(3635,2909)(3606,2886)
	(3579,2861)(3554,2834)(3530,2805)
	(3508,2773)(3487,2740)(3468,2704)
	(3450,2666)(3433,2626)(3417,2583)
	(3402,2539)(3388,2493)(3375,2445)
	(3364,2396)(3353,2348)(3343,2300)
	(3335,2253)(3327,2209)(3320,2167)
	(3315,2130)(3311,2097)(3307,2070)
	(3304,2047)(3303,2030)(3300,2004)
\blacken\path(3283.953,2126.648)
             (3300.000,2004.000)
             (3343.557,2119.770)
             (3309.628,2087.446)
             (3283.953,2126.648)
\path(6450,2004)(6450,2005)(6450,2008)
	(6451,2017)(6451,2032)(6452,2052)
	(6454,2077)(6456,2104)(6459,2130)
	(6462,2156)(6465,2180)(6469,2201)
	(6474,2220)(6479,2237)(6485,2252)
	(6492,2266)(6500,2279)(6508,2290)
	(6517,2301)(6527,2312)(6538,2322)
	(6549,2332)(6561,2341)(6574,2349)
	(6588,2357)(6602,2363)(6616,2369)
	(6629,2373)(6643,2376)(6656,2378)
	(6669,2378)(6681,2377)(6692,2375)
	(6703,2372)(6713,2366)(6723,2359)
	(6733,2350)(6742,2339)(6750,2326)
	(6757,2311)(6764,2295)(6769,2278)
	(6774,2260)(6777,2241)(6779,2221)
	(6779,2202)(6778,2184)(6776,2166)
	(6773,2148)(6768,2132)(6763,2116)
	(6755,2102)(6747,2088)(6736,2074)
	(6723,2060)(6707,2046)(6689,2032)
	(6668,2017)(6645,2001)(6620,1985)
	(6595,1969)(6572,1956)(6552,1944)(6525,1929)
\blacken\path(6615.330,2013.502)
             (6525.000,1929.000)
             (6644.468,1961.052)
             (6598.429,1969.794)
             (6615.330,2013.502)
\path(6375,1704)(6375,1702)(6375,1696)
	(6374,1686)(6373,1671)(6372,1650)
	(6370,1623)(6367,1589)(6364,1551)
	(6360,1508)(6355,1462)(6350,1413)
	(6344,1363)(6337,1313)(6330,1264)
	(6322,1215)(6313,1169)(6303,1125)
	(6293,1083)(6282,1044)(6271,1007)
	(6258,972)(6244,939)(6229,908)
	(6213,880)(6196,852)(6178,826)
	(6157,801)(6136,777)(6113,754)
	(6094,737)(6075,720)(6054,704)
	(6033,687)(6010,671)(5986,656)
	(5961,640)(5935,624)(5907,609)
	(5878,594)(5848,579)(5816,564)
	(5784,549)(5749,535)(5714,520)
	(5677,506)(5639,493)(5600,479)
	(5560,466)(5518,453)(5476,441)
	(5433,429)(5389,417)(5344,406)
	(5298,395)(5252,384)(5205,374)
	(5158,365)(5110,355)(5062,346)
	(5014,338)(4965,330)(4916,322)
	(4867,315)(4817,308)(4767,302)
	(4717,296)(4666,290)(4615,284)
	(4563,279)(4517,275)(4470,270)
	(4423,266)(4375,263)(4326,259)
	(4276,256)(4226,253)(4174,250)
	(4122,247)(4069,245)(4016,243)
	(3961,241)(3906,239)(3850,238)
	(3794,238)(3737,237)(3680,237)
	(3623,237)(3566,238)(3508,239)
	(3451,241)(3393,243)(3336,246)
	(3280,249)(3224,252)(3168,256)
	(3113,260)(3059,265)(3006,270)
	(2954,276)(2903,282)(2854,289)
	(2805,296)(2758,303)(2712,311)
	(2667,319)(2624,328)(2583,338)
	(2542,347)(2503,358)(2466,368)
	(2430,379)(2396,391)(2362,403)
	(2331,416)(2300,429)(2263,447)
	(2228,465)(2195,485)(2164,506)
	(2135,529)(2108,553)(2082,579)
	(2058,607)(2036,637)(2015,669)
	(1995,703)(1976,739)(1959,778)
	(1943,819)(1927,862)(1913,908)
	(1899,955)(1886,1004)(1874,1055)
	(1863,1106)(1853,1158)(1844,1209)
	(1836,1259)(1829,1307)(1822,1353)
	(1817,1394)(1812,1431)(1809,1464)
	(1806,1491)(1804,1512)(1802,1529)(1800,1554)
\blacken\path(1839.474,1436.775)
             (1800.000,1554.000)
             (1779.665,1431.990)
             (1806.699,1470.268)
             (1839.474,1436.775)
\path(4800,1704)(4800,1701)(4800,1695)
	(4799,1683)(4799,1666)(4798,1643)
	(4797,1615)(4795,1584)(4794,1550)
	(4792,1516)(4789,1482)(4787,1449)
	(4784,1417)(4780,1388)(4777,1361)
	(4773,1335)(4768,1312)(4763,1291)
	(4758,1271)(4752,1252)(4745,1234)
	(4738,1216)(4731,1203)(4724,1190)
	(4717,1177)(4708,1164)(4699,1151)
	(4689,1138)(4677,1126)(4665,1113)
	(4651,1100)(4635,1088)(4618,1076)
	(4600,1064)(4580,1052)(4559,1041)
	(4536,1030)(4511,1020)(4485,1009)
	(4458,1000)(4429,990)(4398,982)
	(4366,973)(4333,966)(4298,958)
	(4262,952)(4224,945)(4184,940)
	(4143,934)(4100,929)(4069,926)
	(4036,922)(4002,919)(3967,917)
	(3931,914)(3893,911)(3855,909)
	(3815,907)(3774,905)(3732,904)
	(3688,902)(3644,901)(3599,901)
	(3553,900)(3506,900)(3458,900)
	(3410,900)(3362,901)(3314,902)
	(3265,904)(3216,906)(3168,908)
	(3120,910)(3072,913)(3025,916)
	(2978,920)(2933,924)(2888,928)
	(2844,933)(2802,938)(2760,943)
	(2720,949)(2681,955)(2643,962)
	(2606,968)(2571,976)(2537,983)
	(2504,991)(2473,999)(2443,1008)
	(2401,1021)(2362,1035)(2325,1050)
	(2291,1067)(2258,1085)(2226,1105)
	(2196,1127)(2167,1150)(2139,1176)
	(2111,1204)(2085,1234)(2058,1267)
	(2032,1301)(2007,1337)(1982,1375)
	(1958,1413)(1936,1451)(1915,1487)
	(1895,1522)(1879,1554)(1864,1582)
	(1852,1605)(1844,1623)(1830,1652)
\blacken\path(1909.186,1556.976)
             (1830.000,1652.000)
             (1855.153,1530.891)
             (1866.519,1576.354)
             (1909.186,1556.976)
\path(1724,2005)(1722,2008)(1718,2015)
	(1710,2027)(1700,2043)(1689,2062)
	(1676,2084)(1663,2105)(1651,2126)
	(1640,2146)(1630,2164)(1622,2181)
	(1615,2197)(1609,2213)(1604,2228)
	(1599,2242)(1595,2258)(1591,2273)
	(1588,2290)(1586,2307)(1584,2324)
	(1583,2343)(1582,2361)(1583,2380)
	(1584,2399)(1586,2417)(1588,2436)
	(1591,2453)(1595,2470)(1600,2487)
	(1605,2502)(1612,2517)(1618,2532)
	(1626,2547)(1635,2562)(1644,2577)
	(1655,2591)(1666,2605)(1679,2618)
	(1692,2630)(1705,2642)(1719,2652)
	(1732,2660)(1746,2667)(1760,2673)
	(1773,2677)(1786,2679)(1799,2680)
	(1812,2679)(1825,2677)(1838,2673)
	(1852,2667)(1866,2660)(1879,2652)
	(1893,2642)(1906,2630)(1919,2618)
	(1932,2605)(1943,2591)(1954,2577)
	(1963,2562)(1972,2547)(1980,2532)
	(1987,2517)(1993,2502)(1998,2487)
	(2003,2470)(2007,2453)(2010,2436)
	(2012,2417)(2014,2399)(2015,2380)
	(2016,2361)(2015,2343)(2014,2324)
	(2012,2307)(2010,2290)(2007,2273)
	(2003,2258)(1999,2242)(1994,2228)
	(1989,2213)(1983,2197)(1976,2181)
	(1968,2164)(1958,2146)(1947,2126)
	(1935,2105)(1922,2084)(1909,2062)
	(1898,2043)(1874,2005)
\blacken\path(1912.714,2122.478)
             (1874.000,2005.000)
             (1963.444,2090.439)
             (1918.855,2076.021)
             (1912.714,2122.478)
\path(6520,1854)(7720,1854)
\blacken\path(7600.000,1824.000)
             (7720.000,1854.000)
             (7600.000,1884.000)
             (7636.000,1854.000)
             (7600.000,1824.000)
\put(650,1929){\makebox(0,0)[lb]{$N\!:=\!0$}}
\put(1350,1629){\makebox(0,0)[lb]{{\it new}1}}
\put(2850,1629){\makebox(0,0)[lb]{{\it new}2}}
\put(4350,1629){\makebox(0,0)[lb]{{\it new}3}}
\put(5850,1629){\makebox(0,0)[lb]{{\it new}4}}
\put(7550,1629){\makebox(0,0)[lb]{{\it true}}}
\put(4500,2734){\makebox(0,0)[lb]{$=\!a$, $N\!:=\!N\!+\!1$}}
\put(4500,3204){\makebox(0,0)[lb]{$=a$, $N\!:=\!N\!+\!3$}}
\put(6575,2434){\makebox(0,0)[lb]{{\it any character}}}
\put(6850,1929){\makebox(0,0)[lb]{{\it return N}}}
\put(3600,25){\makebox(0,0)[lb]{$\neq\! a$, $N\!:=\!N\!+\!4$}}
\put(1500,2734){\makebox(0,0)[lb]{$\neq\! a$, $N\!:=\!N\!+\!1$}}
\put(2250,1929){\makebox(0,0)[lb]{$=\!a$}}
\put(3750,1929){\makebox(0,0)[lb]{$=\!a$}}
\put(5250,1929){\makebox(0,0)[lb]{$=\!b$}}
%\put(-90,1589){\makebox(0,0)[lb]{{\it match\_pos_s}}}
\put(3000,1179){\makebox(0,0)[lb]{$\neq a$, $N\!:=\!N\!+\!2$}}
\put(3225,695){\makebox(0,0)[lb]{$\neq\!a$ {\it and} $\neq\!b$,
$N\!:=\!N\!+\!3$}}
\end{picture}
}
\end{centering}
\caption{The finite automaton with counter $N$ which corresponds to
${\it Match\_Pos_s}$.\label{FAfig}}
\end{figure}
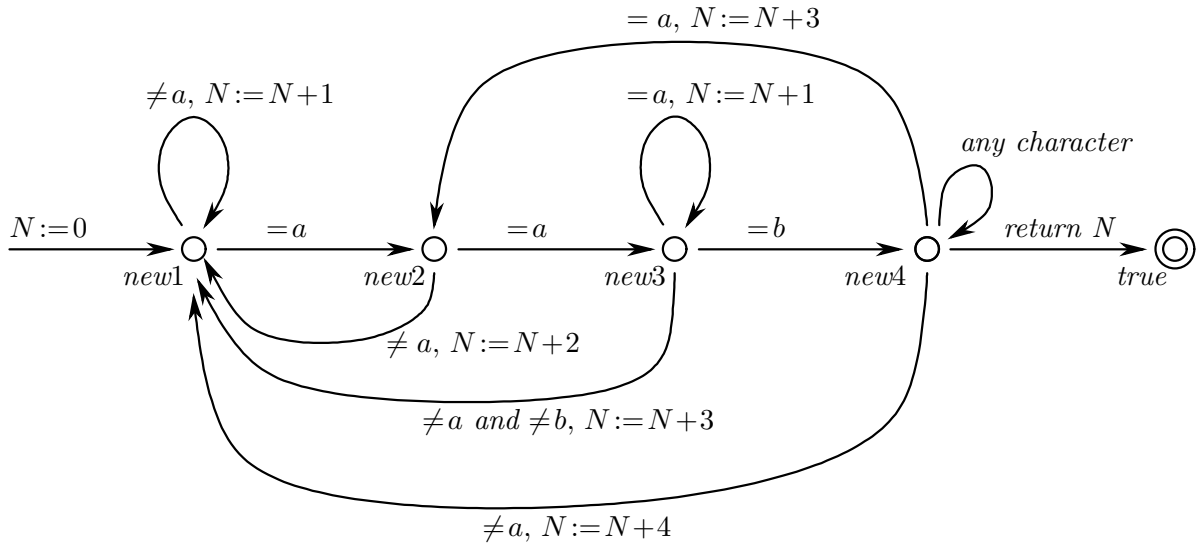

The ${\it Match\_Pos_s}$ program is semideterministic and it corresponds
to the finite automaton with one counter
depicted in Fig.~\ref{FAfig}.
The predicates correspond to the states of the automaton and the
clauses correspond to the transitions. The predicate {\it new}1
corresponds to the initial state, because the program is intended to
be used for goals of the form ${\it match\_pos_s}(S,N)$,
where $S$ is bound to a list of characters, and by clause 1
${\it match\_pos_s}(S,N)$ calls ${\it new}1(S,N)$.
Notice that this finite automaton is deterministic except
for the state corresponding to the predicate {\it new}4, where the
automaton can either (i)~accept the input string by returning the value
of $N$ and moving to the final state {\it true}, even if the input string
has not been completely scanned (see clause~49), or (ii)~move to the state
corresponding to {\it new}2, if the symbol of the input string which is
scanned is $a$ (see clause~55), or (iii)~move to the state corresponding to
{\it new}1, if the symbol of the input string which is scanned is different
from $a$ (see clause~56).

\subsection{Multiple Pattern Matching}
\label{mmatch_ex}

Given a list  {\it Ps} of patterns and a string $S$ we want to
compute the position, say $N$, of any occurrence in $S$
of a pattern which is a member of the list {\it Ps}.
For any given {\it Ps} and $S$ the following program
computes $N$ in a nondeterministic way:

\medskip

\fpr{{\it Mmatch} \hfill (initial, nondeterministic)}{
\Feq{1.} {\it mmatch}([P|{\it Ps}],S,N)  \If {\it match\_pos}(P,S,N)
\Nex{2.} {\it mmatch}([P|{\it Ps}],S,N)  \If
{\it mmatch}({\it Ps},S,N)
\eeq }

\medskip

\noindent
The atom ${\it mmatch}({\it Ps},S,N)$ holds iff
there exists a pattern in the list {\it Ps} of patterns
which occurs in the string $S$ at position $N$.
The predicate {\it match\_pos} is defined
as in program {\it Match\_Pos} of Section~\ref{match_pos_ex},
and its clauses are not listed here.
We consider the following mode for the program {\it Mmatch\/}:
 $\{\mathit{mmatch}(+,+,?),$ ${\it match\_pos}(+,+,?)$, ${\it
append}(?,?,+)$, ${\it length}(+,?)\}$.

We want to specialize this multi-pattern matching program \wrt the list
$[[a,a,a],$ $[a,a,b]]$  of patterns.
Thus, we introduce the following definition clause:

\noindent
\bpr
\Feq{3.}{\it mmatch_s}(S,N)  \leftarrow
{\it mmatch}([[a,a,a],[a,a,b]],S,N)
\eeq
\epr

\noindent
The mode of the new predicate is
${\it mmatch_s}(+,?)$ because $S$
is an input argument of ${\it mmatch}$
and $N$ is not an input argument. Thus,  our Determinization Strategy starts
off with the following initial values:
${\it Defs} = {\it Cls} =
\{3\}$, ${\it TransfP} = {\it Mmatch}$, and
$M_s = M \cup \{{\it mmatch_s}(+,?)\}$.

The output of the Determinization Strategy is the
following program ${\it Mmatch_s}$:

\medskip

\fpr{${\it Mmatch_s}$\hfill (specialized, semideterministic)}{
\Feq{4.} {\it mmatch_s}(S,N) \leftarrow {\it new}1(S,N)
\Nex{5.} {\it new}1([a|S],M) \leftarrow {\it new}2(S,M)
\Nex{6.} {\it new}1([C|S],s(N)) \leftarrow C\!\neq\! a,\
            {\it new}1(S,N)
\Nex{7.} {\it new}2([a|S],M) \leftarrow {\it new}3(S,M)
\Nex{8.} {\it new}2([C|S],s(s(N))) \leftarrow
            C\!\neq\!a,\  {\it new}1(S,N)
\Nex{9.} {\it new}3([a|S],M) \leftarrow {\it new}4(S,M)
\Nex{10.} {\it new}3([b|S],M) \leftarrow {\it new}5(S,M)
\Nex{11.} {\it new}3([C|S],s(s(s(N)))) \leftarrow
            C\!\neq\! a,\  C\!\neq\! b,\  {\it new}1(S,N)
\Nex{12.} {\it new}4(S,0) \leftarrow
\Nex{13.} {\it new}4([a|S],s(N)) \leftarrow {\it new}4(S,N)
\Nex{14.} {\it new}4([b|S],s(N)) \leftarrow {\it new}5(S,N)
\Nex{15.} {\it new}4([C|S],s(s(s(s(N))))) \leftarrow \!
             C\!\neq\! a,\  C\!\neq\! b,\ {\it new}1(S,N)
\Nex{16.} {\it new}5(S,0) \leftarrow
\Nex{17.} {\it new}5([a|S],s(s(s(N)))) \leftarrow {\it new}2(S,N)
\Nex{18.} {\it new}5([C|S],s(s(s(s(N))))) \leftarrow C\!\neq\! a,\
            {\it new}1(S,N)
\eeq}

\medskip
\noindent
Similarly to the single-pattern string matching example
of the previous Section~\ref{match_pos_ex}, this specialized,
semideterministic program corresponds to a finite
automaton with counters. This finite automaton is deterministic,
except for the states corresponding to the predicates  {\it new}4 and {\it
new}5 where any remaining portion of the input word is accepted. A similar
derivation cannot be performed  by usual partial deduction techniques
without a prior transformation into {\em failure continuation passing
style}~\cite{Smi91}.

\subsection{From Regular Expressions to Finite Automata}
\label{reg_expr_ex}

In this example we show the derivation of
a deterministic finite automaton by specializing a general parser for
regular  expressions w.r.t.~a given regular expression. The initial program
{\it Reg\_Expr} for testing whether or not a string belongs to the language
denoted by a regular expression over the alphabet $\{a,b\}$, is the one
given below.

\medskip

\fpr{{\it Reg\_Expr}\hfill (initial, nondeterministic)}{
\Feq{1.}{\it in\_language}(E,S) \If {\it string}(S), \ {\it accepts}(E,S)
\Nex{2.}{\it string}([~])\If
\Nex{3.}{\it string}([a|S])\If {\it string}(S)
\Nex{4.}{\it string}([b|S])\If {\it string}(S)
\Nex{5.}{\it accepts}(E,[E]) \If {\it symbol}(E)
\Nex{6.} {\it accepts}(E_1 E_2,S) \If
        {\it append}(S_1,S_2,S),\
        {\it accepts}(E_1,S_1),\
        {\it accepts}(E_2,S_2)
\Nex{7.} {\it accepts}(E_1\!+\!E_2,S) \If {\it accepts}(E_1,S)
\Nex{8.} {\it accepts}(E_1\!+\!E_2,S) \If {\it accepts}(E_2,S)
\Nex{9.} {\it accepts}(E^*,[~])
\Nex{10.} {\it accepts}(E^*,S) \If
        {\it ne\_append}(S_1,S_2,S),\
        {\it accepts}(E,S_1),\
        {\it accepts}(E^*,S_2)
\Nex{11.} {\it symbol}(a) \If
\Nex{12.} {\it symbol}(b) \If
\Nex{13.} {\it ne\_append}([A],Y,[A|Y]) \If
\Nex{14.} {\it ne\_append}([A|X],Y,[A|Z]) \If {\it ne\_append}(X,Y,Z)
\eeq
}

\medskip
\noindent
We have that
${\it in\_language}(E,S)$ holds iff
$S$ is a string in $\{a,b\}^*$ and $S$ belongs to the language
denoted by the regular expression $E$. In this {\it Reg\_Expr} program
we have used the predicate ${\it ne\_append}(S_1,S_2,S)$ which holds iff
the non-empty string $S$ is the concatenation of
the {\em nonempty} string $S_1$ and the string $S_2$.
The use of the atom ${\it ne\_append}(S_1,S_2,S)$ in clause
10 ensures that we have a {\em terminating}
program, that is, a program for which we cannot have an
infinite derivation when starting from a ground goal.
Indeed, if in clause 10 we replace ${\it ne\_append}(S_1,S_2,S)$
by ${\it append}(S_1,S_2,S)$, then we may construct
an infinite derivation because from a goal of the form
${\it accepts}(E^*,S)$ we can derive a new goal of the form
$({\it accepts}(E,[~]),\ {\it accepts}(E^*,S))$.

We consider the following mode for the program {\it Reg\_Expr\/}:\\
$\{{\it in\_language}(+,+),$ ${\it string}(+),$ ${\it accepts}(+,+),$
${\it symbol}(+),$ ${\it ne\_append}(?,?,+),$ ${\it append}(?,?,+)\}$.

We use our Determinization Strategy
to specialize the program {\it Reg\_Expr\/} \wrt the atom \\
${\it in\_language}((aa^* (b\!+\!bb))^*,S)$.
Thus, we begin by introducing the definition clause:

\oeq{15.} {\ {\it in\_language_s}(S) \leftarrow
{\it in\_language}((aa^* (b\!+\!bb))^*,S)}

\noindent
The mode for this new predicate is  ${\it in\_language_s}(+)$ because
$S$ is an input argument of ${\it in\_language}$.
The output of the Determinization
Strategy is the following specialized program ${\it Reg\_Expr_s}$:

\medskip
\fpr{${\it Reg\_Expr_s}$\hfill(specialized, semideterministic)}{
\Feq{16.} {\it in\_language_s}(S) \If \mathit{new}1(S)
\Nex{17.} \mathit{new}1([~]) \If
\Nex{18.} \mathit{new}1([a|S]) \If \mathit{new}2(S)
\Nex{19.} \mathit{new}2([a|S]) \If \mathit{new}3(S)
\Nex{20.} \mathit{new}2([b|S]) \If \mathit{new}4(S)
\Nex{21.} \mathit{new}3([a|S]) \If \mathit{new}3(S)
\Nex{22.} \mathit{new}3([b|S]) \If \mathit{new}4(S)
\Nex{23.} \mathit{new}4([~]) \If
\Nex{24.} \mathit{new}4([a|S]) \If \mathit{new}2(S)
\Nex{25.} \mathit{new}4([b|S]) \If \mathit{new}1(S)
\eeq}

\medskip
\noindent
This specialized program corresponds to a
deterministic finite automaton.

\subsection{Matching Regular Expressions}
\label{reg_match_ex}

The following nondeterministic program defines a relation
${\it re\_match}(E,S)$, where $E$ is a regular expression and $S$ is a
string, which holds iff there exists a substring $P$ of
$S$ such that $P$ belongs to the language  denoted
by $E$:

\medskip

\fpr{{\it Reg\_Expr\_Match} \hfill (initial, nondeterministic)}{
\Feq{1.} {\it re\_match}(E,S)  \If
        {\it append}(Y,{\it R},S),\
        {\it append}({\it L},P,Y),\
        {\it accepts}(E,P)
\eeq }

\medskip

\noindent
The predicates {\it append} and {\it accepts}
are defined as in the programs {\it Naive\_Match} (see
Section~\ref{limitations_pd}) and {\it Reg\_Expr} (see
Section~\ref{reg_expr_ex}), respectively, and their clauses are not listed
here. We consider the following mode for the program {\it
Reg\_Expr\_Match\/}: $\{{\it append}(?,?,+),$ ${\it accept}(+,+),$ ${\it
re\_match}(+,+)\}$.

We want to specialize the program~{\it Reg\_Expr\_Match}
\wrt the regular expression
$a a^*  b$.
Thus, we introduce the following definition clause:

\noindent
\bpr
\Feq{2.}{\it re\_match_s}(S)  \leftarrow
{\it re\_match}(a a^*  b, \ S)
\eeq
\epr

\noindent
The mode of this new predicate is
${\it re\_match_s}(+)$ because $S$
is an input argument of ${\it re\_match}$.
The output of the Determinization Strategy is the
following program:

\medskip

\fpr{${\it Reg\_Expr\_Match}_s$\hfill
(specialized, semideterministic)}{
\Feq{3.} {\it re\_match_s}(S) \leftarrow {\it new}1(S)
\Nex{4.} {\it new}1([a|S]) \leftarrow {\it new}2(S)
\Nex{5.} {\it new}1([C|S]) \leftarrow C\!\neq\! a,\
            {\it new}1(S)
\Nex{6.} {\it new}2([a|S]) \leftarrow {\it new}3(S)
\Nex{7.} {\it new}2([C|S]) \leftarrow
            C\!\neq\!a,\  {\it new}1(S)
\Nex{8.} {\it new}3([a|S]) \leftarrow {\it new}4(S)
\Nex{9.} {\it new}3([b|S]) \leftarrow {\it new}3(S)
\Nex{10.} {\it new}3([C|S]) \leftarrow
            C\!\neq\! a,\ C\!\neq\! b,\ {\it new}1(S)
\Nex{11.} {\it new}4(S) \leftarrow
\eeq}

\medskip
\noindent
Similarly to the single-pattern string matching example
of Section~\ref{limitations_pd}, this specialized,
semideterministic program corresponds to a deterministic finite automaton.

\subsection{Specializing Context-free Parsers
to Regular Grammars}
\label{cfg_parse_ex}

Let us consider the following program
for parsing context-free languages:

\medskip
\fpr{{\it CF\_Parser}\hfill (initial, nondeterministic)}{
\Feq{1.}{\it string\_parse}(G,A,W) \If {\it string}(W),
        \ {\it parse}(G,A,W)
\Nex{2.}{\it string}([~])\If
\Nex{3.}{\it string}([0|W])\If {\it string}(W)
\Nex{4.}{\it string}([1|W])\If {\it string}(W)
\Nex{5.} {\it parse}({\it G},[~],[~]) \If
\Nex{6.} {\it parse}({\it G},[A|X],[A|Y]) \If
        {\it terminal}(A),\
        {\it parse}(G,X,Y)
\Nex{7.} {\it parse}(G,[A|X],Y) \nc \If  \nc
        {\it nonterminal}(A),\
        {\it member}(A \rightarrow B,G),\
\bdy        {\it append}(B,X,Z),\
        {\it parse}(G,Z,Y)
\Nex{8.} {\it member}(A,[A|X]) \If
\Nex{9.} {\it member}(A,[B|X]) \If \  {\it member}(A,X)
\eeq}

\medskip

\noindent
together with the clauses for the
predicate {\it append} defined as in program {\it Match}\_{\it Pos} (see
Section~\ref{match_pos_ex}), and the unit clauses stating that 0 and 1 are
terminals and $s,u,v,$ and $w$ are nonterminals. The
first argument of {\it parse\/} is a
context-free grammar, the second argument is  a list of terminal
and nonterminal symbols, and the
third argument is a word represented as a list
of terminal symbols.
We assume that a context-free grammar is represented as a list of
productions of the form $x\rightarrow y$,
where $x$ is a nonterminal symbol and
$y$ is a list of terminal
and nonterminal symbols.
We have that ${\it parse}(G,[s], W)$ holds iff
from the symbol $s$ we can derive the word $W$ using  the grammar $G$.
We consider the following mode for the program {\it
CF\_Parser\/}:
$\{{\it string\_parse}(+,+,+), {\it string}(+),
{\it parse}(+,+,+), {\it terminal}(+), {\it nonterminal}(+),$
${\it member}(?,+), {\it append}(+,+,?)  \}$.

We want to specialize our parsing program \wrt
the following regular grammar:

\smallskip
{\centering
\begin{tabular}{lll}
$s \rightarrow 0\,u$ &  $s\rightarrow 0\,v$ & $s \rightarrow 0\,w$
\\
$u \rightarrow 0$ & $u \rightarrow 0\,u$ & $u \rightarrow 0\,v$
\\
$v \rightarrow 0$ & $v \rightarrow 0\,v$ & $v \rightarrow 0\,u$
\\
$\!w \rightarrow 1$ & $\!w \rightarrow 0\,w$ &
\\
\end{tabular}\par
}
\smallskip

\noindent
To this aim we apply our Determinization Strategy starting
from the following definition clause:

\smallskip
\begin{tabular}{rrllll}
10. & ${\it string\_parse_s}(W)  \If{\it parse}([\!\!\!\!\!$
&  $s \rightarrow [0,u]$,\   & $s \rightarrow [0,v]$,\ & $s
\rightarrow [0,w]$,\\

&& $u \rightarrow [0]$,\ & $u \rightarrow [0,u]$,\ & $u \rightarrow
[0,v]$,\\

&& $v \rightarrow [0]$,\ & $v \rightarrow [0,v]$,\ & $v \rightarrow
[0,u]$,\\

&& $\!w \rightarrow [1]$,\ & $\!w \rightarrow [0,w]$ &&
$\!\!\!\!\!],\ [s],\  W )$\\

\end{tabular}
\smallskip

\noindent The mode for this new predicate
is ${\it string\_parse_s}(+)$.
The output of the Determinization Strategy is the following
specialized program ${\it CF\_Parser_s}$:

\medskip
\fpr{${\it CF\_Parser_s}$\hfill (specialized, semideterministic)}{
\Feq{11.} {\it string\_parse_s}(W) \If {\it new}1(W)
\Nex{12.} {\it new}1([0|W]) \If {\it new}2(W)

\Nex{13.} {\it new}2([0|W]) \If {\it new}3(W)
\Nex{14.} {\it new}2([1|W]) \If {\it new}4(W)

\Nex{15.} {\it new}3([~]) \If
\Nex{16.} {\it new}3([0|W]) \If {\it new}5(W)
\Nex{17.} {\it new}3([1|W]) \If {\it new}4(W)

\Nex{18.} {\it new}4([~]) \If

\Nex{19.} {\it new}5([~]) \If
\Nex{20.} {\it new}5([0|W]) \If {\it new}3(W)
\Nex{21.} {\it new}5([1|W]) \If {\it new}4(W)
\eeq
}

\medskip
\noindent
This program corresponds to a deterministic finite automaton.

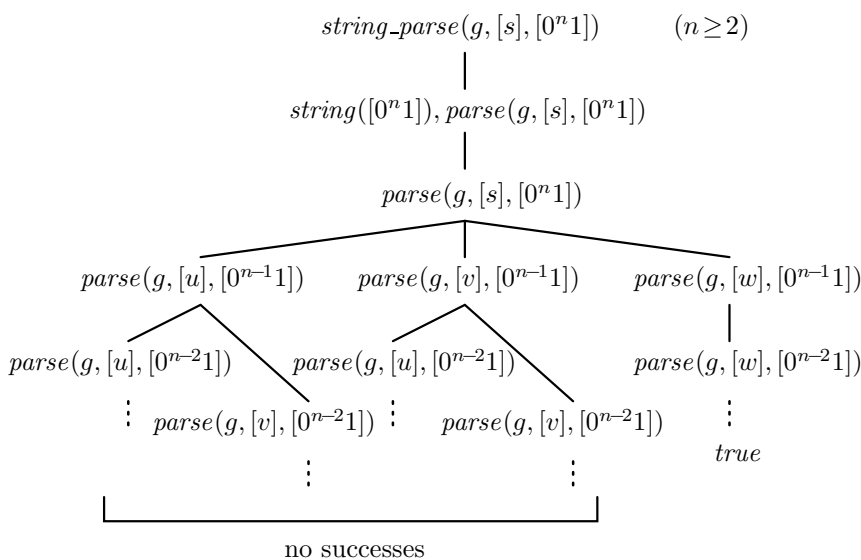
\begin{figure}

\begin{centering}

\setlength{\unitlength}{0.00083333in}
\begingroup\makeatletter\ifx\SetFigFont\undefined%
\gdef\SetFigFont#1#2#3#4#5{%
   \reset@font\fontsize{#1}{#2pt}%
   \fontfamily{#3}\fontseries{#4}\fontshape{#5}%
   \selectfont}%
\fi\endgroup%
{\renewcommand{\dashlinestretch}{30}
\begin{picture}(5238,3375)(0,-10)
\thicklines
\path(2850,2650)(2850,2425)
\dottedline{68}(4500,975)(4500,825)
\path(2850,3150)(2850,2925)
\path(2850,2100)(1200,1875)
\path(2850,2100)(2850,1875)
\path(2850,2100)(4500,1875)
\path(4500,1575)(4500,1350)
\path(1200,1575)(750,1350)
\path(2850,1575)(2400,1350)
\dottedline{68}(750,975)(750,825)
\dottedline{68}(2400,975)(2400,825)
\path(2850,1575)(3525,975)
\path(1200,1575)(1875,975)
\dottedline{68}(1875,600)(1875,450)
\dottedline{68}(3525,600)(3525,450)
\thicklines   %%%%%%%%%%%%%%%%%%%%%%%%%%%%%%
\path(600,375)(600,225)(3675,225)(3675,375)
\put(1750,2700){\makebox(0,0)[lb]{{{{\SetFigFont{10}{13.2}{\rmdefault}
{\mddefault}{\updefault}${\it string}([0^n1]),
        {\it parse}(g,[s],[0^n1])$}}}}}

\put(1950,3225){\makebox(0,0)[lb]{{{{\SetFigFont{10}{13.2}{\rmdefault}
{\mddefault}{\updefault}${\it string\_parse}(g,[s],[0^n1])$}}}}}

\put(3900,1650){\makebox(0,0)[lb]{{{{\SetFigFont{10}{13.2}{\rmdefault}
{\mddefault}{\updefault}${\it parse}(g,[w],[0^{n\!-\!1}1])$}}}}}

\put(3900,1125){\makebox(0,0)[lb]{{{{\SetFigFont{10}{13.2}{\rmdefault}
{\mddefault}{\updefault}${\it parse}(g,[w],[0^{n\!-\!2}1])$}}}}}
\put(4400,600){\makebox(0,0)[lb]{{{{\SetFigFont{10}{13.2}{\rmdefault}{
\mddefault}{\updefault}{\it true}}}}}}            %%%%%%%%%%%%%%%%%%%
\put(0,1125){\makebox(0,0)[lb]{{{{\SetFigFont{10}{13.2}{\rmdefault}
{\mddefault}{\updefault}${\it parse}(g,[u],[0^{n\!-\!2}1])$}}}}}
\put(2325,2175){\makebox(0,0)[lb]{{{{\SetFigFont{10}{13.2}{\rmdefault}
{\mddefault}{\updefault}${\it parse}(g,[s],[0^n1])$}}}}}
\put(4150,3225){\makebox(0,0)[lb]{{{{\SetFigFont{10}{13.2}{\rmdefault}
{\mddefault}{\updefault}($n\!\geq\!2$)}}}}}
\put(1770,1125){\makebox(0,0)[lb]{{{{\SetFigFont{10}{13.2}{\rmdefault}
{\mddefault}{\updefault}${\it parse}(g,[u],[0^{n\!-\!2}1])$}}}}}
\put(450,1650){\makebox(0,0)[lb]{{{{\SetFigFont{10}{13.2}{\rmdefault}{
\mddefault}{\updefault}${\it parse}(g,[u],[0^{n\!-\!1}1])$}}}}}  %%%%%
\put(2175,1650){\makebox(0,0)[lb]{{{{\SetFigFont{10}{13.2}{\rmdefault}
{\mddefault}{\updefault}${\it parse}(g,[v],[0^{n\!-\!1}1])$}}}}}
\put(2700,750){\makebox(0,0)[lb]{{{{\SetFigFont{10}{13.2}{\rmdefault}{
\mddefault}{\updefault}${\it parse}(g,[v],[0^{n\!-\!2}1])$}}}}} %%%%%%
\put(900,750){\makebox(0,0)[lb]{{{{\SetFigFont{10}{13.2}{\rmdefault}
{\mddefault}{\updefault}${\it parse}(g,[v],[0^{n\!-\!2}1])$}}}}}
\put(1725,0){\makebox(0,0)[lb]{{{{\SetFigFont{10}{13.2}{\rmdefault}
{\mddefault}{\updefault}no successes}}}}} \end{picture} }

\end{centering}
\caption{Derivation tree $T_1$
for ${\it string\_parse}(g,[s],[0^n1])$.\label{T1}}
\end{figure}

Now, we would like to discuss the improvements we
achieved in this example by applying our Determinization Strategy. Let
us consider the {\it derivation tree} $T_1$ (see Fig.~\ref{T1})
generated by the initial program {\it CF\_Parser} starting from the
goal ${\it string\_parse}(g,[s],[0^n1])$, where $g$ denotes the
grammar \wrt which we have specialized the {\it CF\_Parser} program
and $[0^n1]$ denotes the list $[0,\dots,0,1]$ with $n$ occurrences of
$0$. The nodes of $T_1$ are labeled by the
goals derived from ${\it string\_parse}(g,[s],[0^n1])$.
In particular, the root of the derivation
tree is labeled by ${\it string\_parse}(g,[s],[0^n1])$
and a node labeled by a goal $G$ has $k$ children labeled
by the goals $G_1, \dots , G_k$ which are derived
from $G$ (see Section~\ref{op_sem}). The tree $T_1$ has a number of nodes
which is $O(2^n)$. Thus, by using the initial program {\it CF\_Parser}
it takes $O(2^n)$ number of steps to search for
a derivation from the root goal ${\it string\_parse}(g,[s],[0^n1])$
to the goal {\it true}.
(Indeed, this is the case if one uses a Prolog compiler.)  In
contrast, by using the specialized program ${\it CF\_Parser_s}$, it
takes $O(n)$ steps to search for a derivation  from the goal ${\it
string\_parse_s}([0^n1])$ to {\it true}, because the derivation
tree $T_2$ has a number of nodes which is $O(n)$ (see Fig.~\ref{T2}).

The improvement of performance is due to the fact that our
Determinization Strategy is able to avoid repeated
derivations by introducing new definition clauses
whose bodies have goals from which common subgoals are derived.
Thus, after performing folding steps which use these definition
clauses, we reduce the search space during program execution.

For instance, our strategy introduces the
predicate {\it new}2 defined by the following clauses:

\bpr
\feq {\it new}2(W) \If {\it string}(W),\ {\it parse}(g,[u],W)
\nex {\it new}2(W) \If {\it string}(W),\ {\it parse}(g,[v],W)
\nex {\it new}2(W) \If {\it string}(W),\ {\it parse}(g,[w],W)
\eeq
\epr

\noindent
whose bodies are goals from which common subgoals are derived
for $W\! =\! [0^{n-1}1]$ and $n\!\geq\!2$.
Indeed, for instance, ${\it parse}(g,[u],[0^{n-2}1])$ can be derived
from both ${\it parse}(g,[u],[0^{n-1}1])$ and ${\it
parse}(g,[v],[0^{n-1}1])$ (see Fig.~\ref{T1}).
The reader may verify that by using the specialized program
$\mathit{CF\_Parser_s}$ no repeated goal is derived from
${\it string\_parse_s}(g,[s],[0^{n}1])$.

The ability of our Determinization Strategy of putting together
the computations performed by the initial program in
different branches of the computation tree, so that common repeated
subcomputations are avoided, is based on the ideas which motivate
the {\it tupling\/} strategy~\cite{Pet77}, first proposed
as a transformation
technique for functional languages.

\begin{figure}
\begin{centering}

\setlength{\unitlength}{0.00083333in}
\begingroup\makeatletter\ifx\SetFigFont\undefined%
\gdef\SetFigFont#1#2#3#4#5{%
  \reset@font\fontsize{#1}{#2pt}%
  \fontfamily{#3}\fontseries{#4}\fontshape{#5}%
  \selectfont}%
\fi\endgroup%
{\renewcommand{\dashlinestretch}{30}
\begin{picture}(2528,2250)(0,-10)
\thicklines
\path(975,2025)(975,1850)
\path(975,1500)(975,1325)
\dottedline{68}(975,375)(975,225)
\path(975,975)(975,800)
\put(2400,2100){\makebox(0,0)[lb]{($n\!\geq\!2$)}}
\put(975,2100){\makebox(0,0)[cb]{${\it string\_parse_s}(g,[s],[0^n1])$}}
\put(975,1575){\makebox(0,0)[cb]{${\it new}1([0^n1])$}}
\put(975,1050){\makebox(0,0)[cb]{${\it new}2([0^{n\!-\!1}1])$}}
\put(975,525){\makebox(0,0)[cb]{${\it new}3([0^{n\!-\!2}1])$}}
\put(975,0){\makebox(0,0)[cb]{\it true}}
\end{picture}
}

\end{centering}
\caption{Derivation tree $T_2$
for ${\it string\_parse_s}([0^n1])$.\label{T2}}
\end{figure}

\section{Experimental Evaluation}
\label{statistics}

The Determinization Strategy has been implemented in the
MAP program transformation system~\cite{Ren97}.
All  program specialization examples presented in
Sections~\ref{limitations_pd}, \ref{semidet}, and~\ref{examples}
have been worked out in a fully automatic way by the MAP system. We have
compared the specialization times and the speedups obtained by the MAP
system with those obtained by ECCE, a system for (conjunctive) partial
deduction~\cite{Leu00}. All experimental results reported in this
section have been obtained by using SICStus Prolog 3.8.5 running on a
Pentium II under Linux.

In Table~1 we consider
the examples of Sections~\ref{limitations_pd}, \ref{semidet},
and~\ref{examples}, and we show the times
taken (i)~for performing partial
deduction by using the ECCE system, (ii)~for performing conjunctive partial
deduction by using the ECCE system, and (iii)~for applying the
Determinization Strategy by using the MAP system. The
{\em static input} shown in Column~2 of Table~1 is the goal
\wrt which we have specialized the programs of Column~1. For
running the ECCE system suitable choices among the available  unfolding
strategies and generalization strategies should be made. We have used the
choices suggested by the system itself for partial deduction and
conjunctive partial deduction, and we made some changes only when
specialization was not performed within a reasonable amount of time. For
running the MAP system the only information to be provided by the user is
the mode for the program to be specialized. The system assumes that the
program satisfies this mode and no mode analysis is performed.

\begin{table}[ht]
\begin{centering}
\vspace{0.3cm}
%\noindent \raggedright
\begin{tabular}{|l|c|c|c|c|}
\hline
Program&
Static Input&
ECCE&
ECCE&
MAP\\
{\par}&
&
(PD)&
(CPD)&
(Det)\\
\hline
\hline
\textit{Naive\_Match}&
\( \mathit{naive}\_\mathit{match}([aab],S) \)&
360&
370&
70\\
\hline
\textit{Naive\_Match}&
\( \mathit{naive}\_\mathit{match}([aaaaaaaaab],S) \)&
420&
2120&
480\\
\hline
\emph{Match\_Pos}&
\( \mathit{match}\_\mathit{pos}([aab],S,N) \)&
540&
360&
100\\
\hline
\emph{Match\_Pos}&
\( \mathit{match}\_\mathit{pos}([aaaaaaaaab],S,N) \)&
650&
910&
500\\
\hline
\emph{Mmatch}&
\( \mathit{mmatch}([[aaa],[aab]],S,N) \)&
1150&
1400&
280\\
\hline
\emph{Mmatch}&
\( \mathit{mmatch}([[aa],[aaa],[aab]],S,N) \)&
1740&
2040&
220\\
\hline
\emph{Reg\_Expr}&
\( in\_language((aa^{*}(b\! +\! bb))^{*},S) \)&
6260&
138900&
420\\
\hline
\emph{Reg\_Expr}&
\( in\_language(a^{*}(b\! +\! bb\! +\! bbb),S) \)&
3460&
5430&
230\\
\hline
\emph{Reg\_Expr\_Match}&
\( \mathit{re}\_\mathit{match}(aa^{*}b,S) \)&
970&
5290&
210\\
\hline
\emph{Reg\_Expr\_Match}&
\( \mathit{re}\_\mathit{match}(a^{*}(b+bb),S) \)&
1970&
11200&
300\\
\hline
\emph{CF\_Parser}&
\( \mathit{string}\_\mathit{parse}(g,[s],W) \)&
23400&
32700&
1620\\
\hline
\emph{CF\_Parser}&
\( \mathit{string}\_\mathit{parse}(g_{1},[s],W) \)&
31200&
31800&
2000\\
\hline
\end{tabular}\vspace{0.3cm}

\caption{Specialization Times (in milliseconds).}
\end{centering}
\end{table}

The experimental results of Table~1 show that the MAP implementation of
the Determinization Strategy is much faster than the
ECCE implementation of both partial deduction and conjunctive
partial deduction. We believe that, essentially, this is due to the fact
that ECCE employs very sophisticated techniques, such as those based on
{\em homeomorphic embeddings}, for controlling the unfolding and
the generalization steps, and ensuring the termination of the
specialization process. For a fair comparison, however, we should recall
that Determinization may not terminate on examples different from those
considered in this paper.

\medskip

We have already mentioned in Section~\ref{limitations_pd} that
the performance of the programs derived by the Determinization Strategy may
be further improved by applying post-processing transformations
which exploit the semideterminism of the programs. In particular, we may:
(i) reorder the clauses so that unit clauses appear before non-unit
clauses, and (ii) remove disequations by introducing cuts instead.
The reader may verify that these transformations
preserve the operational semantics.
{F}or a systematic
treatment of cut introduction, the reader may refer to~\cite{Dev90,SaTa85}.
As an example we now show the program obtained from ${\it
Match\_Pos_s}$ (see Section~\ref{semidet}) after the above
post-processing transformations have been performed.

\medskip

\fpr{\({\it Match\_Pos_{cut}}\)\hfill (specialized, with cuts)}{
\Feq{}  {\it match\_pos_s}(S,N) \leftarrow {\it new}1(S,N)
\Nex{} {\it new}1([a|S],M) \leftarrow \ !,\ {\it new}2(S,M)
\Nex{} {\it new}1([C|S],s(N)) \leftarrow {\it new}1(S,N)
\Nex{} {\it new}2([a|S],M) \leftarrow \ !,\ {\it new}3(S,M)
\Nex{}  {\it new}2([C|S],s(s(N))) \leftarrow
        {\it new}1(S,N)
\Nex{} {\it new}3([a|S],s(M)) \leftarrow \ !,\ {\it new}3(R,S)
\Nex{} {\it new}3([b|S],M) \leftarrow \ !,\ {\it new}4(R,S)
\Nex{} {\it new}3([C|S],s(s(s(N)))) \leftarrow
        {\it new}1(S,N)
\Nex{} {\it new}4(S,0) \leftarrow
\Nex{} {\it new}4([a|S],s(s(s(M)))) \leftarrow \ !,\ {\it new}2(S,M)
\Nex{} {\it new}4([C|S],s(s(s(s(N))))) \leftarrow
        {\it new}1(S,N)
\eeq}

\bigskip

\noindent
In Table~2 below we report the speedups obtained by
partial deduction, conjunctive partial deduction, Determinization,
and Determinization followed by disequation removal and
cut introduction.
Every speedup is computed as the ratio between the timing of
the initial program and the timing of the
specialized program.
These timings were obtained by running the various programs
several times (up to 10,000) on significantly large input
lists (up to 4,000 items).

\begin{table}[ht]
\begin{centering}
\vspace{0.3cm}
{%\noindent \raggedright
\begin{tabular}{|l|c|c|c|c|c|}
\hline
Program&
Static Input&
Speedup&
Speedup&
Speedup&
Speedup\\
{\par}&
&
(PD)&
(CPD)&
(Det)&
(Det $\! \& \!$ Cut)\\
\hline
\hline
\textit{Naive\_Match}&
\( \mathit{naive}\_\mathit{match}([aab],S) \)&
3.1&
\( 5.8\!\times\! 10^{3} \)&
\( 3.0\!\times\! 10^{3} \)&
\( 6.8\!\times\! 10^{3} \)\\
\hline
\textit{Naive\_Match}&
\( \mathit{naive}\_\mathit{match}([aaaaaaaaab],S) \)&
3.3&
\( 6.9\!\times\! 10^{3} \)&
\( 5.8\!\times\! 10^{3} \)&
\( 12.4\!\times\! 10^{3} \)\\
\hline
\emph{Match\_Pos}&
\( \mathit{match}\_\mathit{pos}([aab],S,N) \)&
1.6&
\( 3.6\!\times\! 10^{3} \)&
\( 1.8\!\times\! 10^{3} \)&
\( 4.0\!\times\! 10^{3} \)\\
\hline
\emph{Match\_Pos}&
\( \mathit{match}\_\mathit{pos}([aaaaaaaaab],S,N) \)&
2.1&
\( 5.3\!\times\! 10^{3} \)&
\( 2.9\!\times\! 10^{3} \)&
\( 8.1\!\times\! 10^{3} \)\\
\hline
\emph{Mmatch}&
\( \mathit{mmatch}([[aaa],[aab]],S,N) \)&
1.7&
\( 4.5\!\times\! 10^{3} \)&
\( 3.5\!\times\! 10^{3} \)&
\( 6.2\!\times\! 10^{3} \)\\
\hline
\emph{Mmatch}&
\( \mathit{mmatch}([[aa],[aaa],[aab]],S,N) \)&
1.6&
\( 2.5\!\times\! 10^{3} \)&
\( 3.9\!\times\! 10^{3} \)&
\( 5.4\!\times\! 10^{3} \)\\
\hline
\emph{Reg\_Expr}&
\( in\_language((aa^{*}(b\! +\! bb))^{*},S) \)&
29.8&
\( 6.2\!\times\! 10^{3} \)&
\( 2.3\!\times\! 10^{5} \)&
\( 3.9\!\times\! 10^{5} \)\\
\hline
\emph{Reg\_Expr}&
\( in\_language(a^{*}(b\! +\! bb\! +\! bbb),S) \)&
\( 1.3\!\times\! 10^{4} \)&
\( 3.3\!\times\! 10^{4} \)&
\( 4.6\!\times\! 10^{4} \)&
\( 5.7\!\times\! 10^{4} \)\\
\hline
\emph{Reg\_Expr\_Match}&
\( \mathit{re}\_\mathit{match}(aa^{*}b,S) \)&
\( 5.7\!\times\! 10^{2} \)&
\( 2.7\!\times\! 10^{4} \)&
\( 1.5\!\times\! 10^{6} \)&
\( 3.0\!\times\! 10^{6} \)\\
\hline
\emph{Reg\_Expr\_Match}&
\( \mathit{re}\_\mathit{match}(a^{*}(b+bb),S) \)&
\( 2.1\!\times\! 10^{2} \)&
\( 3.4\!\times\! 10^{3} \)&
\( 2.5\!\times\! 10^{5} \)&
\( 4.1\!\times\! 10^{5} \)\\
\hline
\emph{CF\_Parser}&
\( \mathit{string}\_\mathit{parse}(g,[s],W) \)&
1.5&
1.5&
87.1&
87.1\\
\hline
\emph{CF\_Parser}&
\( \mathit{string}\_\mathit{parse}(g_{1},[s],W) \)&
1.1&
1.1&
61.3&
61.3\\
\hline
\end{tabular}\par}
\vspace{0.3cm}
\caption{Speedups.}
\end{centering}
\end{table}

%%%%%%%%%%%%%%%%%%%

\noindent
To clarify the
content of Table~2 let us remark that:

\smallskip
\noindent
Column 1 shows the names of the initial programs
with reference to Sections~\ref{limitations_pd}, \ref{semidet},
and~\ref{examples}.

\smallskip
\noindent
Column 2 shows the static input. The argument $[aab]$
denotes the list
$[a,a,b]$. Similar notation has been used for the other static input
arguments. The argument $g$ of the first ${string\_parse}$ atom
denotes the regular grammar considered in Example~\ref{cfg_parse_ex}.
The argument $g_1$ of the last ${string\_parse}$ atom denotes
the regular grammar:\\
$\{s \rightarrow 0\,u  $,
\ $s \rightarrow 1\,v  $,
\ $u \rightarrow 0     $,
\ $u \rightarrow 0\,v  $,
\ $u \rightarrow 0\,w  $,
\ $v \rightarrow 1     $,
\ $v \rightarrow 0\,v  $,
\ $v \rightarrow 1\,u  $,
\ $w \rightarrow 1     $,
\ $w \rightarrow 1\,w\}$.

\smallskip
\noindent
Column 3, called Speedup (PD), shows the speedups we
have obtained after the application of partial deduction.

\smallskip
\noindent
Column 4, called Speedup (CPD), shows the speedups we
have obtained after the application of conjunctive partial deduction.

\smallskip
\noindent
Column 5, called Speedup (Det), shows the speedups we
have obtained after the application of the Determinization Strategy.

\smallskip
\noindent
Column 6, called Speedup (Det \& Cut), shows the speedups  we have
obtained after the application of the Determinization Strategy
followed by the removal of disequations and the introduction of cuts.

\medskip

Let us now discuss our experimental results of Table~2.
In all examples the best speedups are those obtained after the application
of the Determinization Strategy followed by the removal of disequations and
the introduction of cuts (see column Det \& Cut).

As expected, conjunctive partial deduction gives higher speedups than
partial deduction.

In some cases, conjunctive partial deduction gives better results
than Determinization (see the first 5 rows of columns CPD and Det).
This happens in examples where most nondeterminism is avoided
by eliminating intermediate lists (see, for instance, the example of
Section~\ref{limitations_pd}). In those examples, in fact, the
Determinization Strategy may be less advantageous than conjunctive partial
deduction because it introduces disequations which may be costly
to check at runtime. However, as already mentioned, all disequations may
be eliminated by introducing cuts (or, equivalently, if-then-else
constructs) and the programs derived after disequation removal and cut
introduction are indeed more efficient than those derived by conjunctive
partial deduction (see column Det \& Cut).

{F}or some  programs (see, for instance, the entries for
{\it Reg\_Expr\/} and {\it CF\_Parser})
the speedups of the (Det) column are equal to the speedups of the (Det \&
Cut) column. The reason for this fact is the absence of disequations in the
specialized program, so that the introduction of cuts does not improve
efficiency.

We would like to notice that further post-processing techniques
are applicable. For instance, similarly to the
familiar case of finite automata, we may eliminate
clauses corresponding to $\varepsilon$-transitions where no input
symbols are consumed (such as clause 9 in program ${\it
Match\_Pos_s}$), and we may also minimize the number of predicate symbols
(this corresponds to the minimization of the number of states). We do
not present here these post-processing techniques because they are outside
the scope of the paper.

In summary, the experimental results of Table~2 confirm that
in the examples we have considered, the
Determinization Strategy followed by the removal of disequations
in favour of cuts, achieves greater speedups
than (conjunctive) partial deduction.
However, it should be noticed that, as already mentioned, 
Determinization does not guarantee termination, while
(conjunctive) partial deduction does, and in order to
terminate in all cases, (conjunctive) partial deduction
employs generalization techniques that
may reduce speedups. 
In the next section we further discuss the issue of 
devising a generalization technique that ensures the 
termination of the Determinization Strategy.   

\section{Concluding Remarks and Related Work}
\label{discussion}

We have proposed a specialization technique  for logic
programs based on an automatic strategy, called Determinization Strategy,
which makes use of the following trans\-formation rules: (1) definition
introduction, (2) definition elimination, (3) un\-fold\-ing, (4) folding,
(5) subsumption, (6) head generalization, (7) case split, (8) equation
elimination, and (9) disequation replacement. (Actually, we make use of the
safe versions of the rules 4, 6, 7, and 8.)
We have also shown that our strategy may reduce the
amount of non\-de\-ter\-minism in the specialized programs and
it may achieve exponential gains in time complexity.

To get these results, we allow new predicates  to be
introduced by  {\em one or more} non-recursive definition
clauses  whose bodies  may contain  {\it more than one}
atom. We also allow folding steps using these
definition clauses. By a folding step several clauses
are replaced by a single clause, thereby reducing
nondeterminism.

The use of the subsumption rule is motivated by the desire of
increasing efficiency by avoiding redundant computations.
Head generalizations are used for deriving clauses with equal
heads and thus, they allow us to perform folding
steps. The case split rule is very important for reducing nondeterminism
because it replaces a clause, say $C$, by several clauses which correspond
to exhaustive and mutually exclusive instantiations of the head of $C$.
To get exhaustiveness and
mutual exclusion, we allow the introduction of
disequalities. To further increase program efficiency, in a
post-processing phase these disequalities may be removed in favour of cuts.

We assume that the initial program to be specialized is associated with
a mode of use for its predicates.
Our Determinization Strategy makes use of this mode information for
directing the various transformation steps, and in particular,
the applications of the unfolding and case split rules.
Moreover, if our strategy terminates,
it derives specialized programs
which are semideterministic \wrt the given mode. This notion has been
formally defined in Section~\ref{semidet}. Although semideterminism is
not in itself a guarantee for efficiency improvement, it is often the
case that efficiency is increased because
nondeterminism is reduced and redundant computations are avoided.

We have shown that the transformation rules we  use
for program specialization, are correct \wrt
the declarative semantics of logic programs
based on the least Herbrand model.
The proof of this correctness result is
similar to the proofs of the correctness results which are
presented in~\cite{GeK94,Ro&99a,TaS84}.

We have also considered an operational semantics
for our logic language where a disequation $t_1\!\neq\!t_2$
holds iff $t_1$ and $t_2$ are not unifiable.
This operational semantics is sound, but not complete \wrt the
declarative semantics. Indeed, if a
goal operationally succeeds in a program,
then it is true in the least Herbrand model of the program,
but not vice versa.
Thus, the proof of correctness of our transformation rules \wrt
the operational semantics cannot be based on
previous results and it is much more elaborate. Indeed, it
requires some restrictions, related to the modes of the predicates, both on
the programs to be specialized and on the applicability of the
transformation rules.

In Section~\ref{rules_stra_pe_sect} we have
extensively discussed the fact that our specialization technique is more
powerful than partial deduction~\cite{Jo&93,LlS91}.
The main reason of the greater power of our technique is that it uses
more powerful transformation rules. In particular, partial deduction
corresponds to the use the definition introduction, definition
elimination, unfolding, and folding transformation rules,
with the restriction that we may only fold a single atom
at a time in the body of a clause.

Our extended rules allow us to introduce and transform
new predicates defined in terms of {\it disjunctions of
conjunctions of atoms} (recall that a set of clauses with
the same head is equivalent to a single clause whose premise
is the disjunction of the bodies of the clauses in the given set).
In this respect, our technique improves over
{\em conjunctive partial deduction}~\cite{De&99},
which is a specialization technique
where new predicates are defined in terms of
conjunctions of atoms.

We have implemented the Determinization Strategy in the MAP transformation
system~\cite{Ren97} and we have tested this implementation by performing
several specializations of string matching and parsing programs.
We have also compared the results obtained by using the MAP system
with those obtained by using the ECCE system for (conjunctive)
partial deduction~\cite{Leu00}.
Our computer experiments confirm that the Determinization Strategy pays off
\wrt both partial deduction and conjunctive partial deduction.

Our transformation technique works for programs
where the only negative literals which are allowed in the body of a
clause, are disequations between terms.
The extension of the Determinization Strategy to normal logic programs
would require an extension of the transformation rules and,
in particular, it would be necessary to use
a {\em negative unfolding} rule, that is, a rule for unfolding
a clause \wrt a (possibly nonground) negative literal different from a
disequation. The correctness of unfold/fold transformation systems which
use the negative unfolding rule has been studied in contexts rather
different from the one considered here (see, for instance, the
work on transformation of {\em first order programs}~\cite{Sat92}) and its
use within the Determinization Strategy requires further work.

The Determinization Strategy may fail to terminate for two
reasons: (i)~the Unfold-Simplify subsidiary strategy may apply the
unfolding rule infinitely often, and (ii)~the while-do loop of the
Determinization Strategy may not terminate, because at each iteration the
Define-Fold subsidiary strategy may introduce new predicates.

The termination of the Unfold-Simplify strategy can be
guaranteed by applying the techniques for finite unfolding already
developed for (conjunctive) partial deduction (see, for
instance,~\cite{De&99,Leu98a,Ma&94b}). Indeed, the unfolding rule used in
this paper is similar to the unfolding rule used in partial deduction.

The introduction of an infinite number of new predicates
can be avoided by extending various methods based on {\em
generalization}, such as those used in (conjunctive) partial
deduction~\cite{De&99,Gall93,Le&98a,PrP93b}.
Recall that in conjunctive partial deduction we may generalize
a predicate definition essentially by means of two techniques:
(i)~the replacement of a term by a variable, which is then taken as an
argument of a new predicate definition, and (ii)~the splitting of a
conjunction of literals into subconjunctions (together with the
introduction of a new predicate for each subconjunction). It has
been shown that the use of~(i) and~(ii) in a suitably controlled way,
allows conjunctive partial deduction to terminate in all cases.
However, termination is guaranteed at the expense of a possibly
incomplete specialization or a possibly incomplete elimination of the
intermediate data structures.

In order to avoid the introduction of an infinite number of new
predicate definitions while applying the Determinization Strategy,
we may follow an approach similar to the one used in the case of
conjunctive partial deduction. However, besides the
generalization techniques (i) and (ii) mentioned above, we may also need
(iii)~the splitting of the set of clauses defining a predicate into
subsets (together with the introduction of a new  predicate for
each subset). Similarly to the case of conjunctive partial deduction,
it can be shown that suitably controlled applications of the
generalization techniques (i), (ii), and (iii) guarantee the termination
of the Determinization Strategy at the expense of deriving programs which
may fail to be semideterministic.

We leave it for further research the issue of controlling generalization,
so that we achieve the termination of the specialization process and at
the same time we maximize the reduction of nondeterminism.

In the string matching examples we have worked out, our strategy
is able to automatically derive programs which behave like
Knuth-Morris-Pratt algorithm, in the sense that they generate a
finite automaton from any given pattern and a general pattern matcher.
This was done also in the case of programs for
matching sets of patterns and programs for matching
regular expressions.

In these examples the improvement over similar
derivations performed by partial
deduction techniques~\cite{Fuj87,Gall93,Smi91}
consists in the fact that we have started from
naive, nondeterministic initial programs, while the corresponding
derivations by partial deduction described in the literature, use
initial programs which are deterministic.
Our derivations also improve over the derivations performed by using {\em
supercompilation} with {\em perfect driving}~\cite{GlK93,Tur86}
and {\em generalized partial computation}~\cite{Fu&91},
which start from initial functional programs which already
incorporate some ingenuity.

A formal derivation of the Knuth-Morris-Pratt algorithm for
pattern matching has also been presented in~\cite{Bi&89}.
This derivation follows the {\it calculational\/} approach
which consists in applying equivalences of higher order functions.
On the one hand the calculational derivation is more general than ours,
because it takes into consideration a generic pattern,
not a fixed one (the string $[a,a,b]$ in our
Example~\ref{limitations_pd}), on the other hand the calculational
derivation is more specific than ours, because it  deals with
single-pattern string matching only, whereas our strategy is able to
automatically derive programs in a much larger class which
also  includes multi-pattern matching, matching with regular
expressions, and parsing.

The use of the case split rule is a form of
reasoning {\em by cases}, which is a very well-known technique
in mechanical theorem proving (see, for instance, the
Edinburgh LCF theorem prover~\cite{Go&79}). Forms of
reasoning by cases have been
incorporated in program specialization techniques such
as the already mentioned supercompilation with
perfect driving~\cite{GlK93,Tur86} and
generalized partial computation~\cite{Fu&91}. However,
the strategy presented in this paper is the first
fully automatic transformation technique
which uses case reasoning to reduce nondeterminism
of logic programs.

Besides specializing programs and reducing nondeterminism,
our strategy is able to eliminate intermediate data structures. Indeed,
the initial programs of our examples in Section~\ref{examples} all have
intermediate lists, while the specialized programs do not have them.
Thus, our strategy can be regarded as an extension of
the transformation strategies for the elimination of
intermediate data structures (see the {\em deforestation}
technique~\cite{Wad90} for
the case of functional programs and the strategy for
{\em eliminating unnecessary variables}~\cite{PrP95a}
for the case of logic programs).
Moreover, our strategy derives specialized programs which
avoid repeated subcomputations (see the Context-free Parsing
example of Section~\ref{cfg_parse_ex}). In this respect
our strategy is similar to the {\em tupling strategy} for
functional programs~\cite{Pet77}.

{F}inally, our specialization strategy is
related to the program derivation techniques called
{\em finite differencing}~\cite{PaK82}
and {\it incrementalization}~\cite{Liu00}.
These techniques use program
invariants to avoid costly, repeated calculations
of function calls.
Our specialization strategy implicitly discovers
and exploits program invariants
when using the folding rule.
It should be noticed, however, that it is difficult to
establish in a rigorous way the formal connection
between the basic ideas underlying our specialization
strategy and the above mentioned program derivation methods based on
program invariants. These methods, in fact, are presented in a very
different framework.

This paper is an improved version of~\cite{Pe&97a}.

\newpage

\section*{Appendix A. Proof of Theorem~\ref{corr_th_op}}

{F}or the reader's convenience,
we rewrite the statement of Theorem~\ref{corr_th_op}.

\medskip
\noindent
{\bf Theorem~\ref{corr_th_op} (Correctness of the Rules \wrt the
Operational Semantics)}
Let $P_0,\dots,P_n$  be a transformation sequence constructed
by using the transformation rules
\ref{def_intro_r}--\ref{diseq_repl_r} and let $p$ be a
non-basic predicate in $P_n$.
Let $M$ be a mode for $P_0\cup \mathit{Defs_n}$ such that:
(i) $P_0\cup \mathit{Defs_n}$ is safe \wrt $M$,
(ii) $P_0\cup \mathit{Defs_n}$ satisfies $M$, and
(iii) the applications of the unfolding,
folding, head generalization, and case split rules
during the construction of $P_0,\dots,P_n$ are all safe \wrt $M$.
Suppose also that:
\begin{enumerate}
\item  {\em if} the folding rule is applied
for the derivation of
a clause $C$ in program $P_{k+1}$ from clauses $C_1,\dots, C_m$ in
program  $P_{k}$ using clauses $D_1,\dots, D_m$ in $\mathit{Defs_k}$,
with $0\!\leq\!k\!<\!n$,\\
{\em then}  for every $i \in \{1,\dots,m\}$ there exists
$j\in\{1,\dots,n\!-\!1\}$ such that $D_i$ occurs in
$P_j$ and $P_{j+1}$ is derived from $P_j$ by unfolding $D_i$.

\item during the transformation
sequence $P_0,\dots,P_n$ the definition elimination rule {\em
either} is never applied {\em or} it is applied \wrt predicate $p$
once only, when deriving $P_n$ from $P_{n-1}$.
\end{enumerate}
Then:
(i) $P_n$ is safe \wrt $M$,
(ii) $P_n$ satisfies $M$, and
(iii) for each atom $A$ which has predicate $p$ and
satisfies mode $M$, $A$ succeeds in $P_0\cup \mathit{Defs_n}$
iff $A$ succeeds in $P_n$.

\medskip

The proof of Theorem~\ref{corr_th_op} will be divided in four parts,
corresponding to Propositions~\ref{safety_preserv},
\ref{mode_preserv}, \ref{part_corr}, and
\ref{completeness} presented below.

Proposition \ref{safety_preserv} ({\em Preservation of Safety})
shows that program
$P_n$ derived according to the hypotheses of Theorem~\ref{corr_th_op},
is safe \wrt mode $M$ (that is, Point (i) of the
thesis of Theorem~\ref{corr_th_op}). Proposition \ref{mode_preserv} ({\em
Preservation of Modes}) shows that $P_n$ satisfies $M$ (that is, Point
(ii) of the thesis of Theorem~\ref{corr_th_op}).
Propositions \ref{part_corr} ({\em Partial Correctness}) and
\ref{completeness} ({\em Completeness}) show the {\em if} part and the
{\em only-if} part, respectively, of Point (iii) of the thesis of
Theorem~\ref{corr_th_op}. For proving these propositions we will use
various notions and lemmata which we
introduce below.

\subsection*{A1. Preservation of Safety}

In this section we prove that, if the transformation rules are
applied according to the restrictions indicated in Theorem \ref{corr_th_op},
then from a program which is safe \wrt a given mode we derive a
program which is safe \wrt the same mode.

\begin{proposition}[Preservation of Safety]\label{safety_preserv}
{\rm
Let $P_0,\dots,P_n$  be a transformation sequence constructed
by using the transformation rules
\ref{def_intro_r}--\ref{diseq_repl_r}.
Let $M$ be a mode for $P_0\cup \mathit{Defs_n}$ such that:
(i) $P_0\cup \mathit{Defs_n}$ is safe \wrt $M$  and
(ii) the applications of the unfolding,
head generalization, and case split rules
during the construction of $P_0,\dots,P_n$ are safe \wrt~$M$.
Then, for $k=0,\ldots,n$, the program $P_k$ is safe \wrt~$M$.
} %end rm
\end{proposition}

\begin{proof} The proof proceeds by induction on $k$.
During the proof we will omit the reference to mode $M$. In particular,
we will simply say that a program (or a clause) is safe,
instead of saying that a program (or a clause) is safe \wrt $M$.

{F}or $k=0$ the thesis follows directly from the hypothesis
that $P_0\cup \mathit{Defs_n}$ is safe and thus, $P_0$
is safe.
Let us now assume that, for $k<n$, program $P_k$ is safe.
We will show that also $P_{k+1}$ is safe. We consider the
following cases, corresponding to the rule which is applied to derive
$P_{k+1}$ from $P_k$.

\begin{description}
\item Case 1: $P_{k+1}$ is derived by applying the definition
introduction rule. $P_{k+1}$ is safe because $P_k$
is safe and, by hypothesis, every definition clause in
${\it Defs}_n$ is safe.

\item
Case 2: $P_{k+1}$ is derived by applying the definition
elimination rule. Then $P_{k+1}$ is safe because
$P_k$ is safe and $P_{k+1} \subseteq P_k$.

\item
Case 3: $P_{k+1}$ is derived by a safe application of the unfolding rule
(see Definition \ref{safe_unfold}).
Let us consider a clause $D_i$ in $P_{k+1}$
which has been derived by unfolding a clause $C$
in $P_k$ of the form: $H\leftarrow G_1, A, G_2$
\wrt the atom $A$. Then there exists a
clause $C_i$ in $P_k$ such that
(i) $A$ is unifiable with ${\it hd}(C_i)$
via the mgu $\vartheta_i$, and
(ii) clause $D_i$ in $P_{k+1}$ of the form
$(H\leftarrow G_1, {\it bd}(C_i), G_2)\vartheta_i$.

Let us now show that $D_i$ is safe.
We take a variable $X$ occurring in a disequation
$t_1\!\neq\!t_2$ in the body of $D_i$, and
we prove that  $X$ is either an input variable
of ${\it hd}(D_i)$ or a local variable of
$t_1\!\neq\!t_2$ in $D_i$.
We have that $t_1\!\neq\!t_2$ is of the form
$(u_1\!\neq\!u_2)\vartheta_i$,
where $u_1\!\neq\!u_2$ is a disequation occurring in
$G_1, {\it bd}(C_i), G_2$.
We consider two cases:

\noindent
{\em Case} A: $u_1\!\neq\!u_2$ occurs in $G_1$ or $G_2$.
Since $t_1\!\neq\!t_2$ is of the form
$(u_1\!\neq\!u_2)\vartheta_i$, there exists a variable
$Y\in {\it vars}(u_1\!\neq\!u_2)$ such that $X \in {\it vars}(Y\vartheta)$.
By the inductive hypothesis, $C$ is safe and thus,
$Y$ is either an input variable of ${\it hd}(C)$ or a
local variable of $u_1\!\neq\!u_2$ in $C$.
We have that: (i) if $Y$ is an input variable of ${\it hd}(C)$ then $X$ is
an input variable of ${\it hd}(D_i)$, and
(ii) if  $Y$ is a local variable of $u_1\!\neq\!u_2$ in $C$
then $X=Y=Y\vartheta_i$ and $X$ is a local variable of
$t_1\!\neq\!t_2$ in $D_i$.

\noindent
{\em Case} B: $u_1\!\neq\!u_2$ occurs in ${\it bd}(C_i)$. From the
definition of safe unfolding we have that $X$ is either: (B.1) an input
variable of $H \vartheta_i$ or (B.2) a local variable of
$u_1\!\neq\!u_2$ in $C_i$.
In case (B.1) $X$ is an input variable
of ${\it hd}(D_i)$, which is equal to $H \vartheta_i$.
In case (B.2) $X$ does not occur in $\vartheta_i$
and, since ${\it vars}(C) \cap {\it vars}(C_i) =\emptyset$,
$X$ is a local
variable of $(u_1\!\neq\!u_2)\vartheta_i$, which is equal to
$t_1\!\neq\!t_2$, in $D_i$.

\item
Case 4: $P_{k+1}$ is derived by applying the folding rule.
Let us consider a clause $P_{k+1}$ of the form:

\smallskip
\hspace*{5mm}
$C$. $H\leftarrow G_1, {\it newp}(X_1,\dots,X_h)\vartheta, G_2$

\smallskip

which has been derived by folding the following clauses in $P_k$:

\medskip

\hspace*{5mm}
$\left\{
\begin{array}{ll}
C_1.\ H  \leftarrow \ G_1,  (A_1,K_1)\vartheta, G_2\\
 \cdots & \\
C_m.\ H  \leftarrow \ G_1,  (A_m,K_m)\vartheta, G_2 &
\end{array}
\right.$

\medskip

\noindent
using the following definition clauses in $\mathit{Defs_k}$:

\medskip
\hspace*{5mm}
$\left\{
\begin{array}{ll}
D_1.\ {\it newp}(X_1,\dots,X_h)  \leftarrow A_1,K_1\\
  \cdots & \\
D_m.\ {\it newp}(X_1,\dots,X_h)  \leftarrow A_m,K_m &
\end{array}
\right.$

\medskip
Now we take a variable $X$ occurring in a disequation
$t_1\!\neq\!t_2$ in the body of $C$, and
we prove that  $X$ is either an input variable
of $H$ or a local variable of
$t_1\!\neq\!t_2$ in $C$.

The disequation $t_1\!\neq\!t_2$ occurs in $G_1$ or $G_2$
and, by the hypothesis that $P_k$ is safe,
either $X$ is an input variable
of $H$ or, for $i = 1, \ldots, m$, $X$ is a local variable of
$t_1\!\neq\!t_2$ in $C_i$.
If for $i = 1, \ldots, m$,
$X$ is a local variable of $t_1\!\neq\!t_2$ in $C_i$,
then $X$ is a local variable of $t_1\!\neq\!t_2$ in $C$,
because by the definition of the folding rule (see Rule \ref{fold_r})
$X$ does not occur in ${\it newp}(X_1,\dots,X_h)\vartheta$.

\item
Case 5: $P_{k+1}$ is derived by applying the subsumption rule.
$P_{k+1}$ is safe
because $P_{k+1} \subseteq P_{k}$.

\item
Case 6: $P_{k+1}$ is derived by a safe application
of the head generalization
rule (see Definition \ref{safe_head_gen}).
Let {\it GenC\/} be a clause in $P_{k+1}$ of the form:

\smallskip
\hspace*{5mm}
$H \leftarrow Y\!=\!t, {\it Body}$
\smallskip

derived from a clause $C$ in $P_k$ of the form:

\hspace*{5mm}
$H\{Y/t\} \leftarrow {\it Body}$

where $\{Y/t\}$ is a
substitution such that $Y$ occurs in $H$ and
$Y$ does not occur in $C$.

Let us now prove that {\it GenC\/} is safe.
Let  $X$ be a variable occurring in a disequation
$t_1\!\neq\!t_2$ in {\it Body}.
By inductive hypothesis $C$ is safe and thus,
$X$ is either an input variable
of $H\{Y/t\}$ or a local variable of
$t_1\!\neq\!t_2$ in $C$.
If $X$ is an input variable of $H\{Y/t\}$,
then it is also an input variable of $H$,
because from the definition of safe head generalization
it follows that $H$ and $H\{Y/t\}$
have the same input variables.
If $X$ is a local variable of
$t_1\!\neq\!t_2$ in $C$,
then $X$ is a local variable of
$t_1\!\neq\!t_2$ in ${\it GenC\/}$,
because $X$ does not occur in
$Y\!=\!t$.

\item
Case 7: $P_{k+1}$ is derived by a safe application of the case split rule
(see Definition \ref{safe_cs})
to a clause $C$ in $P_k$. Let us consider the following two
clauses in $P_{k+1}$:

\smallskip

\hspace*{5mm}
$C_1$. $(H  \leftarrow $ ${\it Body}) \{X/t\}$

\hspace*{5mm}
$C_2$. $H  \leftarrow $ $X\! \neq\! t, {\it Body}$.

\smallskip
\noindent
derived by safe case split from $C$.
Let us now show that $C_1$ and $C_2$ are safe.
Let us consider clause $C_1$ and
let $Y$ be a variable occurring in a
disequation $t_1\!\neq\!t_2$ in ${\it Body}\{X/t\}$.
$t_1\!\neq\!t_2$ is of the form $(u_1\!\neq\!u_2)\{X/t\}$
where $u_1\!\neq\!u_2$ occurs in {\it Body}.
We consider two cases.

{\it Case} A: $Y\in {\it vars}(t)$.
By the definition of safe case split, either $Y$ is an input variable
of $H$ or $Y$ does not occur in $C$.
If $Y$ is an input variable of $H$,
then $Y$ is an input variable of $H\{X/t\}$, and
if $Y$ does not occur in $C$,
then $Y$ is a local variable of
$(u_1\!\neq\!u_2)\{X/t\}$ in $C_1$.

{\it Case} B: $Y\not\in {\it vars}(t)$.
We have that $Y$ occurs in $u_1\!\neq\!u_2$, and thus,
from the inductive hypothesis that $C$ is safe,
it follows that $Y$ is either an input variable of
$H$ or a local variable of $u_1\!\neq\!u_2$ in $C$.
If $Y$ is an input variable of
$H$, then $Y$ is either an input variable of
$H\{X/t\}$,
and if $Y$ a local variable of $u_1\!\neq\!u_2$ in $C$,
then it is a local variable of
$(u_1\!\neq\!u_2)\{X/t\}$ in $C_1$.

Thus, $C_1$ is a safe clause.

Let us now consider clause $C_2$ and
let $Y$ be a variable occurring in a
disequation $t_1\!\neq\!t_2$ in $X\! \neq\! t, {\it Body}$.
If $t_1\!\neq\!t_2$ occurs in ${\it Body}$ then
from the inductive hypothesis that $C$ is safe,
it follows that $Y$ is either an input variable of
$H$ or a local variable of $t_1\!\neq\!t_2$ in $C_2$.
If $t_1\!\neq\!t_2$ is $X\! \neq\! t$, then
by the definition of safe case split (i) $X$ is an input variable
of $H$, and (ii) for every variable $Y\in {\it vars}(t)$,
either (ii.1) $Y$ is an input variable of $H$ or
(ii.2) $Y$ does not occur in $H, {\it Body}$, and thus, $Y$ is
a local variable of $X\! \neq\! t$ in $C_2$.

Thus, $C_2$ is a safe clause.

\item
Case 8: $P_{k+1}$ is derived by applying the equation
elimination rule to a clause $C_1$ in $P_k$ of the form:
$H\leftarrow G_1,\ t_1 \!=\!t_2,\  G_2$. We consider two cases:

{\it Case} A: $t_1$ and $t_2$
are unifiable via the most general unifier $\vartheta$.
We derive the clause: $C_2.\ (H\leftarrow G_1, G_2)\vartheta$.
We can show that clause $C_2$
is safe similarly to Case 3 (A).

{\it Case} B: $t_1$ and $t_2$ are not unifiable.
In this case $P_{k+1}$ is safe because
$P_{k+1}$ is  $P_k - \{C_1\}$ and, by inductive hypothesis
all clauses in $P_k$ are safe.

\item
Case 9: $P_{k+1}$ is derived by applying the disequation
replacement rule to clause $C$ in $P_k$.
Let us consider the cases 9.1--9.5 of Rule \ref{diseq_repl_r}.
Cases 9.1 and 9.3--9.5 are straightforward, because
they consist in the deletion of a disequation in ${\it bd}(C)$
or in the deletion of clause $C$. Thus, in these cases
the safety of program $P_{k+1}$ derives directly from the
safety of $P_{k}$.

Let us now consider case 9.2. Suppose that clause
$C$ is of the form:
$H\leftarrow  G_1,\ f(t_1,\ldots,t_m)\!\neq\!f(u_1,\ldots,u_m), \ G_2$,
and it is replaced by the following $m\ (\geq0)$ clauses:

\hspace*{5mm}
$C_1$. $H\leftarrow G_1, t_1\!\neq\! u_1, G_2$

\hspace*{5mm}
$\ldots$

\hspace*{5mm}
$C_m$. $H\leftarrow G_1, t_m\!\neq\! u_m, G_2$

We now prove that, for $j=0,\ldots, m$, $C_j$ is safe.
Indeed, for $j=0\ldots m$, if we consider a variable $X$
occurring in $t_j\!\neq\! u_j$ then, by the inductive hypothesis,
either (i) $X$ is an input variable of $H$ or
(ii) $X$ is a local variable
of $f(t_1,\ldots,t_m)\!\neq\!f(u_1,\ldots,u_m)$ in $C$, and thus,
$X$ is a local variable of $t_j\!\neq\! u_j$ in $C_j$.

In the case where $X$ occurs in a disequation in
$G_1$ or $G_2$, it follows directly from the inductive hypothesis
that $X$ is either an input variable of $H$ or
a local variable of that disequation in $C_j$.

Thus, $C_j$ is safe.
\end{description}
\vspace*{-8mm}
\end{proof}

\subsection*{A2. Preservation of Modes}

Here we show that, if the program $P_0\cup {\it Defs_n}$
satisfies a mode $M$ and we apply our transformation rules according to the
restrictions indicated in Theorem~\ref{corr_th_op}, then the derived
program $P_n$ satisfies $M$.

In this section and in the rest of the paper,
we will use the following notation and terminology.
Let us consider two non-basic atoms $A_1$ and $A_2$
of the form $p(t_1,\ldots,t_m)$ and $p(u_1,\ldots,u_m)$,
respectively. By $A_1\!=\!A_2$ we denote the conjunction of
equations: $t_1\!=\!u_1,\ldots,t_m\!=\!u_m$.
By ${\it mgu}(A_1,A_2)$ we denote a relevant mgu of
two unifiable non-basic atoms $A_1$ and $A_2$.
Similarly, by ${\it mgu}(t_1,t_2)$ we denote a relevant mgu of
two unifiable terms $t_1$ and $t_2$.
The {\it length} of the derivation
$G_0 \longmapsto_{P}G_1\longmapsto_{P}\ldots\longmapsto_{P} G_n$ is $n$.
Given a program $P$ and a mode $M$ for $P$, we say that a derivation
$G_0\longmapsto_P G_1 \longmapsto_P \dots \longmapsto_P G_n$ is {\em consistent with}
$M$ iff for $i=0,\ldots, n-1$, if the leftmost atom of $G_i$ is a non-basic
atom $A$ then $A$ satisfies $M$.

The following properties of the operational semantics
can be proved by induction on the length of the derivations.

\begin{lemma} \label{answer_subs}
{\rm
Let $P$ be a program and $G_1$ a goal.
If $G_1$ succeeds in $P$ with answer substitution $\vartheta$, then  for
all goals $G_2$,
\ $(G_1,G_2) \ \longmapsto^*_P \ G_2\vartheta.$

}%end rm
\end{lemma}

\begin{lemma} \label{eq_rearr1}
{\rm
Let $P$ be a safe program \wrt mode $M$,
let {\it Eqs} be a conjunction of equations, and let $G_1$ be a goal
without occurrences of disequations.
{F}or all goals $G_2$, if there exists
a goal $(A^\prime,G^\prime)$ such that $A^\prime$ is a non-basic atom which
does not satisfy $M$ and

\smallskip

$({\it Eqs},\, G_1,\, G_2)\ \longmapsto^*_P\ (A^\prime,G^\prime)$

\smallskip

\noindent
then there exists a goal $(A^{\prime\prime},G^{\prime\prime})$
such that $A^{\prime\prime}$ is a non-basic atom which
does not satisfy $M$ and

\smallskip

$(G_1,\,  {\it Eqs},\, G_2) \ \longmapsto^*_P \
(A^{\prime\prime},G^{\prime\prime}).$
}%end rm
\end{lemma}

\begin{lemma}\label{mode_preserv_lemma}
{\rm
Let $P_0,\dots,P_n$  be a transformation sequence constructed
by using the transformation rules
\ref{def_intro_r}--\ref{diseq_repl_r}.
Let $M$ be a mode for $P_0\cup \mathit{Defs_n}$ such that:
(i) $P_0,\cup \mathit{Defs_n}$ is safe \wrt $M$,
(ii) $P_0\cup \mathit{Defs_n}$ satisfies $M$,  and
(iii) the applications of the unfolding, folding,
head generalization, and case split rules
during the construction of $P_0,\dots,P_n$ are safe \wrt $M$.
Then, for $k=0,\ldots,n$,
 for all goals $G$, if all derivations from $G$ using $P_0\cup \mathit{Defs_n}$
 are consistent with $M$, then all derivations from $G$ using $P_k$  are
 consistent with $M$.
} %end rm
\end{lemma}

\begin{proof}
By Proposition \ref{safety_preserv} we have that,
for $k=0,\ldots,n$, the program
$P_k$ is safe \wrt $M$.

\noindent
The proof proceeds by induction on $k$.

\noindent
The base case $(k=0)$ follows from the fact that all derivations from $G$
using $P_0$ are also derivations using $P_0\cup \mathit{Defs_n}$.

\noindent
In order to prove the step case, we prove the following
counterpositive statement:

\smallskip

\noindent
for all goals $(A_0,G_0)$, if there exists a
goal $(A_s,G_s)$ such that
$(A_0,G_0) \longmapsto^*_{P_{k+1}}(A_s,G_s)$ and
$(A_s,G_s)$ does not satisfy $M$,
then  there exists a
goal $(A_t,G_t)$ such that
$(A_0,G_0) \longmapsto^*_{P_k}(A_t,G_t)$
and $A_t$ does not satisfy $M$.

\smallskip

\noindent
We proceed by induction on the length $s$ of
derivation of $(A_s,G_s)$ from $(A_0,G_0)$ using $P_{k+1}$.
As an inductive
hypothesis we assume that, for all $r<s$ and for all goals $\hat{G}$, if
there exists a derivation
$\hat{G} \longmapsto_{P_{k+1}}\ldots\longmapsto_{P_{k+1}}(A_r,G_r)$ of length $r$,
such that $A_r$ does not satisfy $M$, then there exists
$(A^{\prime},G^{\prime})$ such that
$\hat{G}\longmapsto^*_{P_k}(A^{\prime},G^{\prime})$ and $A^{\prime}$ does not satisfy
$M$.

\noindent
Let us consider the derivation $(A_0,G_0)
\longmapsto_{P_{k+1}}\ldots\longmapsto_{P_{k+1}} (A_s,G_s)$ of length $s$, such
that $A_s$ does not satisfy $M$.

\noindent
If $s\!=\!0$ then $G$ is $(A_s,G_s)$ and
$(A_0,G_0) \longmapsto^*_{P_k}(A_s,G_s)$ where $A_s$ does not satisfy $M$.

\noindent
If $s>0$ then we may assume $A_0\!\neq\!{\it true}$, and
we have the following cases.

\smallskip

\noindent
{\it Case} 1: $A_0$ is the equation $t_1\!=\!t_2$. Thus,
by Point (1) of the operational semantics of Section \ref{op_sem},
the derivation from $(A_0,G_0)$ to $(A_s,G_s)$ using $P_{k+1}$
is of the form:

\smallskip

\noindent
$(A_0,G_0) \longmapsto_{P_{k+1}} G_0\,{\it mgu}(t_1,t_2) \longmapsto_{P_{k+1}}
\ldots\longmapsto_{P_{k+1}} (A_s,G_s)$

\smallskip

\noindent
By the inductive hypothesis there
exists $(A^{\prime},G^{\prime})$ such that $G_0\,{\it mgu}(t_1,t_2)
\longmapsto^*_{P_k}(A^\prime,G^\prime)$ and $A^\prime$ does not satisfy $M$.
Thus, $(A_0,G_0) \longmapsto^*_{P_k}(A^\prime,G^\prime)$.

\smallskip

\noindent
{\it Case} 2: $A_0$ is the disequation $t_1\!\neq\!t_2$.
The proof proceeds as in Case 1, by using Point (2) of the operational
semantics and the inductive hypothesis.

\smallskip

\noindent
{\it Case} 3: $A_0$ is a non-basic atom
which satisfies $M$. (The case where $A_0$
does not satisfy $M$ is subsumed by the case $s\!=\!0$.)
By Point (3) of the operational semantics, the derivation
from $(A_0,G_0)$ to $(A_s,G_s)$ using $P_{k+1}$ is of the form:

\smallskip

\noindent
$(A_0,G_0) \longmapsto_{P_{k+1}}
(bd(E),G_0){\it mgu}(A_0,hd(E)) \longmapsto_{P_{k+1}}
\ldots \longmapsto_{P_{k+1}} (A_s,G_s)$

\smallskip

\noindent
where $E$ is a renamed apart clause in $P_{k+1}$.

\noindent
If $E\in P_k$ then
$(A_0,G_0) \longmapsto_{P_{k}}
(bd(E),G_0){\it mgu}(A_0,hd(E))$ and the thesis
follows directly from the inductive hypothesis.

\noindent
Otherwise, if $E\in (P_{k+1} - P_k)$, we prove that:

\smallskip

there exists a goal $(A_t,G_t)$ such that
$(A_0,G_0) \longmapsto^*_{P_k}(A_t,G_t)$
and $A_t$ does not satisfy $M$ \hfill $(\dagger)$\hspace*{1cm}

\smallskip
\noindent
by considering the following cases,
corresponding to the rule which is applied to derive $E$.

\begin{description}
\item
{\it Case} 3.1: $E$ is derived by applying the definition
introduction rule. Thus, $E \in \mathit{Defs_n}$
and $(\dagger)$ follows from the
inductive hypothesis and the hypothesis that
$P_0\cup \mathit{Defs_n}$ satisfies $M$.

\item
{\it Case} 3.2: $E$ is derived by unfolding a clause $C$ in $P_k$
of the form $H \leftarrow D, G_1, A, G_2$, where {\it D}
is a conjunction of disequations, \wrt the
non-basic atom $A$.
By Proposition \ref{rearr_prop} we may assume that no
disequation occurs in $G_1, A, G_2$.
Let $C_1, \ldots, C_m$, with $m\geq 0$, be the clauses of
$P_k$ such that, for all
$i \in \{1,\dots, m\}$ $A$ is unifiable with the head of
$C_i$ via  the mgu\ $\vartheta_i$.

Thus, $E$ is of the form $(H \leftarrow D, G_1, bd(C_i),
G_2)\vartheta_i$, for some $i \in \{1,\dots, m\}$, and the
derivation from $(A_0,G_0)$ to $(A_s,G_s)$ using $P_{k+1}$
is of the form:

\smallskip

\noindent
\hspace*{5mm}
$(A_0,G_0) \longmapsto_{P_{k+1}}
((D, G_1, bd(C_i), G_2)\vartheta_i, G_0)\eta_i
\longmapsto_{P_{k+1}}\ldots\longmapsto_{P_{k+1}} (A_s,G_s)$

\smallskip

\noindent
where $\eta_i$ is an mgu of $A_0$ and $H\vartheta_i$.
By the inductive hypothesis there exists $(A^{\prime},G^{\prime})$
such that $A^\prime$ does not satisfy $M$ and:

\smallskip

\noindent
\hspace*{5mm}
$((D, G_1, bd(C_i), G_2)\vartheta_i, G_0)\eta_i
\longmapsto^*_{P_k}(A^\prime,G^\prime)$

Since $\vartheta_i$ is ${\it mgu}(A,hd(C_i))$, $\vartheta_i$ is relevant, and
${\it vars}(G_0) \cap {\it vars}((A,hd(C_i))) = \emptyset$,
we have that:

\smallskip

\noindent
\hspace*{5mm}
$(D, G_1, bd(C_i), G_2, G_0)\vartheta_i\eta_i
\longmapsto^*_{P_k}(A^\prime,G^\prime)$

\smallskip

\noindent
and thus, by the definition of the
operational semantics (Point 1), we have that:

\smallskip

\noindent
\hspace*{5mm}
$(A\!=\!hd(C_i), A_0\!=\!H, D, G_1, bd(C_i),
G_2, G_0) \ \longmapsto^*_{P_k}\ (A^\prime,G^\prime)$

\smallskip

\noindent
Then, by properties of mgu's, we have that:

\smallskip

\noindent
\hspace*{5mm}
$(A_0\!=\!H, A\!=\!hd(C_i), D, G_1, bd(C_i),
G_2, G_0) \ \longmapsto^*_{P_k}\ (A^\prime,G^\prime)$

Since $A_0$ satisfies $M$, $C$ is safe, and $C_i$ is
renamed apart, we have that
${\it vars}(D\, {\it mgu}(A_0,H)) \cap {\it
vars}(A,hd(C_i))=\emptyset$. Thus,
$(D\ {\it mgu}(A_0,H)\ {\it mgu}(A \ {\it
mgu}(A_0,H),hd(C_i))) = (D\ {\it mgu}(A_0,H))$ and
we have that:

\smallskip

\noindent
\hspace*{5mm}
$(A_0\!=\!H,  D, A\!=\!hd(C_i), G_1, bd(C_i), G_2, G_0)
\ \longmapsto^*_{P_k}\ (A^\prime,G^\prime)$

Now, by Lemma \ref{eq_rearr1}, there exists a goal
$(A^{\prime\prime},G^{\prime\prime})$ such that:

\smallskip

\noindent
\hspace*{5mm}
$(A_0\!=\!H,  D, G_1,  A\!=\!hd(C_i), bd(C_i), G_2, G_0)
\ \longmapsto^*_{P_k}\ (A^{\prime\prime},G^{\prime\prime})$

where  $A^{\prime\prime}$ is a non-basic atom which
does not satisfy $M$.
There are two cases:

\smallskip

\noindent
{\it Case} A. $(A_0\!=\!H,  D, G_1)\ \longmapsto^*_{P_k}\
(A^{\prime\prime},G^{\prime\prime\prime})$ for some goal
$G^{\prime\prime\prime}$. In this case, by using clause $C\in P_k$,
we have that:

\smallskip

\noindent
\hspace*{5mm}
$(A_0,G_0)\ \longmapsto_{P_k}\ (D, G_1, A, G_2, G_0)\,\mgu(A_0,H)
\ \longmapsto^*_{P_k} (A^{\prime\prime},G^{\prime\prime\prime\prime})$

\noindent
for some goal $G^{\prime\prime\prime\prime}$.

{\it Case} B. There is no $(A^{\prime\prime\prime},G^{\prime\prime\prime})$
such that
$(A_0\!=\!H,  D,
G_1)\,\longmapsto^*_{P_k}\,(A^{\prime\prime\prime},G^{\prime\prime\prime})$
and $A^{\prime\prime\prime}$ does not satisfy $M$.
In this case
$(A_0\!=\!H,  D, G_1, A\!=\!hd(C_i))$
succeeds in $P_k$.
It follows that, for some
substitution $\vartheta$,

\smallskip

\noindent
\hspace*{5mm}
$(A_0\!=\!H,  D, G_1, A\!=\!hd(C_i), bd(C_i), G_2, G_0)$

\hspace*{2.5cm}$\longmapsto^*_{P_k} (A\!=\!hd(C_i), bd(C_i), G_2,
              G_0)\vartheta$
\hfill(by Lemma \ref{answer_subs})

\hspace*{2.5cm}$\longmapsto_{P_k}(bd(C_i), G_2, G_0)\vartheta\
              {\it mgu}(A\vartheta,hd(C_i))$

\hfill(because mgu's are relevant and $C_i$ is renamed apart)

\hspace*{2.5cm}$\longmapsto^*_{P_k}
(A^{\prime\prime},G^{\prime\prime\prime\prime})$

\noindent
for some goal $G^{\prime\prime\prime\prime}$.
Thus,

\hspace*{5mm}
$(A_0\!=\!H,  D, G_1, A, G_2, G_0)$

\hspace*{2.5cm}$\longmapsto^*_{P_k} (A, G_2, G_0)\vartheta$

\hspace*{2.5cm}$\longmapsto_{P_k}(bd(C_i), G_2, G_0)\vartheta\
              {\it mgu}(A\vartheta,hd(C_i))$

\hspace*{2.5cm}$\longmapsto^*_{P_k}
(A^{\prime\prime},G^{\prime\prime\prime\prime})$

and therefore, by using clause $C \in P_k$,

\hspace*{5mm}
$(A_0,G_0)\ \longmapsto^*_{P_k}\
(A^{\prime\prime},G^{\prime\prime\prime\prime})$

where $A^{\prime\prime}$
is a non-basic atom which does not satisfy $M$.
Thus, $(\dagger)$ holds.

\item
{\it Case} 3.3: $E$ is derived by a safe application of the folding
rule (see Definition \ref{safe_fold}). In particular, suppose that from the
following clauses in $P_k$:

\medskip

\hspace*{5mm}
$\left\{
\begin{array}{ll}
C_1.\ H  \leftarrow \ G_1,  (A_1,K_1)\vartheta, G_2\\
 \cdots & \\
C_m.\ H  \leftarrow \ G_1,  (A_m,K_m)\vartheta, G_2 &
\end{array}
\right.$

\medskip

\noindent
and the following definition clauses in $\mathit{Defs_k}$:

\medskip
\hspace*{5mm}
$\left\{
\begin{array}{ll}
D_1.\ {\it newp}(X_1,\dots,X_h)  \leftarrow A_1,K_1\\
  \cdots & \\
D_m.\ {\it newp}(X_1,\dots,X_h)  \leftarrow A_m,K_m &
\end{array}
\right.$

\medskip
\noindent we have derived the clause $E$ of the form:

\smallskip
\hspace*{5mm}
$E.\ H\leftarrow G_1, {\it newp}(X_1,\dots,X_h)\vartheta, G_2$

\smallskip
\noindent where Property $\Sigma$ of Definition \ref{safe_fold}
holds, that is, each input variable of
${\it newp}(X_1,\dots,X_h)\vartheta$, is also an input variable of at least
one of the non-basic atoms occurring in
$(H,G_1,A_1\vartheta,\dots,A_m\vartheta)$.

Thus, the derivation from $(A_0,G_0)$ to $(A_s,G_s)$ using $P_{k+1}$
is of the form:

\hspace*{5mm}
$(A_0,G_0)\longmapsto_{P_{k+1}}
(G_1, {\it newp}(X_1,\dots,X_h)\vartheta, G_2, G_0){\it mgu}(A_0,H)
\longmapsto^*_{P_{k+1}} (A_s,G_s)$

By the inductive hypothesis, there exists a
goal $(A^\prime,G^\prime)$ such that $A^\prime$ does not satisfy $M$
and the following holds:

\hspace*{5mm}
$(G_1, {\it newp}(X_1,\dots,X_h)\vartheta, G_2, G_0){\it mgu}(A_0,H)
\longmapsto^*_{P_{k}} (A^\prime,G^\prime)$

There are two cases:

{\it Case} A:
$G_1{\it mgu}(A_0,H)
\longmapsto^*_{P_{k}} (A^\prime,G^{\prime\prime})$
for some goal $G^{\prime\prime}$.
In this case we have that, for some $i\in \{1,\ldots,m\}$,
and for some goal $G^{\prime\prime\prime}$,

\hspace*{5mm}
\makebox[1.6cm][l]{$(A_0,G_0)$}$\longmapsto_{P_{k}}
(G_1, (A_i,K_i)\vartheta, G_2, G_0){\it mgu}(A_0,H)$\hfill (by
using clause $C_i$ in $P_k$)

\hspace*{5mm}
\hspace*{1.6cm}$\longmapsto^*_{P_{k}} (A^\prime,G^{\prime\prime\prime})$

Thus, $(\dagger)$ holds.

{\it Case} B:
There is no $(A^{\prime\prime},G^{\prime\prime})$
such that
$G_1{\it mgu}(A_0,H)
\longmapsto^*_{P_{k}} (A^{\prime\prime},G^{\prime\prime})$ and
$A^{\prime\prime}$ does not satisfy $M$. In this case
$G_1{\it mgu}(A_0,H)$
succeeds in $P_k$, and thus, for some
substitution $\alpha$,

\hspace*{5mm}
$(A_0,G_0)\longmapsto^*_{P_{k}}
({\it newp}(X_1,\dots,X_h)\vartheta, G_2, G_0)\alpha \longmapsto^*_{P_{k}}
(A^{\prime},G^{\prime})$

By Property $\Sigma$, we have that
${\it newp}(X_1,\dots,X_h)\vartheta\alpha$ satisfies $M$.

It can be shown
the following fact. Let us consider the set of all definition
clauses with head predicate {\it newp} in ${\it Defs}_k$,
for any $k\in\{0,\ldots,n\}$:

\smallskip

\hspace*{5mm}
$\left\{
\begin{array}{ll}
\mathit{newp}(X_1,\dots,X_h)  \leftarrow \mathit{Body}_1\\
  \hspace{1cm} \cdots & \\
{\it newp}(X_1,\dots,X_h)  \leftarrow \mathit{Body_m}&
\end{array}
\right.$

\smallskip

\sloppy
If for a substitution $\beta$ and a goal
$G$, the atom ${\it newp}(X_1,\dots,X_h)\beta$ satisfies $M$ and
${\it newp}(X_1,\dots,X_h)\beta, G\longmapsto^*_{P_k}
(A^{\prime},G^{\prime})$, where
$A^{\prime}$ is a non-basic atom which does not satisfy $M$,
then for some $i\in \{1,\ldots,m\}$ we have that
there exists a goal $(A_t,G_t)$ such that
${Body}_i\beta, G\longmapsto^*_{P_k} (A_t,G_t)$,
where $A_t$
is a non-basic atom which does not satisfy $M$.

\fussy

By using this fact, we have that, for some $i\in \{1,\ldots,m\}$,

\hspace*{5mm}
$(A_0,G_0)\longmapsto^*_{P_{k}}
((A_i,K_i)\vartheta, G_2, G_0)\alpha \longmapsto^*_{P_{k+1}}
(A_t,G_t)$

where $A_t$
is a non-basic atom which does not satisfy $M$ and thus, $(\dagger)$ holds.

\item
{\it Case} 3.4: $E$ is derived by applying the head generalization
rule. In this case $(\dagger)$ follows from
the inductive hypothesis  and from the definition of
the operational semantics (Point 1).

\item
{\it Case} 3.5: $E$ is derived by safe case
split (see Definition \ref{safe_cs}) from a clause $C$ in $P_k$.
By Proposition \ref{rearr_prop},
we may assume that $C$ is
of the form: $H \leftarrow D, B$, where $D$ is a conjunction of
disequations and in $B$ there are no occurrences of disequations.
Thus, $E$ is of one of the following two forms:

\smallskip
\hspace*{5mm}
$C_1$. $(H  \leftarrow D, B) \{X/t\}$

\hspace*{5mm}
$C_2$. $H  \leftarrow X\deq t, D, B$

\smallskip
\noindent
where $X$ is an input variable of $H$, $X$ does not occur in $t$,
and for all variables $Y\in {\it vars}(t)$,
{\em either} $Y$ is an input variable of $H$ {\em or}
$Y$ does not occur in $C$.

{\it Case} A: $E$ is $C_1$.
Thus, the derivation from $(A_0,G_0)$ to $(A_s,G_s)$ using $P_{k+1}$
takes the form:

\hspace*{5mm}
$(A_0,G_0) \, \longmapsto_{P_{k+1}} \,
((D, B)\{X/t\}, G_0)\,{\it mgu}(A_0,H\{X/t\}) \,
\longmapsto^*_{P_{k+1}} \,
(A_s,G_s)$

By the inductive hypothesis, there exists a
goal $(A^\prime,G^\prime)$ such that $A^\prime$ does not satisfy $M$
and the following holds:

\hspace*{5mm}
$((D, B)\{X/t\}, G_0)\,{\it mgu}(A_0,H\{X/t\}) \,
\longmapsto^*_{P_{k}} \,
(A^\prime,G^\prime)$

By properties of mgu's and Point (1) of the operational semantics,
we have that:

\hspace*{5mm}
$A_0\eq H,\, X\eq t,\, D,\, B,\, G_0\, \longmapsto^*_{P_{k}}\,
(A^\prime,G^\prime)$

By the conditions for safe case split, we have
that:

\hspace*{5mm}
$\vars((X\eq t)\, \mgu(A_0,H))\cap
\vars((D,\, B,\, G_0)\, \mgu(A_0,H))=\emptyset$

and therefore:

\hspace*{5mm}
$A_0\eq H,\, D,\, B,\, G_0\, \longmapsto^*_{P_{k}}\,
(A^\prime,G^\prime)$

Thus, by using clause $C\in P_k$,

\hspace*{5mm}
$(A_0,G_0) \,
\longmapsto_{P_{k}} \,
(D,\, B,\, G_0)\mgu(A_0,H)\,
\longmapsto^*_{P_{k}}\,
(A^{\prime},G^{\prime})$

and $(\dagger)$ holds.

{\it Case} B: $E$ is $C_2$.
Thus, the derivation from $(A_0,G_0)$ to $(A_s,G_s)$ using $P_{k+1}$
takes the form:

\hspace*{5mm}
$(A_0,G_0) \, \longmapsto_{P_{k+1}} \,
(X\deq t, D, B, G_0)\,{\it mgu}(A_0,H) \,
\longmapsto^*_{P_{k+1}} \,
(A_s,G_s)$

By the inductive hypothesis, there exists a goal
$(A^\prime,G^\prime)$ such that $A^\prime$ does not satisfy
$M$ and:

\hspace*{5mm}
$(X\deq t, D, B, G_0)\,{\it mgu}(A_0,H) \,
\longmapsto^*_{P_{k}} \,
(A^\prime,G^\prime)$

Since the
answer substitution for any successful disequation is
the identity substitution, we have that:

\hspace*{5mm}
$(D, B, G_0)\,{\it mgu}(A_0,H) \,
\longmapsto^*_{P_{k}} \,
(A^\prime,G^\prime)$

Thus, by using clause $C\in P_k$, we have that

\hspace*{5mm}
$(A_0,G_0) \,
\longmapsto^*_{P_{k}} \,
(A^\prime,G^\prime)$

and $(\dagger)$ holds.

\item
{\it Case} 3.6: $E$ is derived by applying the equation elimination rule.
In this case $(\dagger)$ is a consequence of the inductive hypothesis,
Point (1) of the operational semantics, the safety of $P_k$, and Lemma
\ref{eq_rearr1}.

\item
{\it Case} 3.7: $E$ is derived by applying the disequation replacement
rule.
In this case $(\dagger)$ is a consequence of the inductive hypothesis,
Point (2) of the operational semantics, and the properties of unification.
\end{description}
\vspace*{-8mm}
\end{proof}

{F}rom Lemma \ref{mode_preserv_lemma} and Definition \ref{mode_sat}
we have the following proposition.

\begin{proposition}[Preservation of Modes]\label{mode_preserv}
{\rm
Let $P_0,\dots,P_n$  be a transformation sequence constructed
by using the transformation rules
\ref{def_intro_r}--\ref{diseq_repl_r}.
Let $M$ be a mode for $P_0\cup \mathit{Defs_n}$ such that:
(i) $P_0\cup \mathit{Defs_n}$ is safe \wrt $M$,
(ii) $P_0\cup \mathit{Defs_n}$ satisfies $M$,  and
(iii) the applications of the unfolding, folding,
head generalization, and case split rules
during the construction of $P_0,\dots,P_n$ are safe \wrt $M$.
Then, for $k=0,\ldots,n$, the program
$P_k$ satisfies $M$.
} %end rm
\end{proposition}

\subsection*{A3. Partial Correctness}

{F}or proving the partial correctness of the transformation rules \wrt
the operational semantics (that is, Proposition \ref{part_corr}),  we will
use the following two lemmata.

\begin{lemma} \label{eq_rearr2}
{\rm
Let $P$ be a safe program \wrt mode $M$,
let {\it Eqs} be a conjunction of equations, and let $G_1$ be a goal
without occurrences of disequations.
{F}or all goals $G_2$, if

\smallskip

$({\it Eqs},\, G_1,\, G_2)\, \longmapsto^*_P\, G_2\vartheta$

\smallskip

\noindent
then either

$(G_1,\, {\it Eqs},\, G_2)\, \longmapsto^*_P \, G_2\vartheta$

\noindent
or there exists a goal $(A^{\prime},G^{\prime})$
such that $A^{\prime}$ is a non-basic atom which
does not satisfy $M$ and

\smallskip

$G_1\, \longmapsto^*_P \, (A^{\prime},G^{\prime}).$
}%end rm
\end{lemma}

\begin{lemma} \label{part_corr_lemma}
{\rm
Let $P_0,\dots,P_n$  be a transformation sequence constructed
by using the transformation rules
\ref{def_intro_r}--\ref{diseq_repl_r}.
Let $M$ be a mode for $P_0\cup \mathit{Defs_n}$ such that:
(i) $P_0\cup \mathit{Defs_n}$ is safe \wrt $M$,
(ii) $P_0\cup \mathit{Defs_n}$ satisfies $M$, and
(iii) the applications of the unfolding,
folding, head generalization, and case split rules during
the construction of $P_0,\dots,P_n$ are all safe \wrt $M$.

\noindent
Then, for $k=0,\ldots ,n-1$, for each goal $G$,
if there exists a derivation
$G\longmapsto _{P_{k+1}}\ldots \longmapsto _{P_{k+1}}\mathit {true}$
which is consistent with $M$,
then $G\longmapsto^{*}_{P_{k}\cup\mathit {Defs}_{n}}\mathit{true} $,
that is, $G$ succeeds in $ P_{k}\cup \mathit{Defs}_{n}$.
} %end rm
\end{lemma}

\begin{proof}
By hypotheses (i--iii), and Propositions \ref{safety_preserv} and
\ref{mode_preserv}, for $k=0,\ldots,n$, program  $P_k$ is safe and
satisfies $M$.
Let $G$ be a goal of the form $ (A_{0},G_{0})$, such that
there exists a derivation

\smallskip
$\delta:\ \ (A_{0},G_{0})\longmapsto _{P_{k+1}}\ldots \longmapsto
_{P_{k+1}}\mathit {true} $ %\hfill\hspace*{1cm}
\smallskip

\noindent
which is consistent with $M$. We will prove
that:

\smallskip
$(A_0,G_0)\longmapsto^*_{P_k\cup\mathit{Defs_n}}\mathit{true}$
\smallskip

\noindent
The proof proceeds by induction on the length $s$ of the
derivation $\delta$.

\noindent \emph{Base Case}. For $s=0$, the goal $ (A_{0},G_{0}) $
is \emph{true} and the thesis follows from the fact that \emph{true}
succeeds in all programs.

\smallskip
\noindent \emph{Step Case}. Let us now assume the following

\smallskip
\noindent \emph{Inductive Hypothesis}: for all $ r<s $ and for
all goals $ G $, if there exists a derivation
$ G\longmapsto _{P_{k+1}}\ldots \longmapsto _{P_{k+1}}\mathit {true} $
of length $r$ which is consistent with $ M $, then
$ G\longmapsto ^{*}_{P_{k}\cup \mathit {Defs}_{n}}\mathit {true} $.
\smallskip

\noindent
There are the following three cases.

\smallskip

\noindent
{\it Case} 1: $A_0$ is the equation $t_1\!=\!t_2$. By Point (1) of
the operational semantics of Section \ref{op_sem}, the
derivation $\delta$ is of the form:

\smallskip

$(t_1\!=\!t_2,G_0) \longmapsto_{P_{k+1}} G_0\,{\it mgu}(t_1,t_2) \longmapsto_{P_{k+1}}
\ldots\longmapsto_{P_{k+1}} \true$

\smallskip

\noindent
Thus, the derivation $G_0\,{\it mgu}(t_1,t_2) \longmapsto_{P_{k+1}}
\ldots\longmapsto_{P_{k+1}} \true$ has length $s-1$ and it is consistent with
$M$. By the inductive hypothesis there
exists a derivation $G_0\,{\it mgu}(t_1,t_2)
\longmapsto^*_{P_k}\true$.
Thus, $(A_0,G_0) \longmapsto^*_{P_k}\true$ and $(A_0,G_0)$ succeeds in $P_k\cup
\mathit{Defs_n}$.

\smallskip

\noindent
{\it Case} 2: $A_0$ is the disequation $t_1\!\neq\!t_2$.
The proof proceeds as in Case 1, by using Point (2) of the operational
semantics and the inductive hypothesis.

\smallskip

\noindent
{\it Case} 3: $A_0$ is a non-basic atom which satisfies $M$
(otherwise there is no derivation starting from $(A_0,G_0)$ which is
consistent with $M$). By Point (3) of the operational semantics, the
derivation $\delta$ is of the
form:

\smallskip

$(A_0,G_0) \longmapsto_{P_{k+1}}
(bd(E),G_0){\it mgu}(A_0,hd(E)) \longmapsto_{P_{k+1}}
\ldots \longmapsto_{P_{k+1}} \true$

\smallskip

\noindent
where $E$ is a renamed apart clause in $P_{k+1}$.

\noindent
If $E\in P_k$ then
$(A_0,G_0) \longmapsto_{P_{k}}
(bd(E),G_0){\it mgu}(A_0,hd(E))$ and the thesis
follows directly from the inductive hypothesis.

\noindent
Otherwise, if $E\in (P_{k+1} - P_k)$, we prove that
$(A_0,G_0)$ succeeds in $P_k\cup\mathit{Defs_n}$
by considering the following cases,
which correspond to the rules applied for deriving $E$.

\begin{description}
\item
{\it Case} 3.1: $E$ is derived by applying the definition
introduction rule. Thus, $E$ is a clause in
$\mathit{Defs_n}$ of the form: ${\it newp}(X_1,\ldots,X_h)
\If B$ and the derivation $\delta$
is of the form:

\hspace*{5mm}
$({\it newp}(t_1,\ldots,t_h),G_0) \longmapsto_{\mathit{Defs_n}}
(B\{X_1/t_1,\ldots,X_h/t_h\},G_0) \longmapsto_{P_{k+1}}
\ldots \longmapsto_{P_{k+1}} \true$

By the inductive hypothesis, we have that:

\hspace*{5mm}
$(B\{X_1/t_1,\ldots,X_h/t_h\},G_0) \longmapsto^*_{P_k} \true$

and thus,

\hspace*{5mm}
$({\it newp}(t_1,\ldots,t_h),G_0)
\longmapsto^*_{P_k\cup\mathit{Defs_n}}\true$

\item
{\it Case} 3.2: $E$ is derived by unfolding a clause $C$ in $P_k$
of the form $H \leftarrow D, G_1, A, G_2$, where {\it D}
is a conjunction of disequations, \wrt the
non-basic atom $A$.
By Proposition \ref{rearr_prop} we may assume that no
disequation occurs in $G_1, A, G_2$.
Let $C_1, \ldots, C_m$, with $m\geq 0$, be the clauses of
$P_k$ such that, for all
$i \in \{1,\dots, m\}$ $A$ is unifiable with the head of
$C_i$ via  the mgu\ $\vartheta_i$.

Thus, $E$ is of the form $(H \leftarrow D, G_1, bd(C_i),
G_2)\vartheta_i$, for some $i \in \{1,\dots, m\}$, and the
derivation $\delta$
is of the form:

\smallskip

\hspace*{5mm}
$(A_0,G_0) \longmapsto_{P_{k+1}}
((D, G_1, bd(C_i), G_2)\vartheta_i, G_0)\eta_i
\longmapsto_{P_{k+1}}\ldots\longmapsto_{P_{k+1}} \true$

\smallskip

\noindent
where $\eta_i$ is an mgu of $A_0$ and $H\vartheta_i$.
By the inductive hypothesis we have that:

\smallskip

\hspace*{5mm}
$((D, G_1, bd(C_i), G_2)\vartheta_i, G_0)\eta_i
\longmapsto^*_{P_k}\true$

Since $\vartheta_i$ is ${\it mgu}(A,hd(C_i))$, $\vartheta_i$ is relevant, and
${\it vars}(G_0) \cap {\it vars}((A,hd(C_i))) = \emptyset$,
we have that:

\smallskip

\hspace*{5mm}
$(D, G_1, bd(C_i), G_2, G_0)\vartheta_i\eta_i
\longmapsto^*_{P_k}\true$

\smallskip

\noindent
and thus, by the definition of the
operational semantics (Point 1), we have that:

\smallskip

\hspace*{5mm}
$(A\!=\!hd(C_i), A_0\!=\!H, D, G_1, bd(C_i),
G_2, G_0) \, \longmapsto^*_{P_k} \true$

\smallskip

\noindent
Then, by properties of mgu's, we have that:

\smallskip

\hspace*{5mm}
$(A_0\!=\!H, A\!=\!hd(C_i), D, G_1, bd(C_i),
G_2, G_0) \, \longmapsto^*_{P_k} \true$

Since $A_0$ satisfies $M$, $C$ is safe, and $C_i$ is
renamed apart, we have that
${\it vars}(D\, {\it mgu}(A_0,H)) \cap {\it
vars}(A,hd(C_i))=\emptyset$. Thus,
$(D\ {\it mgu}(A_0,H)\ {\it mgu}(A \ {\it
mgu}(A_0,H),hd(C_i))) = (D\ {\it mgu}(A_0,H))$ and
we have that:

\smallskip

\hspace*{5mm}
$(A_0\!=\!H,  D, A\!=\!hd(C_i), G_1, bd(C_i), G_2, G_0)\,
\longmapsto^*_{P_k}\true$

Now, by Lemma \ref{eq_rearr2}, there are the following two cases.

{\it Case} A.
$(A_0\!=\!H,  D, G_1, A\!=\!hd(C_i), bd(C_i), G_2, G_0)\,
\longmapsto^*_{P_k}\true$

In this case, by Points (1) and (3) of the operational semantics we have
that:

\hspace*{5mm}
$(A_0\!=\!H,  D, G_1, A, G_2, G_0)\,
\longmapsto^*_{P_k}\true$

and thus, by using clause $C$ in $P_k$,

\hspace*{5mm}
$(A_0,G_0) \longmapsto^*_{P_k}\true$

{\it Case} B.
There exists a
goal $(A^{\prime},G^{\prime})$ such that:

\smallskip

\noindent
\hspace*{5mm}
$(A_0\!=\!H,  D, G_1)
\ \longmapsto^*_{P_k}(A^{\prime},G^{\prime})$

where $A^{\prime}$ is a non-basic atom which does not satisfy
the mode $M$.
In this case we have that, for some goal $G^{\prime\prime}$,

\hspace*{5mm}
$A_0\longmapsto^*_{P_k}(A^{\prime},G^{\prime\prime})$

which is impossible because $A_0$ and $P_k$ satisfy $M$.

\item
{\it Case} 3.3: $E$ is derived by a safe application of the folding
rule (see Definition \ref{safe_fold}). In particular, suppose that from the
following clauses in $P_k$:

\medskip

\hspace*{5mm}
$\left\{
\begin{array}{ll}
C_1.\ H  \leftarrow \ G_1,  (A_1,K_1)\vartheta, G_2\\
 \cdots & \\
C_m.\ H  \leftarrow \ G_1,  (A_m,K_m)\vartheta, G_2 &
\end{array}
\right.$

\medskip

\noindent
and the following definition clauses in $\mathit{Defs_k}$:

\medskip
\hspace*{5mm}
$\left\{
\begin{array}{ll}
D_1.\ {\it newp}(X_1,\dots,X_h)  \leftarrow A_1,K_1\\
  \cdots & \\
D_m.\ {\it newp}(X_1,\dots,X_h)  \leftarrow A_m,K_m &
\end{array}
\right.$

\medskip
\noindent we have derived the clause $E$ of the form:

\smallskip
\hspace*{5mm}
$E.\ H\leftarrow G_1, {\it newp}(X_1,\dots,X_h)\vartheta, G_2$

\smallskip
\noindent where Property $\Sigma$ of Definition \ref{safe_fold}
holds, that is, each input variable of
${\it newp}(X_1,\dots,X_h)\vartheta$, is also an input variable of at least
one of the non-basic atoms occurring in
$(H,G_1,A_1\vartheta,\dots,A_m\vartheta)$.

Thus, the derivation $\delta$ is of the form:

\hspace*{5mm}
$(A_0,G_0)\longmapsto_{P_{k+1}}
(G_1, {\it newp}(X_1,\dots,X_h)\vartheta, G_2, G_0){\it mgu}(A_0,H)
\longmapsto^*_{P_{k+1}} \true$

By the inductive hypothesis,  the following holds:

\hspace*{5mm}
$(G_1, {\it newp}(X_1,\dots,X_h)\vartheta, G_2, G_0){\it mgu}(A_0,H)
\longmapsto^*_{P_{k}} \true$

and therefore, for some substitution $\alpha$,

\hspace*{5mm}
$(A_0,G_0)\longmapsto^*_{P_{k}}
({\it newp}(X_1,\dots,X_h)\vartheta, G_2, G_0)\alpha \longmapsto^*_{P_{k}}
\true$

By Property $\Sigma$, we have that
${\it newp}(X_1,\dots,X_h)\vartheta\alpha$ satisfies $M$.

It can be shown
the following fact. Let us consider the set of all definition
clauses with head predicate {\it newp} in ${\it Defs}_k$,
for any $k\in\{0,\ldots,n\}$:

\smallskip

\hspace*{5mm}
$\left\{
\begin{array}{ll}
\mathit{newp}(X_1,\dots,X_h)  \leftarrow \mathit{Body}_1\\
  \hspace{1cm} \cdots & \\
{\it newp}(X_1,\dots,X_h)  \leftarrow \mathit{Body_m}&
\end{array}
\right.$

\smallskip

If for a substitution $\beta$ for a goal $G$, the atom
${\it newp}(X_1,\dots,X_h)\beta, G$ satisfies $M$ and
we have that ${\it newp}(X_1,\dots,X_h)\beta, G \longmapsto^*_{P_k}\true$,
then for some $i\in \{1,\ldots,m\}$ we have that
${Body}_i\beta\longmapsto^*_{P_k}\true$.

By using this fact, we have that, for some $i\in \{1,\ldots,m\}$,

\hspace*{5mm}
$(A_0,G_0)\longmapsto^*_{P_{k}}
((A_i,K_i)\vartheta, G_2, G_0)\alpha \longmapsto^*_{P_{k}}
\true$

\item
{\it Case} 3.4: $E$ is derived by applying the head generalization
rule. In this case $(A_0,G_0)\longmapsto^*_{P_{k}}\true$
follows from the inductive hypothesis and from the definition of
the operational semantics (Point 1).

\item
{\it Case} 3.5: $E$ is derived by safe case
split (see Definition \ref{safe_cs}) from a clause $C$ in $P_k$.
By Proposition \ref{rearr_prop},
we may assume that $C$ is
of the form: $H \leftarrow D, B$, where $D$ is a conjunction of
disequations and in $B$ there are no occurrences of disequations.
Thus, $E$ is of one of the following two forms:

\smallskip
\hspace*{5mm}
$C_1$. $(H  \leftarrow D, B) \{X/t\}$

\hspace*{5mm}
$C_2$. $H  \leftarrow X\deq t, D, B$

\smallskip
\noindent
where $X$ is an input variable of $H$, $X$ does not occur in $t$,
and for all variables $Y\in {\it vars}(t)$,
{\em either} $Y$ is an input variable of $H$ {\em or}
$Y$ does not occur in $C$.

{\it Case} A: $E$ is $C_1$.
Thus, the derivation $\delta$ takes the form:

\hspace*{5mm}
$(A_0,G_0) \, \longmapsto_{P_{k+1}} \,
((D, B)\{X/t\}, G_0)\,{\it mgu}(A_0,H\{X/t\}) \,
\longmapsto^*_{P_{k+1}} \,
\true$

By the inductive hypothesis, we have that:

\hspace*{5mm}
$((D, B)\{X/t\}, G_0)\,{\it mgu}(A_0,H\{X/t\}) \,
\longmapsto^*_{P_{k}} \,
\true$

By properties of mgu's and Point (1) of the operational semantics,
we have that:

\hspace*{5mm}
$(A_0\eq H,\, X\eq t,\, D,\, B,\, G_0)\, \longmapsto^*_{P_{k}}\, \true$

By the conditions for safe case split, we have
that:

\hspace*{5mm}
$\vars((X\eq t)\, \mgu(A_0,H))\cap
\vars((D,\, B,\, G_0)\, \mgu(A_0,H))=\emptyset$

and therefore:

\hspace*{5mm}
$(A_0\eq H,\, D,\, B,\, G_0)\, \longmapsto^*_{P_{k}}\,
\true$

Thus, by using clause $C\in P_k$,

\hspace*{5mm}
$(A_0,G_0) \,
\longmapsto_{P_{k}} \,
(D,\, B,\, G_0)\mgu(A_0,H)\,
\longmapsto^*_{P_{k}}\,
\true$

{\it Case} B: $E$ is $C_2$.
Thus, the derivation $\delta$
takes the form:

\hspace*{5mm}
$(A_0,G_0) \, \longmapsto_{P_{k+1}} \,
(X\deq t, D, B, G_0)\,{\it mgu}(A_0,H) \,
\longmapsto^*_{P_{k+1}} \,
\true$

By the inductive hypothesis, we have that:

\hspace*{5mm}
$(X\deq t, D, B, G_0)\,{\it mgu}(A_0,H) \,
\longmapsto^*_{P_{k}} \,\true$

Since the
answer substitution for any successful disequation is
the identity substitution, we have that:

\hspace*{5mm}
$(D, B, G_0)\,{\it mgu}(A_0,H) \,
\longmapsto^*_{P_{k}} \,\true$

Thus, by using clause $C\in P_k$,

\hspace*{5mm}
$(A_0,G_0) \longmapsto^*_{P_{k}} \true$

\item
{\it Case} 3.6: $E$ is derived by applying the equation elimination rule.
In this case $(A_0,G_0) \longmapsto^*_{P_{k}} \true$
is a consequence of the inductive hypothesis,
Point (1) of the operational semantics, the fact that $P_k$ is safe and
satisfies $M$, and Lemma \ref{eq_rearr2}.

\item
{\it Case} 3.7: $E$ is derived by applying the disequation replacement
rule.
In this case $(A_0,G_0) \longmapsto^*_{P_{k}} \true$
is a consequence of the inductive hypothesis,
Point (2) of the operational semantics, and the properties of unification.
\end{description}
\vspace*{-8mm}
\end{proof}

\begin{proposition}
[Partial Correctness]
\label{part_corr}
{\rm
Let $P_0,\dots,P_n$  be a transformation sequence constructed
by using the transformation rules
\ref{def_intro_r}--\ref{diseq_repl_r}.
Let $M$ be a mode for $P_0\cup \mathit{Defs_n}$ such that:
(i) $P_0\cup \mathit{Defs_n}$ is safe \wrt $M$,
(ii) $P_0\cup \mathit{Defs_n}$ satisfies $M$, and
(iii) the applications of the unfolding,
folding, head generalization, and case split rules during
the construction of $P_0,\dots,P_n$ are all safe \wrt $M$.

\noindent
Then, for $k=0,\ldots,n$, for each non-basic atom $A$ which satisfies mode
$M$, if $A$ succeeds in $P_k$ then $A$ succeeds in $P_0\cup
\mathit{Defs_k}$.
} %end rm
\end{proposition}

\begin{proof}
Suppose that a non-basic atom $A$ which satisfies $M$ has a successful
derivation using $P_k$. By Proposition \ref{mode_preserv},
$P_{k}$ satisfies $M$ and, therefore, $A$ has a successful derivation
using $P_k$ which is consistent with $M$. Thus, the
thesis follows from Lemma \ref{part_corr_lemma}.
\end{proof}

\subsection*{A4. Completeness}
{F}or the proofs of Propositions \ref{safety_preserv} (Preservation of
Safety), \ref{mode_preserv} (Preservation of Modes), and \ref{part_corr}
(Partial Correctness), we have proceeded by induction on the length of
the derivations and by cases on the rule used to derive program $P_{k+1}$
from program $P_k$. For the proof of Proposition \ref{completeness} below
(Completeness), we will proceed by induction \wrt more sophisticated
well-founded orderings. This proof technique is a suitable modification of
the one based on {\em weight consistent} proof trees \cite{GeK94,TaS84}.

The following definition introduces some well-founded orders and other
notions which are needed for the proofs presented in this section.

\begin{definition}
{\rm
(i) Given a derivation $\delta$
of the form $G_0\longmapsto_P G_1 \longmapsto_P \dots \longmapsto_P G_z$,
we denote by $\lambda(\delta)$ the number of goals $G_i$ in $\delta$
such that $G_i$ is of the form $(A,K)$ where $A$ is a non-basic atom.

\noindent
(ii) We define the following functions $\mu$ and $\nu$ which given a
program and a goal return either a non-negative integer or $\infty$
(we assume that, for all non-negative integers $n$, $\infty > n$):

\smallskip

$\mu(P,G)=\left\{
\begin{array}{ll}
\min\{\lambda(\delta)\mid \delta  \mbox{ is a successful
derivation of } G \mbox{ in } P\} & \mbox{if } G \mbox{ succeeds in }
P\\ \infty & \mbox{otherwise}
\end{array}
\right.$

\smallskip

$\nu(P,G)=\left\{
\begin{array}{ll}
\min\{n \mid  n \mbox{ is the length of a successful
derivation of }  G \mbox{ in } P\} & \mbox{if } G
\mbox{ succeeds in } P\\ \infty &
\mbox{otherwise} \end{array}
\right.$

\smallskip

\noindent
(iii) Given a program $P$ and two goals $G_1$ and $G_2$,
we write $G_1\succ_P G_2$ iff $\mu(P,G_1)>\mu(P,G_2)$.
Similarly, we write $G_1\succeq_P G_2$ iff $\mu(P,G_1)\geq\mu(P,G_2)$.

\smallskip

\noindent
(iv) Given two programs $P$ and $Q$, we say that a derivation $G_0\longmapsto_P
G_1 \longmapsto_P \dots \longmapsto_P G_z$ is {\em quasi-decreasing}
w.r.t.~$\succ_Q$ iff for $i=0,\ldots, z-1$, either (1) $G_i \succ_Q
G_{i+1}$ or (2) the leftmost atom of $G_i$ is a basic atom and
$G_i\succeq_Q G_{i+1}$.

\smallskip

\noindent
(v) Let $P$ a program and $G_1,G_2$ be goals.
If there exists a derivation
$\delta$ from $G_1$ to $G_2$ such that $\lambda(\delta)=s$, then we write
$G_1\longmapsto^s_P G_2$.
} %\end rm
\end{definition}

{F}or any program $P$ the relation $\succ_P$ is a well-founded order and, for
all goals $G_1, G_2,$ and $G_3$, we have that $G_1\succ_P G_2$ and $G_2
\succeq_P G_3$ implies $G_1\succ_P G_3$.

\begin{lemma}\label{quasi_decr}
{\rm
Let $P$ be a program and $G$ be a goal. If $G$ succeeds in
$P$ then $G$ has a derivation which is quasi-decreasing \wrt $\succ_P$.
}%end rm
\end{lemma}

\begin{proof}
The derivation $\delta$ from $G$ using $P$ such that
$\lambda(\delta)\leq \lambda(\delta^\prime)$ for all successful
derivations $\delta^\prime$ from $G$, is quasi-decreasing \wrt $\succ_P$.
\end{proof}

\begin{lemma}\label{eq_rearr3}
{\rm
Let $M$ be a mode for program $P$, such that $P$ is safe \wrt $M$ and
$P$ satisfies $M$. Let
{\it Eqs} be a conjunction of equations, and $G_0,G_1,G_2$ be goals.
Suppose also that no disequation occurs in $G_1$ and all derivations from
the goal $(G_0,G_1)$ are consistent with $M$.
Then:

\begin{itemize}
\item[(i)]
$(G_0,\, G_1,\, {\it Eqs},\, G_2)\, \longmapsto^*_P \, \true$
\ \ iff
\ \
$(G_0,\, {\it Eqs},\, G_1,\, G_2)\, \longmapsto^*_P\, \true$

\item[(ii)] $\mu(P,\,(G_0,\, G_1,\, {\it Eqs},\, G_2))=\mu(P,\,(G_0,\, {\it
Eqs},\, G_1,\, G_2))$

\item[(iii)]$\nu(P,\,(G_0,\, G_1,\, {\it Eqs},\, G_2))=\nu(P,\,(G_0,\, {\it
Eqs},\, G_1,\, G_2))$

\end{itemize}
}%end rm
\end{lemma}

\begin{proof}
By induction on the length of the derivations.
\end{proof}

\begin{lemma}\label{subsumption_lemma}
{\rm
Let $M$ be a mode for program $P$, such that $P$ is safe \wrt $M$ and
$P$ satisfies $M$. Let $\vartheta$ be a substitution and $G_0,G_1,G_2$ be goals.
Suppose also that no disequation occurs in $G_2$ and all derivations from
the goal $(G_0,G_2)$ are consistent with $M$.
Then:

\begin{itemize}
\item[(i)]
if \ \ $(G_0,\, G_1,\, G_2)\vartheta\, \longmapsto^*_P \, \true$
\ \ then
\ \
$(G_0,\, G_2)\, \longmapsto^*_P\, \true$

\item[(ii)]
$\mu(P,\,(G_0, G_1, G_2)\vartheta)\geq\mu(P,\,(G_0, G_2))$

\item[(iii)]
$\nu(P,\,(G_0,G_1,G_2)\vartheta)\geq\nu(P,\,(G_0,G_2))$

\end{itemize}

} %end rm
\end{lemma}

\begin{proof}
By induction on the length of the derivations.
\end{proof}

\begin{lemma}\label{diseq_prom_lemma}
{\rm
Let $M$ be a mode for program $P$, such that $P$ is safe \wrt $M$ and
$P$ satisfies $M$. Let {\it Diseqs} be a conjunction of disequations and
$G$ be a goal. Suppose also that ${\it vars}({\it Diseqs})\cap
{\it vars}(G) = \emptyset$. Then:

\begin{itemize}
\item[(i)]
$(G,\, {\it Diseqs})\, \longmapsto^*_P \, \true$
\ \ iff
\ \
$({\it Diseqs},\, G)\, \longmapsto^*_P\, \true$

\item[(ii)]
$\mu(P,\,(G,\, {\it Diseqs}))=\mu(P,\,({\it Diseqs},\, G))$

\item[(iii)]
$\nu(P,\,(G,\, {\it Diseqs}))=\nu(P,\,({\it Diseqs},\, G))$

\end{itemize}

} %end rm
\end{lemma}

\begin{proof}
The proof proceeds by induction on the length of the derivations.
\end{proof}

Let us consider a transformation sequence $P_0,\dots,P_n$
constructed by using the transformation rules
\ref{def_intro_r}--\ref{diseq_repl_r} according to
the hypothesis of Theorem~\ref{corr_th_op}.
{F}or reasons of simplicity we assume that each
definition clause is used for folding, and thus,
by Condition~\ref{fold_cond} of Theorem~\ref{corr_th_op},
it is unfolded during the construction of $P_0,\dots,P_n$.
We can rearrange the sequence $P_0,\dots,P_n$ into a
new sequence $P_0,\dots,P_0\cup {\it Defs_n}, \ldots, P_j,
\ldots,P_l,\ldots,P_n$
such that: (1)  $P_0,\dots,P_0\cup {\it Defs_n}$  is constructed
by applications of the definition introduction rule,
(2) $P_0\cup {\it Defs_n}, \ldots, P_j$ is constructed by
unfolding every clause in ${\it Defs_n}$,
(3) $P_j,\ldots,P_l$ is constructed by applications
of rules~\ref{unf_r}--\ref{diseq_repl_r}, and
(4) either $l=n$ or $l=n-1$ and $P_n$ is derived from
$P_{n-1}$ by an application of the definition elimination rule
\wrt predicate $p$.

Throughout the rest of this section we will refer to the transformation
sequence $P_0,\dots,P_0\cup {\it Defs_n}, \ldots, P_j,
\ldots,P_n$ constructed as indicated above.
We also assume that $M$ is a mode for $P_0\cup \mathit{Defs_n}$ such that:
(i) $P_0\cup \mathit{Defs_n}$ is safe \wrt $M$,
(ii) $P_0\cup \mathit{Defs_n}$ satisfies $M$, and
(iii) the applications of the unfolding,
folding, head generalization, and case split rules during
the construction of $P_0,\dots,P_n$ are all safe \wrt $M$.

Thus, by Propositions \ref{safety_preserv} and \ref{mode_preserv},
for $k=0,\ldots,n$, program $P_k$ is safe and satisfies $M$.

\begin{lemma}\label{tc_def_unf}
{\rm
Let us consider the transformation sequence
$P_0,\dots,P_0\cup {\it Defs_n}, \ldots, P_j$ constructed as indicated
above. Then the following properties hold.

\noindent
(i) For all clauses
${\it newp}(X_1,\ldots,X_h)\If {\it Body}$ in $\mathit{Defs_n}$, for all
substitutions $\vartheta$, and for all goals $G_1,G_2$,
such that
all derivations from $(G_1,\, {\it Body}\vartheta,\, G_2)$ using
$P_j$ are consistent with $M$, we have that:

\noindent
(i.1) $(G_1,\, {\it Body}\,\vartheta,\, G_2)
\succeq_{P_j}(G_1,\, {\it newp}(X_1,\ldots,X_h)\vartheta,\, G_2)$;

\noindent
(i.2) all derivations starting from
$(G_1, {\it newp}(X_1,\ldots,X_h)\vartheta,\, G_2)$ using $P_j$ are
consistent with $M$;

\noindent
(ii) for all non-basic atoms $A$ satisfying $M$, if $A$
succeeds in $P_0\cup \mathit{Defs_n}$ then $A$ succeeds in $P_j$.
} %end rm
\end{lemma}

\noindent
Notice that, by Point (i.1),
if $(G_1,\,{\it Body}\,\vartheta,\, G_2)$ succeeds in $P_j$
then $(G_1,\,{\it newp}(X_1,\ldots,X_h)\vartheta,\, G_2)$ succeeds in
$P_j$.

\begin{proof} By induction on the length of the derivations.
\end{proof}

{F}or the proof of the following Lemma \ref{tc_lemma} we will use the
following property.

\begin{lemma}\label{mode_lemma}
{\rm
Let us consider the transformation sequence
$P_j, \ldots, P_l$ and the mode $M$ for $P_0\cup \mathit{Defs_n}$ as
indicated above.
{F}or  $k=j,\ldots, l$ and for all goals
$G_1$ and $G_2$ such that there exists a derivation
$G_1\longmapsto_{P_k}\ldots \longmapsto_{P_k}G_2$,
if all derivations from $G_1$ using $P_j$ are consistent with $M$
then all derivations from $G_2$ using $P_j$ are consistent with $M$.
} %end rm
\end{lemma}

\begin{proof} The proof proceeds by induction on $k$ and on
the length of the derivation $G_1\longmapsto_{P_k}\ldots
\longmapsto_{P_k}G_2$. We omit the details.
\end{proof}

\begin{lemma}\label{tc_lemma}
{\rm
Let us consider the transformation sequence
$P_j, \ldots, P_l$ and the mode $M$ for $P_0\cup \mathit{Defs_n}$ as
indicated above.
Let $G$ be a goal such that (i) no disequation occurs in $G$ and (ii) all
derivations from $G$ using $P_j$ are consistent with $M$. For  $k=j,\ldots,
l$, if $G$ has a successful derivation in $P_j$,
then $G$ has a successful derivation in $P_k$ which is quasi-decreasing
\wrt $\succ_{P_j}$.
} %end rm
\end{lemma}

\begin{proof}
Let us consider the following ordering on goals:

\smallskip
\noindent
$G_1\rhd G_2$ \ iff \
{\it either} $G_1\succ_{P_j} G_2$
{\it or}
$G_1\succeq_{P_j} G_2$
and $\nu(P_j,G_1)>\nu(P_j,G_2)$.

\smallskip
\noindent
$\rhd$ is a well-founded order.

\noindent
The proof proceeds by induction on $k$.

\smallskip
\noindent
{\it Base Case}. The case  $k=j$ follows from Lemma \ref{quasi_decr}.

\smallskip
\noindent
{\it Step Case}. For $k\geq j$ we assume the following:

\smallskip
\noindent
{\it Inductive Hypothesis} (I1). For each goal
$G^\prime$ such that no disequation occurs in $G^\prime$ and all
derivations from $G^\prime$ using $P_j$ are consistent with $M$, if
$G^\prime$ has a successful derivation in $P_j$,
then $G^\prime$ has a successful derivation in $P_k$ which is
quasi-decreasing \wrt $\succ_{P_j}$.

\smallskip
\noindent
Let us now consider a goal
$G$ of the form $(A_0,G_0)$ such that
no disequation occurs in $(A_0,G_0)$ and all derivations from $(A_0,G_0)$
using $P_j$ are consistent with $M$.
Let us assume that
there exists a derivation of the form:
\smallskip

$\delta:\ \ (A_0,G_0)\longmapsto_{P_k}\ldots\longmapsto_{P_k}\true$

\smallskip
\noindent
which is quasi-decreasing \wrt $\succ_{P_j}$.

\noindent
We wish to show that there exists a
derivation of the form:

\smallskip

$\delta^\prime:\ \ (A_0,G_0)\longmapsto_{P_{k+1}}\ldots\longmapsto_{P_{k+1}}\true$

\smallskip
\noindent
which is
quasi-decreasing \wrt $\succ_{P_j}$.
We prove the existence of such a derivation $\delta^\prime$ by induction on
the well-founded order $\rhd$.

\noindent
We assume the following:

\smallskip

\noindent \emph{Inductive Hypothesis} (I2).
{F}or each goal $\hat{G}$ such that no disequation occurs in $\hat{G}$ and
all derivations from $\hat{G}$ using $P_j$ are consistent with $M$ and
$(A_0,G_0)\rhd\hat{G}$, if there exists a derivation of the form:

\smallskip

$\hat{G}\longmapsto_{P_k}\ldots\longmapsto_{P_k}\true$

\smallskip

\noindent
which is quasi-decreasing \wrt $\succ_{P_j}$,
then there exists a derivation of the form:

\smallskip

$\hat{G}\longmapsto_{P_{k+1}}\ldots\longmapsto_{P_{k+1}}\true$

\smallskip

\noindent
which is quasi-decreasing \wrt $\succ_{P_j}$.

\smallskip
\noindent
Now we proceed by cases.

\smallskip
\noindent
{\it Case} 1: $A_0$ is the equation $t_1\!=\!t_2$. By Point (1) of
the operational semantics of Section \ref{op_sem}, the
derivation $\delta$ is of the form:

\smallskip

$(t_1\!=\!t_2,G_0) \longmapsto_{P_{k}} G_0\,{\it mgu}(t_1,t_2) \longmapsto_{P_{k}}
\ldots\longmapsto_{P_{k}} \true$

\smallskip

\noindent
Let us consider the derivation:

\smallskip
$G_0\,{\it mgu}(t_1,t_2)
\longmapsto_{P_{k}} \ldots\longmapsto_{P_{k}} \true$
\smallskip

\noindent
By Proposition \ref{part_corr}, we have that
both $(t_1\!=\!t_2,G_0)$ and $G_0\,{\it mgu}(t_1,t_2)$
succeed in $P_j$. Moreover, by Point (1) of
the operational semantics
$\nu(P_j,\, (t_1\!=\!t_2,G_0)) > \nu(P_j,\,G_0\,{\it mgu}(t_1,t_2))$.
Thus, $(t_1\!=\!t_2,G_0) \rhd G_0\,{\it mgu}(t_1,t_2)$ and,
by the inductive hypothesis (I2),
there exists a successful derivation of the form:

\smallskip
$G_0\,{\it mgu}(t_1,t_2)
\longmapsto_{P_{k+1}} \ldots\longmapsto_{P_{k+1}} \true$
\smallskip

\noindent
which is quasi-decreasing \wrt $\succ_{P_j}$.
Since $(t_1\!=\!t_2,G_0) \succeq_{P_j} G_0\,{\it mgu}(t_1,t_2)$, the
following derivation:

\smallskip
$(t_1\!=\!t_2,G_0) \longmapsto_{P_{k+1}} G_0\,{\it mgu}(t_1,t_2)
\longmapsto_{P_{k+1}} \ldots\longmapsto_{P_{k+1}} \true$
\smallskip

\noindent
is quasi-decreasing \wrt $\succ_{P_j}$.

\smallskip

\noindent
{\it Case} 2: $A_0$ is a non-basic atom which satisfies $M$
(otherwise there is no derivation starting from $(A_0,G_0)$ which is
consistent with $M$). By Point (3) of the operational semantics, in
$P_k$ there exists a renamed apart clause $C$,
such that the derivation $\delta$ is of the form:

\smallskip

$(A_0,G_0) \longmapsto_{P_{k}}
(bd(C), G_0){\it mgu}(A_0,hd(C)) \longmapsto_{P_{k}}
\ldots \longmapsto_{P_{k}} \true$

\smallskip

\noindent
By Proposition \ref{rearr_prop} we may assume that clause $C$ is
of the form $H \If {\it Diseqs}, B$, where {\it Diseqs} is a conjunction of
disequations and $B$ is a goal without occurrences of
disequations.
Thus, ${\it Diseqs}\, {\it mgu}(A_0,H)$ succeeds and $\delta$ is of the
form:

$(A_0,G_0) \longmapsto_{P_{k}}
({\it Diseqs}, B, G_0){\it mgu}(A_0,H) \longmapsto_{P_{k}}
\ldots \longmapsto_{P_{k}}
(B, G_0){\it mgu}(A_0,H) \longmapsto_{P_{k}}
\ldots \longmapsto_{P_{k}}
\true$

\smallskip

\noindent
If $C\in P_{k+1}$ then
$(A_0,G_0) \longmapsto_{P_{k+1}}
({\it Diseqs}, B, G_0){\it mgu}(A_0,H) \longmapsto_{P_{k+1}}\ldots
\longmapsto_{P_{k+1}} (B, G_0){\it mgu}(A_0,H)$ and the thesis
follows from the inductive hypothesis (I2), because
$(A_0,G_0) \succ_{P_{j}}
(B,G_0){\it mgu}(A_0,H)$ (recall that $\delta$ is
quasi-decreasing \wrt $\succ_{P_{j}}$).

\noindent
Otherwise, if $C\in (P_{k} - P_{k+1})$,
we construct the derivation $\delta^\prime$
by considering the following cases,
which correspond to the rules applied for deriving
$P_{k+1}$ from $P_{k}$.

\begin{description}
\item
{\it Case} 2.1: $P_{k+1}$ is derived by unfolding clause
$C$ in $P_k$ \wrt a non-basic atom, say $A$.
Thus, clause $C$ is
of the form $H \If {\it Diseqs}, G_1, A, G_2$.
Let $C_1, \ldots, C_m$, with $m\geq 0$, be
the clauses of $P_k$ such that, for $i = 1,\dots, m$, $A$ is
unifiable with the head of $C_i$. Thus,
$P_{k+1}=(P_k - \{C\}) \cup \{D_1, \ldots, D_m\}$, where for
$i=1,\ldots,m$, $D_i$ is the clause $(H \If {\it Diseqs}, G_1, bd(C_i),
G_2)\,\mgu(A,hd(C_i))$. For reasons of simplicity we assume that
for $i = 1,\dots, m$,
no disequation occurs in $bd(C_i)$. In the general case where,
for some $i \in \{1,\dots, m\}$,
$bd(C_i)$ has occurrences of disequations, the proof proceeds in a very
similar way, by using Proposition  \ref{rearr_prop}, Lemma
\ref{diseq_prom_lemma}, and the hypothesis that all applications of the
unfolding rule are safe (see Definition \ref{safe_unfold}).

The derivation $\delta$ is of the form:

\smallskip

\hspace*{5mm}
$(A_0,G_0) \longmapsto_{P_{k}}
({\it Diseqs}, G_1, A, G_2, G_0)\mgu(A_0,H)
\longmapsto_{P_{k}}\ldots\longmapsto_{P_{k}} \true$

\smallskip

{F}rom the fact that $\delta$ is quasi-decreasing \wrt $\succ_{P_j}$,
from Point (1) of the operational semantics, and from the definition of
$\succ_{P_j}$, we have that:

\hspace*{5mm}
$(A_0,G_0)\succ_{P_j}(A_0\eq H, {\it Diseqs}, G_1, A, G_2, G_0)$

and the derivation

\hspace*{5mm}
$(A_0\eq H, {\it Diseqs}, G_1, A, G_2, G_0)\,
\longmapsto_{P_{k}}\ldots\longmapsto_{P_{k}}\true$

\smallskip

is quasi-decreasing \wrt $\succ_{P_j}$.

Thus, by Points (1) and (3) of the operational semantics,
there exists a clause in
$P_k$, say $C_i$, such that the derivation
\smallskip

\hspace*{5mm}
$(A_0\eq H, {\it Diseqs}, G_1, A\eq hd(C_i), bd(C_i), G_2, G_0)\,
\longmapsto_{P_{k}}\ldots\longmapsto_{P_{k}}\true$
\smallskip

is quasi-decreasing \wrt $\succ_{P_j}$.
Moreover, we have that:

\hspace*{5mm}
$(A_0,G_0)\succ_{P_j}(A_0\eq H, {\it Diseqs}, G_1, A\eq
hd(C_i), bd(C_i), G_2, G_0)$.

Since all derivations from $(A_0,G_0)$ using $P_j$ are consistent with $M$,
we have that all derivations from $(A_0\eq H, {\it Diseqs}, G_1)$ using
$P_j$ are consistent with $M$, and therefore, by Lemma
\ref{mode_preserv_lemma}, all derivations from $(A_0\eq H, G_1)$ using
$P_k$ are consistent with $M$. Then, since no disequation occurs in $G_1$,
by Lemma \ref{eq_rearr3}, there exists a derivation

\smallskip

\hspace*{5mm}
$(A_0\eq H, {\it Diseqs}, A\eq hd(C_i), G_1, bd(C_i), G_2, G_0)\,
\longmapsto_{P_{k}}\ldots\longmapsto_{P_{k}}\true$

\smallskip

which is quasi-decreasing \wrt $\succ_{P_j}$.
Moreover, we have that:

\hspace*{5mm}
$(A_0,G_0)\succ_{P_j}(A_0\eq H, {\it Diseqs}, A\eq hd(C_i), G_1,
bd(C_i), G_2, G_0)$.

Now, since by Lemma \ref{safety_preserv} all clauses in $P_k$ are safe,
we have that:

\hspace*{5mm}
$\vars({\it Diseqs}\,\mgu(A_0,H))\cap \vars( (A\eq
hd(C_i))\mgu(A_0,H))=\emptyset$

and therefore, by using properties of
mgu's, there exists a derivation

\smallskip

\hspace*{5mm}
$(A\eq hd(C_i), A_0\eq H, {\it Diseqs}, G_1, bd(C_i), G_2, G_0)\,
\longmapsto_{P_{k}}\ldots\longmapsto_{P_{k}}\true$

\smallskip

which is quasi-decreasing \wrt $\succ_{P_j}$.
Let $\vartheta_i$ be $\mgu(A,hd(C_i))$ and $\eta_i$ be
$\mgu(A_0,H\,\vartheta_i))$. By Points (1) and (2) of the operational
semantics, we have that ${\it Diseqs} \,\vartheta_i\,\eta_i$ succeeds and
there exists a derivation of the form

\smallskip

\hspace*{5mm}
$((G_1, bd(C_i), G_2)\,\vartheta_i, G_0)\,\eta_i\,
\longmapsto_{P_{k}}\ldots\longmapsto_{P_{k}}\true$

Moreover, we have that:

\hspace*{5mm}
$(A_0,G_0)\succ_{P_j}
((G_1, bd(C_i), G_2)\,\vartheta_i, G_0)\,\eta_i$ \hfill (*)\hspace*{1cm}

and thus, by the inductive hypothesis (I2),
there exists a derivation of the form

\smallskip

\hspace*{5mm}
$((G_1, bd(C_i), G_2)\,\vartheta_i, G_0)\,\eta_i\,
\longmapsto_{P_{k+1}}\ldots\longmapsto_{P_{k+1}}\true$

\smallskip

which is quasi-decreasing \wrt
$\succ_{P_j}$.

Since ${\it Diseqs} \,\vartheta_i\,\eta_i$ succeeds, by using clause $D_i$
in $P_{k+1}$ for the first step, we can construct the following derivation:

\smallskip

\hspace*{5mm}
$(A_0,G_0) \longmapsto_{P_{k+1}}
(({\it Diseqs}, G_1, bd(C_i), G_2)\,\vartheta_i, G_0)\,\eta_i\,
\longmapsto_{P_{k+1}}\ldots\longmapsto_{P_{k+1}}\true$

\smallskip

which, by property (*), is quasi-decreasing \wrt
$\succ_{P_j}$.

\item
{\it Case} 2.2: $P_{k+1}$ is derived from $P_k$ by a safe application of
the folding rule (see Definition \ref{safe_fold}). In particular, suppose
that clause $C$ is one of the following clauses occurring in $P_k$:

\medskip

\hspace*{5mm}
$\left\{
\begin{array}{ll}
C_1.\ H  \leftarrow \ {\it Diseqs}, G_1,  (A_1,K_1)\vartheta, G_2\\
 \cdots & \\
C_m.\ H  \leftarrow \ {\it Diseqs}, G_1,  (A_m,K_m)\vartheta, G_2 &
\end{array}
\right.$

\medskip

\noindent
where {\it Diseqs} is a conjunction of disequations and no disequation
occurs in $(G_1,G_2)$. We also suppose that the following definition
clauses occur in $\mathit{Defs_k}$:

\medskip
\hspace*{5mm}
$\left\{
\begin{array}{ll}
D_1.\ {\it newp}(X_1,\dots,X_h)  \leftarrow A_1,K_1\\
  \cdots & \\
D_m.\ {\it newp}(X_1,\dots,X_h)  \leftarrow A_m,K_m &
\end{array}
\right.$

\medskip
\noindent
and we have derived a clause $E$ of the form:

\smallskip
\hspace*{5mm}
$E.\ H\leftarrow {\it Diseqs}, G_1, {\it newp}(X_1,\dots,X_h)\vartheta,
G_2$

\smallskip
\noindent where Property $\Sigma$ of Definition \ref{safe_fold}
holds, that is, each input variable of
${\it newp}(X_1,\dots,X_h)\vartheta$, is also an input variable of at least
one of the non-basic atoms occurring in
$(H,G_1,A_1\vartheta,\dots,A_m\vartheta)$.

Thus, $P_{k+1} = (P_k - \{C_1,\ldots,C_m\}) \cup \{E\}$.

We may assume, without loss of generality, that clause $C$ is $C_1$, and
the derivation $\delta$ is of the form:

\hspace*{5mm}
$(A_0,G_0)\longmapsto_{P_k}
({\it Diseqs}, G_1, (A_1,K_1)\vartheta, G_2, G_0){\it mgu}(A_0,H)
\longmapsto_{P_{k}} \ldots\longmapsto_{P_{k}} \true$

Thus, ${\it Diseqs}\, {\it mgu}(A_0,H)$ succeeds and, since $\delta$ is
consistent with $M$, by Lemma \ref{part_corr_lemma}, we have that $(G_1,
(A_1,K_1)\vartheta, G_2, G_0){\it mgu}(A_0,H)$ succeeds in $P_j$.

Moreover, by Lemma \ref{mode_lemma}, all derivations
from $(G_1, (A_1,K_1)\vartheta, G_2, G_0){\it mgu}(A_0,H)$ using $P_j$ are
consistent with $M$.

Thus, by Lemmata \ref{quasi_decr} and \ref{tc_def_unf}, all
derivations  from  $(G_1, {\it
newp}(X_1,\dots,X_h)\vartheta, G_2, G_0){\it mgu}(A_0,H)$ using $P_j$
are consistent
with $M$ and there exists a derivation of the form:

\hspace*{5mm}
$(G_1, {\it newp}(X_1,\dots,X_h)\vartheta, G_2, G_0){\it
mgu}(A_0,H)\longmapsto_{P_j}\ldots \longmapsto_{P_j}\true$

which is quasi-decreasing \wrt $\succ_{P_j}$.

No disequation occurs in
$(G_1, {\it newp}(X_1,\dots,X_h)\vartheta, G_2,
G_0){\it mgu}(A_0,H)$, and thus,
by the
inductive hypothesis (I1), there exists a derivation of the form:

\hspace*{5mm}
$(G_1, {\it newp}(X_1,\dots,X_h)\vartheta, G_2, G_0){\it
mgu}(A_0,H)\longmapsto_{P_k}\ldots \longmapsto_{P_k}\true$

which is quasi-decreasing \wrt $\succ_{P_j}$.

Since $\delta$ is quasi-decreasing \wrt $\succ_{P_j}$, by Lemma
\ref{tc_def_unf}, we also have that:

\hspace*{5mm}
$(A_0,G_0) \rhd
(G_1, {\it newp}(X_1,\dots,X_h)\vartheta, G_2, G_0){\it mgu}(A_0,H)$

Thus, by the Inductive hypothesis (I2), there exists a derivation

\hspace*{5mm}
$(G_1, {\it newp}(X_1,\dots,X_h)\vartheta, G_2, G_0){\it mgu}(A_0,H)
\longmapsto_{P_{k+1}}\ldots \longmapsto_{P_{k+1}}\true$

which is quasi decreasing \wrt $\succ_{P_j}$.

Since ${\it Diseqs}\, {\it mgu}(A_0,H)$ succeeds,
by using clause $E\in P_{k+1}$, we can construct the following
derivation

\hspace*{5mm}
$(A_0,G_0)\longmapsto_{P_{k+1}}
({\it Diseqs}, G_1, {\it newp}(X_1,\dots,X_h)\vartheta, G_2, G_0){\it
mgu}(A_0,H) \longmapsto_{P_{k+1}}\ldots \longmapsto_{P_{k+1}}\true$

which is quasi-decreasing \wrt $\succ_{P_j}$ because:

\hspace*{5mm}
\makebox[16mm][l]{$(A_0,G_0)$} $\succ_{P_j}
({\it Diseqs}, G_1, (A_1,K_1)\vartheta, G_2, G_0){\it mgu}(A_0,H)$ \hfill
(because $\delta$ is quasi-decreasing)

\hspace*{5mm}
\makebox[16mm][l]{}
$\succeq_{P_j}({\it Diseqs}, G_1, {\it newp}(X_1,\dots,X_h)\vartheta,
G_2, G_0){\it mgu}(A_0,H)$ \hfill (by Lemma \ref{tc_def_unf})

\item
{\it Case} 2.3: $P_{k+1}$ is derived by deleting clause $C$ from $P_k$ by
applying the subsumption rule. Thus, clause $C$ is of the form
$(H\If {\it Diseqs}, G_1, G_2)\vartheta$ and there exists a clause $D$ in
$P_{k}$ of the form $H\If {\it Diseqs},  G_1$. By Proposition
\ref{rearr_prop} we may assume that no disequation occurs in $G_1$.

Thus, the derivation ($\delta$) is of the form:

\hspace*{5mm}
$(A_0,G_0)\longmapsto_{P_k}
(({\it Diseqs}, G_1, G_2)\vartheta, G_0){\it mgu}(A_0,H\vartheta)
\longmapsto_{P_{k}} \ldots\longmapsto_{P_{k}} \true$

Since all derivations starting from $(A_0,G_0)$ using $P_k$ are consistent
with $M$ and, by using clause $D$,
$(A_0,G_0)\longmapsto_{P_k} ({\it Diseqs},
G_1, G_0){\it mgu}(A_0,H)$, we have that all derivations starting
from $({\it Diseqs}, G_1, G_0){\it mgu}(A_0,H)$ using $P_k$ are
consistent with $M$. Moreover, no disequation occurs in $G_0$ and
therefore, by Lemma \ref{subsumption_lemma}, there exists a derivation

\hspace*{5mm}
$(A_0,G_0)\longmapsto_{P_k}
({\it Diseqs}, G_1, G_0){\it mgu}(A_0,H)
\longmapsto_{P_{k}} \ldots\longmapsto_{P_{k}} \true$

which is quasi-decreasing \wrt $\succ_{P_j}$.
Thus, $({\it Diseqs} \, {\it mgu}(A_0,H))$ succeeds and
there exists a derivation

\hspace*{5mm}
$(G_1,  G_0){\it mgu}(A_0,H)
\longmapsto_{P_{k}} \ldots\longmapsto_{P_{k}} \true$

which is quasi-decreasing \wrt $\succ_{P_j}$. Since
$(A_0,G_0)\rhd
(G_1, G_0){\it mgu}(A_0,H)$, by the inductive
hypothesis (I2), there exists a derivation

\hspace*{5mm}
$(G_1, G_0){\it mgu}(A_0,H)
\longmapsto_{P_{k+1}} \ldots\longmapsto_{P_{k+1}} \true$

which is quasi-decreasing \wrt $\succ_{P_j}$. Since $D$ belongs to
$P_{k+1}$ and $({\it Diseqs} \, {\it mgu}(A_0,H))$ succeeds, there exists
a derivation

\hspace*{5mm}
$(A_0,G_0)\longmapsto_{P_{k+1}}
({\it Diseqs}, G_1, G_0){\it mgu}(A_0,H)
\longmapsto_{P_{k+1}} \ldots\longmapsto_{P_{k+1}} \true$

which is quasi-decreasing \wrt $\succ_{P_j}$.

\item
{\it Case} 2.4: $P_{k+1}$ is derived from $P_k$ by applying the
head generalization rule to clause $C$. Thus, $C$ is
of the form $H\{X/t\}\If {\it Body}$ and  $P_{k+1} = (P_k - \{C\}) \cup
\{{\it GenC}\}$, where clause {\it GenC} is of the form
$H \If X\eq t, {\it Body}$.

In this case we can show that we can construct the derivation
$\delta^\prime$ which is quasi-decreasing \wrt $\succ_{P_j}$, by  using
(i) Point (1) of the operational semantics,
(ii) the inductive hypothesis (I2) and
(iii) the fact that, for all goals of the form $(t_1\eq t_2,G)$,
where $t_1$ and $t_2$ are unifiable terms, and for all programs $P$,
$\mu(P,\,(t_1\eq t_2,G))\eq \mu(P,\,G\mgu(t_1,t_2))$.

\item
{\it Case} 2.5: $P_{k+1}$ is derived from $P_k$ by applying the
safe case split rule  (see Definition \ref{safe_cs}) to clause $C$.
By Proposition \ref{rearr_prop}, we may assume that $C$ is a clause of the
form $H \leftarrow {\it Diseqs}, B$, where {\it Diseqs} is a conjunction of
disequations and $B$ is a  goal  without occurrences of disequations. We
also assume that from $C$ we have derived two clauses of the form:

\smallskip
\hspace*{5mm}
$C_1$. $(H  \leftarrow {\it Diseqs}, B) \{X/t\}$

\hspace*{5mm}
$C_2$. $H  \leftarrow X\deq t, {\it Diseqs}, B$

\smallskip
\noindent
where $X$ is an input variable of $H$, $X$ does not occur in $t$,
and for all variables $Y\in {\it vars}(t)$,
{\em either} $Y$ is an input variable of $H$ {\em or}
$Y$ does not occur in $C$.

We have that $P_{k+1}= (P_k - \{C\}) \cup \{C_1,C_2\}$. The derivation
$\delta$ is of the form:

\hspace*{5mm}
$(A_0,G_0) \longmapsto_{P_{k}} ({\it Diseqs}, B, G_0)\,{\it mgu}(A_0,H)
\longmapsto_{P_{k}} \ldots \longmapsto_{P_{k}}
\true$

Thus, $({\it Diseqs}\,{\it mgu}(A_0,H))$ succeeds and, since $\delta$ is
quasi-decreasing, we have that $(A_0,G_0) \rhd (B,G_0)\,{\it mgu}(A_0,H)$.
The goal $(B,G_0)\,{\it mgu}(A_0,H)$ has no occurrences of disequations
and, by the inductive hypothesis (I2), there exists a derivation

\hspace*{5mm}
$(B, G_0)\,{\it mgu}(A_0,H)
\longmapsto_{P_{k+1}} \ldots \longmapsto_{P_{k+1}}
\true$

which is quasi-decreasing \wrt $\succ_{P_j}$.
Since $({\it Diseqs}\,{\it mgu}(A_0,H))$ succeeds,
there exists a derivation

\hspace*{5mm}
$({\it Diseqs}, B, G_0)\,{\it mgu}(A_0,H)
\longmapsto_{P_{k+1}} \ldots \longmapsto_{P_{k+1}}
\true$

which is quasi-decreasing \wrt $\succ_{P_j}$.

Since $X$ is an input variable of $H$, there exists a binding $X/u$ in
${\it mgu}(A_0,H)$ where $u$ is a ground term. We consider the following
two cases.

{\it Case} A: $t$ and $u$ are unifiable, and thus,
$u$ is an instance of $t$.
In this case $A_0$ and $H\{X/t\}$ are unifiable
and, by the hypotheses on $X/t$, we have that:

\hspace*{5mm}
$({\it Diseqs}, B, G_0)\,{\it mgu}(A_0,H) =
(({\it Diseqs}, B)\{X/t\}, G_0)\,{\it mgu}(A_0,H\{X/t\})$

Thus, we can construct a derivation of the form:

\hspace*{5mm}
$(A_0,G_0) \longmapsto_{P_{k+1}}
(({\it Diseqs}, B)\{X/t\}, G_0)\,{\it mgu}(A_0,H\{X/t\})
\longmapsto_{P_{k+1}} \ldots \longmapsto_{P_{k+1}}
\true$

which is quasi-decreasing \wrt $\succ_{P_j}$.

{\it Case} B: $t$ and $u$ are not unifiable. Thus,
$(X\deq t)\mgu(A_0,H)$ succeeds and the following
derivation is quasi-decreasing \wrt $\succ_{P_j}$.

\hspace*{5mm}
\makebox[16mm][l]{$(A_0,G_0)$}$\longmapsto_{P_{k+1}}
(X\deq t, {\it Diseqs}, B, G_0)\,{\it mgu}(A_0,H)$

\hspace*{5mm}
\makebox[16mm][l]{}$\longmapsto_{P_{k+1}} ({\it Diseqs}, B, G_0)\,{\it
mgu}(A_0,H)\longmapsto_{P_{k+1}}
\ldots \longmapsto_{P_{k+1}}
\true$

\item
{\it Case} 2.6: $P_{k+1}$ is derived from $P_k$ by applying the
equation elimination rule to clause $C$.
In this case the existence of a derivation

\hspace*{5mm}
$(A_0,G_0) \longmapsto_{P_{k+1}} \ldots\longmapsto_{P_{k+1}} \true$

which is quasi-decreasing \wrt $\succ_{P_j}$, can be proved by using
(i) the inductive hypothesis (I2), (ii) Point (1) of the operational
semantics, (iii) the fact that $P_k$ is safe and satisfies $M$, and
(iv) Lemma \ref{eq_rearr3}.

\item
{\it Case} 2.7: $P_{k+1}$ is derived from $P_k$ by applying the
disequation replacement rule to clause $C$.
In this case the existence of a derivation

\hspace*{5mm}
$(A_0,G_0) \longmapsto_{P_{k+1}} \ldots\longmapsto_{P_{k+1}} \true$

which is quasi-decreasing \wrt $\succ_{P_j}$, can be proved by using
(i) the inductive hypothesis (I2), (ii) Point (2) of the operational
semantics, and (iii) the properties of
unification.
\end{description}
\vspace*{-8mm}
\end{proof}

\begin{lemma}\label{tc_uf}
{\rm
Let us consider the transformation sequence
$P_j, \ldots, P_l$ and the mode $M$ for $P_0\cup \mathit{Defs_n}$ as
indicated above.
{F}or $k=j,\ldots, l$,
for each non-basic atom $A$ which satisfies mode $M$,
if $A$ succeeds in $P_j$ then $A$ succeeds in $P_k$.
} %end rm
\end{lemma}

\begin{proof}
It follows from Lemma \ref{tc_lemma}, because if an atom $A$
satisfies $M$ and succeeds in $P_j$, then $A$ has a successful derivation
in $P_j$ which is consistent with $M$ and quasi-decreasing \wrt
$\succ_{P_j}$. Indeed, by Proposition \ref{mode_preserv}, $P_j$ satisfies
$M$, and thus, all derivations starting from $A$ are consistent with $M$.
% Moreover, by Lemma \ref{quasi_decr}, there exists a derivation starting
% from $A$ which is quasi-decreasing \wrt $\succ_{P_j}$.
\end{proof}

\begin{lemma}\label{tc_def_elim}
{\rm
If program
$P_n$ is derived from program $P_{n-1}$
by an application of the definition elimination rule
\wrt a non-basic predicate $p$, then
for each  atom $A$ which has predicate $p$,
if $A$ succeeds in $P_0\cup \mathit{Defs_n}$ then
$A$ succeeds in $P_n$.
} %end rm
\end{lemma}

\begin{proof}
If $A$ has predicate $p$ then $p$ depends on all clauses which
are used for any derivation starting from $A$.
Thus, every derivation from $A$ using
$P_0\cup \mathit{Defs_n}$ is also a derivation using $P_n$.
\end{proof}

\begin{proposition}[Completeness]\label{completeness}
{\rm
Let $P_0,\dots,P_n$  be a transformation sequence constructed
by using the transformation rules
\ref{def_intro_r}--\ref{diseq_repl_r} and let $p$ be a
non-basic predicate in $P_n$.
Let $M$ be a mode for $P_0\cup \mathit{Defs_n}$ such that:
(i) $P_0\cup \mathit{Defs_n}$ is safe \wrt $M$,
(ii) $P_0\cup \mathit{Defs_n}$ satisfies $M$, and
(iii) the applications of the unfolding,
folding, head generalization, and case split rules during
the construction of $P_0,\dots,P_n$ are all safe \wrt $M$.
Suppose also that:
\begin{enumerate}
\item  \label{c1}{\em if} the folding rule is applied
for the derivation of
a clause $C$ in program $P_{k+1}$ from clauses $C_1,\dots, C_m$ in
program  $P_{k}$ using clauses $D_1,\dots, D_m$ in $\mathit{Defs_k}$,
with $0\!\leq\!k\!<\!n$,\\
{\em then}  for every $i \in \{1,\dots,m\}$ there exists
$j\in\{1,\dots,n\!-\!1\}$ such that $D_i$ occurs in
$P_j$ and $P_{j+1}$ is derived from $P_j$ by unfolding $D_i$.

\item \label{c2}during the transformation
sequence $P_0,\dots,P_n$ the definition elimination rule {\em
either} is never applied {\em or} it is applied \wrt predicate $p$
once only, when deriving $P_n$ from $P_{n-1}$.
\end{enumerate}
Then for each atom $A$ which has predicate $p$ and
satisfies mode $M$, if $A$ succeeds in $P_0\cup \mathit{Defs_n}$
then $A$ succeeds in $P_n$.
} %end rm
\end{proposition}

\begin{proof}
Let us consider a transformation sequence
$P_0,\dots,P_n$ constructed by using the transformation rules
\ref{def_intro_r}--\ref{diseq_repl_r} according to
conditions \ref{c1} and \ref{c2}.

As already mentioned, we can rearrange the sequence
$P_0,\dots,P_n$ into a new sequence $P_0,\dots,P_0\cup {\it Defs_n},
\ldots, P_j, \ldots,P_l,\ldots,P_n$
such that: (1)  $P_0,\dots,P_0\cup {\it Defs_n}$  is constructed
by applications of the definition introduction rule,
(2) $P_0\cup {\it Defs_n}, \ldots, P_j$ is constructed by
unfolding every clause in ${\it Defs_n}$,
(3) $P_j,\ldots,P_l$ is constructed by applications
of rules~\ref{unf_r}--\ref{diseq_repl_r}, and
(4) either (4.1) $l=n$ or (4.2) $l=n-1$ and $P_n$ is derived from
$P_{n-1}$ by an application of the definition elimination rule
\wrt predicate $p$.

Thus, Proposition \ref{completeness}
follows from Lemmata \ref{tc_def_unf},
\ref{tc_uf}, and~\ref{tc_def_elim}.
\end{proof}

\section*{Appendix B. Proof of Proposition~\ref{mutual_excl}}

{F}or the proof of Proposition~\ref{mutual_excl} we need the following two
lemmata.

\begin{lemma}
\label{diseq_success}
{\rm
Let us consider a program $P$ and a conjunction $D$ of disequations.
$D$ succeeds in $P$
{\em iff} every ground instance of $D$ holds. } %end rm
\end{lemma}
\begin{proof} Let us consider the conjunction
$(r_1\!\neq\!s_1, \ldots,$ $r_k\!\neq\!s_k)$
of disequations.
Every ground instance of $(r_1\!\neq\!s_1, \ldots,$ $r_k\!\neq\!s_k)$
holds iff
for $i=1,\ldots,k$, and for every ground substitution $\sigma$,
$r_i\sigma\!\neq\! s_i\sigma$ holds iff
for $i=1,\ldots,k$, and for every ground substitution $\sigma$,
$r_i\sigma$ is a ground term different from $s_i\sigma$ iff
for $i=1,\ldots,k$, it does not exist a ground substitution $\sigma$
such that  $r_i\sigma$ and $s_i\sigma$ are the same ground term iff
for $i=1,\ldots,k$, $r_i$ and $s_i$ are not unifiable iff
$(r_1\!\neq\!s_1, \ldots,$
$r_k\!\neq\!s_k)$ succeeds in $P$.
\end{proof}

\begin{lemma} \label{at_most_one}
{\rm
Let $P$ be a program which is safe \wrt mode $M$ and satisfies
mode $M$. Let the non-unit clauses of $P$ be pairwise mutually
exclusive \wrt mode $M$. Given any non-basic atom $A_0$ which
satisfies $M$, and any basic goal $G_0$, there exists at most one goal
$(A_1,G_1)$ such that $A_1$ is a non-basic atom and
$(A_0,G_0)\Rightarrow_P (A_1,G_1)$.
} %end rm
\end{lemma}

\begin{proof}
By the definition of the $\Rightarrow_P$ relation (see
Section~\ref{determ_sec}),
we need to prove that for any non-basic atom $A_0$ which
satisfies $M$, and any basic goal $G_0$, there exists at most one goal
$(A_1,G_1)$ where $A_1$ is a non-basic atom, such that:
(i) $(A_0,G_0)\longmapsto_P^* (A_1,G_1)$, and (ii) the relation
$\longmapsto_P^*$
is constructed by first applying exactly once Point (3) of our
operational semantics, and then applying to the resulting goal
Points (1) and (2) of our operational semantics, as many times as
required to evaluate the leftmost basic atoms, if any.

Since  the non-unit clauses of $P$ are
pairwise mutually exclusive \wrt $M$, for any given non-basic atom
$A_0$ which satisfies $M$, there exists at most one non-unit clause,
say $C$, of $P$ such that $A_0$ unifies with ${\it hd}(C)$ via an mgu,
say $\mu$, and ${\it grd}(C)\mu$ succeeds in $P$. In fact, suppose
to the contrary, that there were two such non-unit clauses, say  $C_1$
and $C_2$. Suppose that, for $j\!=\!1,2$, clause $C_j$ is renamed
apart and it is of the form:

$C_j$.\ \ $p(t_j,u_j) \leftarrow {\it grd_j},\ K_j$,

\noindent
where: (i) $t_j$ is a tuple of terms denoting the input arguments of
$p$ and (ii) the goal ${\it grd}_j$ is the guard of $C_j$, that is, a
conjunction of disequations such that the leftmost atom of the goal
$K_j$ is not a disequation.

Suppose that for $j\!=\!1,2$, ${\it
hd}(C_j)$ unifies with $A_0$ via the mgu $\vartheta_j$. Since $A_0$
satisfies $M$, for $j\!=\!1,2$, the input variables of ${\it hd}(C_j)$
are bound by $\vartheta_j$ to ground terms. Since $t_1$ and $t_2$ have
a common ground instance, namely $t_1 \vartheta_1 (=t_2 \vartheta_2)$,
they have a relevant mgu $\vartheta$ whose domain is a subset of
${\it vars}(t_1,t_2)$, and there exists a ground substitution
$\sigma$ with domain ${\it vars}(t_1,t_2)$ such that
$t_1 \vartheta_1\!=\!t_1 \vartheta\sigma (=\!t_2 \vartheta_2\!=\!t_2
\vartheta\sigma)$. Moreover, since the clauses $C_1$ and $C_2$ are
renamed apart, we have that:

(Property $\alpha$) \ for $j\!=\!1,2$, if we restrict $\vartheta\sigma$
to ${\it vars}(t_j)$ then $\vartheta_j\!=\!\vartheta\sigma$.

\noindent
By hypothesis, both ${\it grd}_1\vartheta_1$ and ${\it
grd}_2\vartheta_2$ succeed in $P$. Thus, by Lemma~\ref{diseq_success},
every ground instance of ${\it grd}_1\vartheta_1$ and ${\it
grd}_2\vartheta_2$ holds. (Recall that the goals ${\it
grd}_1\vartheta_1$ and ${\it grd}_2\vartheta_2$ are ground goals,
except for the local variables of each disequation occurring in
them.)

Since $P$ is safe \wrt $M$, for $j\!=\!1,2$, every variable occurring
in a disequation of ${\it grd}_j$ either occurs in $t_j$ or it is a
local variable of that disequation in $C_j$. Thus, by Property
($\alpha$), ${\it grd}_1\vartheta_1\!=\!{\it grd}_1\vartheta\sigma$
and ${\it grd}_2\vartheta_2\!=\!{\it grd}_2\vartheta\sigma$. Since
every ground instance of ${\it grd}_1\vartheta_1$ and ${\it
grd}_2\vartheta_2$ holds, we have that every ground instance of $({\it
grd}_1\vartheta\sigma, {\it grd}_2\vartheta\sigma)$ holds. In other
words, there exists a ground substitution $\sigma$ whose domain is
${\it vars}(t_1,t_2)$, such that every ground instance of $({\it
grd}_1, {\it grd}_2) \vartheta\sigma$ holds. By definition, this means
that $({\it grd}_1, {\it grd}_2)\vartheta$ is satisfiable \wrt ${\it
vars}(t_1,t_2)$. This contradicts the fact that the non-unit clauses
of $P$ are mutually exclusive \wrt $M$.

We conclude that for any given non-basic atom $A_0$ which
satisfies $M$, $A_0$ unifies via an mgu, say $\mu$, with the
head of at most one non-unit clause, say $C$, of $P$ such that ${\it
grd}(C)\mu$ succeeds in $P$.

Now there are two cases: (Case i) $A_0$ unifies with
the head of the clauses in $\{C, D_1,\ldots, D_n\}$, where
$n\!\geq\!0$, $C$ is a non-unit clause, and clauses $D_1,\ldots, D_n$
are all unit clauses, and (Case ii) $A_0$ unifies with
the head of the clauses in $\{D_1,\ldots, D_n\}$, where $n\!\geq\!0$
and these clauses are all unit clauses.

Let us consider Case (i). Let clause $C$ be of the
form: $H \If K$ for some non-basic goal $K$. For any
basic goal $G_0$, by applying once
Point (3) of our operational semantics, we have that:
$(A_0,G_0) \longmapsto_P
(K, G_0)\mu$. Thus, $(K, G_0)\mu$ is of the form $({\it Bs},G_2)$ where
{\it Bs} is a conjunction of basic atoms and the leftmost atom of $G_2$
is non-basic.
Since for any basic atom $B$ and goal $G_3$,
there exists at most one goal $G_4$ such that $(B,G_3) \longmapsto_P G_4$,
by using Points (1) and (2) of our operational semantics,
we have that there exists at most one goal $(A_1,G_1)$
such that $({\it Bs}, G_2) \longmapsto_P^*(A_1,G_1)$, where the
atom $A_1$ is non-basic.

Every other derivation starting from $(A_0,G_0)$ by applying Point
(3) of our operational semantics using a clause in  $\{D_1,\ldots,
D_n\}$, is such that if for some goal $G_5$ we have that
$(A_0,G_0)\longmapsto_P^* G_5$, then $G_5$ is a basic goal, because
from a basic goal we cannot derive a non-basic one.
This concludes the proof of the Lemma in Case~(i).

The proof in Case~(ii) is analogous to that of the last part of Case~(i).
\end{proof}

Now we give the proof of Proposition~\ref{mutual_excl}.

\begin{proof}
Take a non-basic atom $A$ which satisfies $M$. Every non-basic atom
$A_0$ such that $A \longmapsto_P^* (A_0,G_0)$ for some goal $G_0$,
satisfies $M$ because $P$ satisfies $M$. Since $P$ is linear, $G_0$
is a basic goal.
By Lemma~\ref{at_most_one} there exists at most one
goal  $(A_1,G_1)$ where $A_1$ is a non-basic atom,
such that $(A_0,G_0)\Rightarrow_P (A_1,G_1)$.
Thus, there exists at most one non-unit
clause $C$  in $P$
such that $(A_0,G_0)\Rightarrow_C (A_1,G_1)$.
This means that $P$ is
semideterministic \wrt $M$.
\end{proof}

\section*{Appendix C. Proof of
Proposition~\ref{thm:termination_of_partition}}

\begin{proof}
It is enough to show that the {\rm while-do}
statement in the Partition procedure terminates. To see this, let us
first consider the set ${\it NonunitCls}_{in}$ which is the value of
the set ${\it NonunitCls}$ at the beginning of the execution of the
{\rm while-do} statement. ${\it NonunitCls}_{in}$ can be partitioned
into maximal sets of clauses such that: (i) two clauses which belong
to two distinct sets, are mutually exclusive, and (ii) if two clauses,
say $C_0$ and $C_{n+1}$, belong to the same set, then there exists a
sequence of clauses $C_0, C_1,\ldots,C_{n+1}$, with $n\!\geq\!0$, such that
for $i=0,\ldots,n$, clauses $C_i$ and $C_{i+1}$ are not mutually
exclusive.

{F}or our termination proof it is enough to show the
termination of the Partition procedure when starting from exactly one
maximal set, say $K$, of the partition of ${\it NonunitCls}_{in}$. This is
the case because during the execution of the Partition procedure, the
replacement of a clause, say $C_2$, by the clauses, say $C_{21}$ and
$C_{22}$, satisfies the following property: if clauses $C_2$ and $D$
are mutually exclusive then $C_{21}$ and $D$ are mutually exclusive
and also $C_{22}$ and $D$ are mutually exclusive.

Let every clause of $K$ be renamed apart and written
in a  form, called {\it equational form}, where the input arguments
are generalized to new variables and these new variables are bound by
equations in the body. The equational form of a clause $C$ will be
denoted by $C^{\it eq}$.
{F}or instance, given
the clause $C$: $p(f(X),r(Y,Y),r(X,U)) \leftarrow {\it Body}$, with
mode $p(+,+,?)$ for $p$, we have that $C^{\it eq}$ is: $p(V,W,r(X))
\leftarrow V\!=\!f(X), W\!=\!r(Y,Y), {\it Body}$.

Let $K^{\it eq}$ be the set $\{ C^{\it eq} \mid C \in K\}$.  Thus,
$K^{\it eq}$ has the following form:

\smallskip

~~~~~~~~$\left\{
\begin{array}{ll}
p(v_1,u_1) \leftarrow {\it Eqs}_1,
                 {\it Diseqs}_1, {\it Body}_1\\
  \hspace{.5cm} \cdots & \\
p(v_n,u_n) \leftarrow {\it Eqs}_n,
                 {\it Diseqs}_n,{\it Body}_n &
\end{array}
\right.$

\smallskip \noindent
where, for $i=0,\ldots,n$: (1)~$v_i$ denotes
a tuple of variables which are the input
arguments of $p$, (2)~$u_i$ denotes a tuple of arguments of $p$
which are {\it not\/} input arguments,
(3)~${\it Eqs}_i$ denotes a conjunction of equations of the form
$X\!=\!t$, which bind the variables in $v_i$, (4)~${\it Diseqs}_i$
denotes a conjunction of disequations, and (5)~${\it Body}_i$
denotes a conjunction of atoms which are different from disequations
(recall that the clauses in ${\it NonunitCls}_{in}$ are in normal
form). Equations may occur also in ${\it Body}_i$, but they do not
bind any input variable of $p(v_i,u_i)$.

Let us now introduce the following set $T = \{ t \mid t$ is a
term or a subterm occurring in ${\it Eqs}_i$ or ${\it Diseqs}_i$ for
some $i=1,\ldots,n\}$.

Every execution of the body of the while-do statement of the
Partition procedure works  by replacing a safe clause,
say $C_2$, by two new safe clauses, say $C_{21}$ and $C_{22}$.
We will prove the termination of the Partition procedure by: (i)
mapping the replacements it performs, onto the corresponding
replacements of the clauses written in equational form in the set
$K^{\it eq}$, and (ii) showing that the set $K^{\it eq}$ cannot
undergo an infinite number of such replacements.

Let us then consider the equational forms
$C^{\it eq}_{2}$, $C^{\it eq}_{21}$, and $C^{\it eq}_{22}$ of
the clauses $C_2$, $C_{21}$, and $C_{22}$, respectively. We have that:
(i) ${\it bd}(C^{\it eq}_{21})$ has one more equation of the form
$X\!=\!r$ w.r.t.~${\it bd}(C^{\it eq}_{2})$, and (ii) ${\it
bd}(C^{\it eq}_{22})$ has one more disequation of the form
$X\!\neq\!r$ w.r.t.~${\it bd}(C^{\it eq}_{2})$. We also have that
there exists only a finite number of pairs $\langle X,r\rangle$,
because $X$ is a variable symbol occurring in $K^{\it eq}$ and $r$ is
a term occurring in the finite set $T \cup \{ t \mid  t $ is a term or
a subterm occurring in an mgu of a {\it finite} number of elements of
$T \}$. (We have considered mgu's of a {\rm finite} number of elements
of $T$, rather than mgu's of two elements only, because a finite
number of clause heads in $K$ may have the same common instance.)

Thus, in order to conclude the proof, it remains to show that before
the replacement of $C_2$ by $C_{21}$ and $C_{22}$, neither $X\!=\!r$
nor $X\!\neq\!r$ occurs in ${\it bd}(C^{\it eq}_2)$. Here and in the
rest of the proof, the notion of occurrence of an equation or
a disequation is modulo renaming of the local variables. Indeed,
\\
$-$ in Case (1):
(1.1) $X\!\neq\!r$ does not occur in ${\it bd}(C^{\it
eq}_2)$ because $X/r$ is a binding of an mgu of the input
arguments of ${\it hd}(C_1)$ and ${\it hd}(C_2)$, and clauses $C_1$
and $C_2$ are not mutually exclusive, and thus, $X\!\neq\!r$
does not occur in ${\it bd}(C_2)$,
and (1.2) $X\!=\!r$ does not
occur in ${\it bd}(C^{\it eq}_2)$ because $X/r$ is, by construction, a
binding of an mgu between the input arguments of the heads of the
clauses $C_1$ and $C_2$ and these
clauses are obtained as a result of the {\it Simplify\/} function
which eliminates every occurrence of the variable $X$ from $C_2$, and
\\
$-$ in Case (2):
(2.1) $X\!=\!r$ does not occur in ${\it bd}(C^{\it eq}_2)$ because,
by hypothesis, a variant
of $X\!\neq\!r$ occurs in ${\it bd}(C_1)$ and clauses $C_1$ and $C_2$
are not mutually exclusive, and (2.2)
$X\!\neq\!r$ does not occur in ${\it bd}(C^{\it eq}_2)$  because
$X\!\neq\!r$ does not occur in ${\it bd}(C_2)$ (indeed, we
choose $X\!\neq\!r$ precisely to satisfy this condition).\end{proof}

\section*{Acknowledgments}

We would like to thank D. De Schreye, S. Etalle, J. Gallagher,
R. Gl\"{u}ck, N. D. Jones, M. Leuschel,
B. Martens, and M. H. S{\o}rensen for stimulating discussions about
partial deduction and logic program specialization.
We also acknowledge very constructive and useful comments by the anonymous
referees.
This work has been partially supported by the EC under the
HCM Project `Logic Program Synthesis and Transformation'
and the Italian Ministry for Education, University, and Research.

%\bibliographystyle{plain}
%\bibliography{Transformation}

\end{document}